\documentclass[pre,superscriptaddress,preprintnumbers,amsmath,amssymb]{revtex4-1}
\usepackage{graphicx,color}
\usepackage{bm,amssymb}
\usepackage{epsfig}
\usepackage{xcolor}
\usepackage{comment}
\usepackage{pgfplots}
\usepackage{enumerate}
\usepackage{bm}

\usepackage{enumerate}

\newcommand{\be}{\begin{equation}}
\newcommand{\ee}{\end{equation}}
\newcommand{\bea}{\begin{eqnarray}}
\newcommand{\eea}{\end{eqnarray}}

\begin{document}
\title{Condensation transition in the late-time position of a Run-and-Tumble particle}

\author{Francesco Mori }
\affiliation{LPTMS, CNRS, Univ. Paris-Sud, Universit\'e Paris-Saclay, 91405 Orsay, France}
  \author{Pierre Le Doussal}
  \affiliation{Laboratoire de Physique de l'Ecole Normale Sup\'erieure, PSL University, CNRS, Sorbonne Universit\'es, 24 rue Lhomond, 75231 Paris, France}
\author{Satya N. Majumdar}
\affiliation{LPTMS, CNRS, Univ. Paris-Sud, Universit\'e Paris-Saclay, 91405 Orsay, France}
\author{Gr\'egory Schehr }
\affiliation{Sorbonne Universit\'e, Laboratoire de Physique Th\'eorique et Hautes Energies, CNRS, UMR 7589, 4 Place Jussieu, 75252 Paris Cedex 05, France}
\date{\today}
\begin{abstract}

We study the position distribution $P(\vec{R},N)$ 
of a run-and-tumble particle (RTP) in arbitrary dimension $d$,
after $N$ runs. We assume 
that the constant speed $v>0$ of the particle during each running phase 
is independently drawn from a probability distribution $W(v)$ and that 
the direction of the particle is chosen isotropically after each tumbling.
The position distribution is clearly isotropic,
$P(\vec{R},N)\to P(R,N)$ where $R=|\vec{R}|$. 
We show that, under certain conditions on $d$ and 
$W(v)$ and for large $N$, a 
condensation transition occurs at some critical value of $R=R_c\sim O(N)$
located in the large deviation regime of $P(R,N)$.
For $R<R_c$ (subcritical fluid phase), all runs are roughly
of the same size in a typical trajectory. In contrast, an RTP trajectory with $R>R_c$ is typically dominated by a `condensate', i.e., a
large single run
that subsumes a finite fraction of the total displacement  (supercritical condensed phase). 
Focusing on the family of speed distributions 
$W(v)=\alpha(1-v/v_0)^{\alpha-1}/v_0$, parametrized by $\alpha>0$, we 
show that, for large $N$, $P(R,N)\sim \exp\left[-N\psi_{d,\alpha}(R/N)\right]$ and we compute exactly the rate function $\psi_{d,\alpha}(z)$ for 
any $d$ and $\alpha$. We show that the transition manifests itself as 
a singularity of this rate function at $R=R_c$ and that its order depends 
continuously on $d$ and $\alpha$. We 
also compute
the distribution of the condensate size for $R>R_c$.
Finally, we study the model when the total duration $T$ of the RTP, 
instead of the total number of runs, is fixed. 
Our analytical predictions are confirmed by numerical simulations, 
performed using a constrained Markov chain Monte Carlo technique, with 
precision $\sim 10^{-100}$.

\end{abstract}
\maketitle

\section{Introduction}

In recent years there has been a surge of interest in the study of
simple stochastic models of self-propelled particles in the
context of active matter, 
both theoretically and experimentally, 
for reviews see~\cite{cates12,soft,BDL16,Ramaswamy2017,Marchetti2018}.
This class of stochastic models can describe a wide range of 
artificial and natural 
systems, e.g., vibrated granular matter \cite{WWS17}, active gels 
\cite{R10,NVG19}, bacterial motion~\cite{berg_book,cates12,TC08}, 
animal movements~\cite{R10,VCB95,HBS04,VZ12} etc. At variance with its passive counterpart 
(for instance the standard Brownian motion, whose movement is driven by 
the random collisions with the surrounding fluid), active particles can 
absorb energy directly from the environment and convert it into persistent 
self-propelled motion. As a result, active motion 
violates time-reversal symmetry and 
these models belong to the category of out-of-equilibrium 
stochastic processes. In 
order to describe theoretically the persistence of the particle motion, 
one needs to introduce in the model a stochastic noise with non-vanishing 
time correlations. This can be done in several ways. For instance, in 
the active Ornstein-Uhlenbeck (AOU) model, the noise is chosen to be a 
Ornstein-Uhlenbeck process whose temporal correlation decays exponentially 
with time~\cite{FNC16,Bonilla19,SRG2019,WKL20}. Another possibility is 
to include the noise in the 
rotational degree of freedom of the particle, as done for the active 
Brownian particle (ABP) 
model~\cite{BDL16,seifert16,Franosch18,BMR18,BMR19,Limmer18,SDC20,MM20,SBS21},
where the orientation angle of the particle itself performs
a Brownian motion. 
Finally,
yet another variant is the so called  
run-and-tumble particle (RTP) model~\cite{kac74,W02,TC08,cates12}, where the 
active particle is driven by a telegraphic noise with exponential time 
correlations. In this paper, we will focus on this latter version, i.e.,
the RTP model.

\begin{figure}
\includegraphics[scale=1.0]{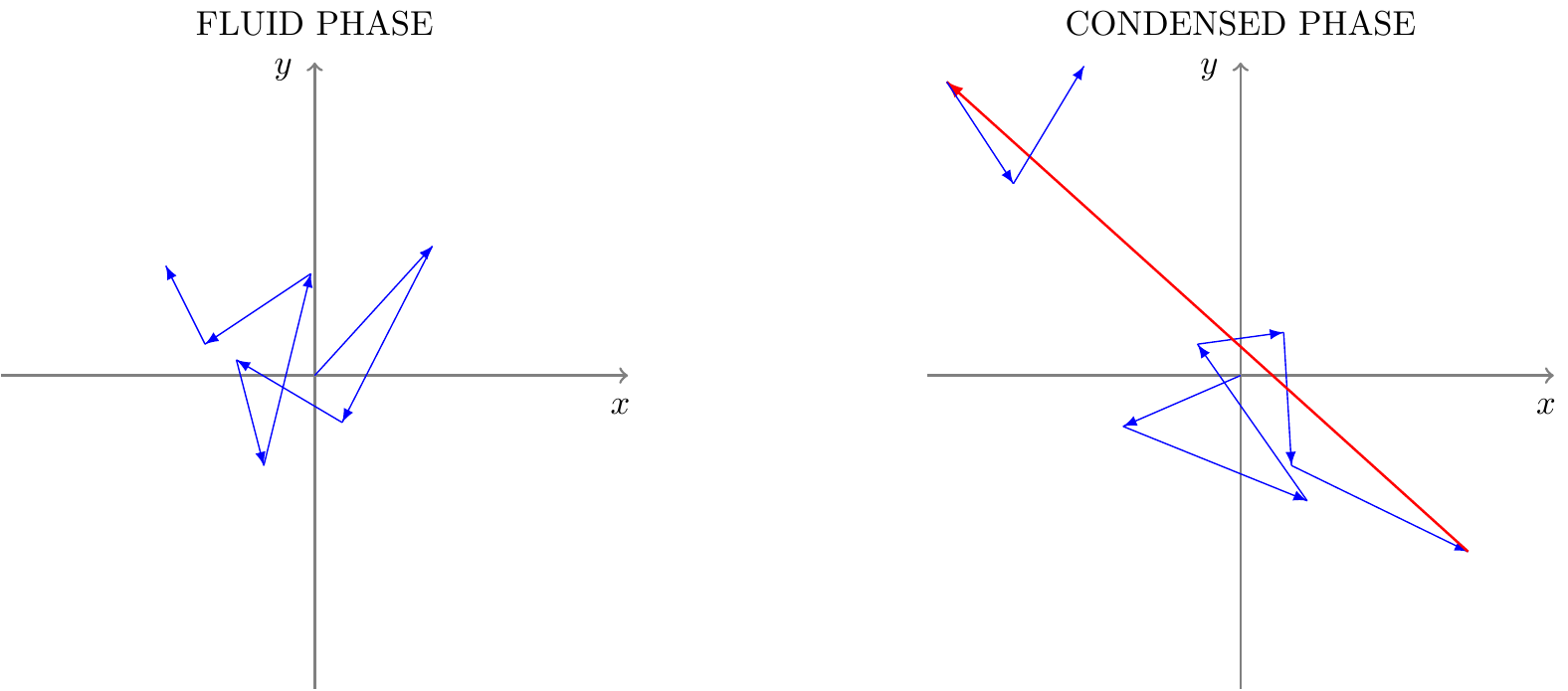}

\caption{{\bf Left panel:} 
Typical trajectory of a run-and-tumble particle (RTP) in two dimensions 
in the fluid phase. The particle starts at the origin, it chooses a 
direction uniformly at random and starts moving in that direction with 
constant speed $v_1$, drawn from the probability density function 
$W(v)$. After some random time, the particle tumbles, i.e., it changes 
its orientation at random and it chooses a new velocity $v_2$, drawn 
independently from $W(v)$. Then, it continues to move ballistically in 
this new direction, until it tumbles again, and so on. The different 
runs contribute to the displacement by roughly the same amount. {\bf 
Right panel:} Typical trajectory of an RTP in the condensed phase. One 
single run (colored in red) dominates the trajectory.}
\label{fig:RTP}
\end{figure}

Originally known as the persistent random walk~\cite{kac74,S87,Orshinger90,W02,HJ95,ML17}, the RTP 
model has been employed in recent years to describe the motion of a class of 
bacteria, e.g. \emph{E. coli}~\cite{berg_book,cates12,TC08,CT15,Solon15}, which 
typically move alternating between phases of straight motion with 
constant velocity (runs) and almost instantaneous changes of direction 
(tumblings), as shown in Fig.~\ref{fig:RTP}. This model is known to 
exhibit complex and interesting features not just in the many-particle
setting with 
interactions~\cite{cates12,TC08,BDL16,CT15,Solon15,SEB16}, but
even at the single-particle 
level~\cite{MAD12,Angelani15,MJK18,DM18,EM18,DKM19,GM2019,
SAP19,LMS19,SK19,MLDM20a,MLDM20,DDK20,HMS20,BMRS20,Bressloff20,SBS20,LM20,PTV2020,DMS21,BMS2021}.

In the single-particle case, the RTP model can be described as follows. 
The particle starts initially from the origin in
a $d$-dimensional continuous space. It chooses a 
direction isotropically and a speed $v_1>0$, drawn from the probability 
density function (PDF) $W(v)$, and starts moving in that direction 
ballistically with speed $v_1$. After a random time $\tau_1$, which is 
exponentially distributed with rate $\gamma$, the particle tumbles, 
i.e., it chooses a new random direction, and starts moving in the new 
direction with the new speed $v_2$, independently drawn from $W(v)$. 
Then, after running during an exponentially distributed time $\tau_2$, 
it tumbles 
again, and so on (see Fig. \ref{fig:RTP}). One can either observe the 
trajectory for a fixed duration $T$ (\emph{fixed-$T$ ensemble}) or wait 
until the particle undergoes exactly $N$ complete running phases 
(\emph{fixed-$N$ ensemble}). Even if at short times these two ensembles 
are quite different, it is reasonable to expect that they display 
similar behaviors when both $T$ and $N$ are large. The RTP dynamics is 
thus parametrized by three quantities: (i) the tumbling rate $\gamma$ 
that characterizes the time scale (the motion persists in a given
direction during a typical time $\gamma^{-1}$), (ii) the spatial
dimension $d$ in which the RTP lives and (iii) the speed
distribution $W(v)$ which is 
normalized to unity, i.e., $\int_0^{\infty} W(v)\, dv=1$.
Note that in the canonical and perhaps the most well studied
RTP model, the speed of the particle is a constant $v_0>0$ 
and does not vary from one run to another, 
corresponding to the choice $W(v)=\delta(v-v_0)$. 
Nevertheless, RTP models with 
generic $W(v)$ have also been studied~\cite{GM2019,MLDM20,MLDM20a,BMS2021}.

One of the simplest natural questions that one can ask about a self-propelled
active particle is: how does the position distribution $P(\vec R, T)$
evolve with time $T$? Here $T$ stands either for the real time $T$ or
the number of steps $N$, e.g., in the fixed-$N$ RTP model. While
for the AOU model, the position distribution is trivially Gaussian
at all times since the driving noise is Gaussian, for the
other two models ABP and RTP, the PDF $P(\vec R, T)$ is nontrivial.
For times $T\ll T^*$, where $T^*$ is the persistence time of the
driving noise (e.g.,
$T^*=\gamma^{-1}$ in RTP), the noise correlation plays a stronger role.
A typical manifestation of this, for instance, in the ABP model starting
from an anistropic initial condition, is that the position
distribution at short times remains strongly anisotropic--a
signature of activity of the process at short times~\cite{BMR18,MM20}. 
However, for times $T\gg T^*$, the diffusion
takes over and the particle behaves more like a Brownian motion at
late times. As a result, the process becomes more and more isotropic
as time progresses beyond $T^*$, i.e., $P(\vec R, T)\to P(R, T)$,
where $R= |\vec R|$.
Moreover, due to its convergence to a Brownian motion via
the central limit theorem (CLT),
this position distribution $P(R,T)$ 
has a Gaussian shape near its peak 
at late times~\cite{cates12}. Since the anisotropy
in the position distribution is lost at late times, one can
ask: is there any other remnant signature of `activity' in the
position distribution $P(R, T)$ at lates times $T\gg T^*$? 
It turns out that
indeed one can still find signatures of activity
in $P(R,T)$ at late times, but one needs to investigate the atypical
{\em large deviation tails} of $P(R,T)$, thus going beyond the Gaussian shape
near the peak. The non-Gaussian large deviation tails
of $P(R,T)$ at late times has been computed both
in the ABP model~\cite{seifert16,BMR19} and in a class
of RTP models~\cite{GM2019,PTV2020,DMS21}. In both cases,
the rate functions characterizing the large deviation behavior
were found to carry clear signatures of activity at late times.
Thus, to detect the signature of activity of the particle at late times, one needs to
investigate the rare events where the particle is far away from its
starting point. A relevant motivation to study such rare events is that
many biological phenomena, e.g., insemination, occur when a single
active particle reaches for the first time a faraway target. Let us
remark in passing that
another method to detect the signature of activity of a particle 
at late times is to confine it in an external potential--
the resulting stationary state position distribution
is highly non-Boltzmann and carries the signatures
of activity~\cite{MJK18,DKM19,TDV16,DD19,LDMS20}.

The position distribution of an RTP in the canonical model $W(v)=\delta(v-v_0)$ was first 
computed in Ref. \cite{S87} in two dimensions. Later, in Ref. \cite{MAD12}, this result was extended to arbitrary dimension $d$. 
However, these authors did not investigate the large-deviation regime, 
which was first studied in detail in Ref. \cite{PTV2020}. Remarkably, it 
was observed that in dimensions $d>5$ and with speed distribution 
$W(v)=\delta(v-v_0)$, the system undergoes a dynamical phase transition 
as one increases the total displacement $R$ of the particle. This turns 
out to be a \emph{condensation} transition, in the sense that above a 
certain distance $R$ from the origin, the total displacement of the 
particle is dominated by a single very long run (see the right panel in 
Fig. \ref{fig:RTP}). Moreover, in \cite{GM2019}, a similar condensation 
transition was observed for a one-dimensional RTP with a half-Gaussian 
speed distribution $W(v)=\sqrt{2/\pi}\, e^{-v^2/2}\, \theta(v)$ 
(where $\theta(v)$ is the Heaviside step function), 
when the particle is driven by a constant force. In 
both cases, the transition occurs in $P(R,T)$ by
varying the total displacement $R$ beyond a critical value $R_c$ (typically
of $O(T)$)--thus
the total distance $R$ plays the role of a control parameter.
These two examples 
suggest that condensation could be a general feature of the RTP model.
Unfortunately, in the canonical RTP model with fixed speed $v_0$,
this condensation occurs only in $d>5$, which is clearly not
accessible physically. The motivation behind our present work
is to investigate if it is possible to observe this interesting
condensation transition in $P(R,T)$ in a physically accessible
dimension, e.g., in $d=1, 2$ or $3$. 
One of our main results in this paper is to show that indeed this 
can be achieved by appropriately
choosing the speed distribution $W(v)$.

Traditionally condensation transition is well known to occur 
in the momentum/energy space, e.g., the Bose-Einstein condensation
in an ideal Bose gas in $d>2$ where a macroscopic number of particles
condense in the single particle ground state below a critical
temperature.  
However, condensation transition has
also been observed to occur even in real space in a variety
of situations--for reviews see \cite{EH05,M2008}. These include
traffic models~\cite{Krug91,Evans96,OEC98}, models of diffusion, aggregation
and fragmentation~\cite{MKB98,MKB2000},
mass transport models such as 
Zero Range type processes~\cite{MEZ2005,EMZ06,EHM06,EM08,EMPT10,SEM2014,SEM2014b,SEM2016,GB2017},  
macroeconomic models~\cite{BJJ2002}, 
network models~\cite{DMS03}, discrete nonlinear 
Shr\"odinger equation~\cite{RCK2000,GIL19}, financial models \cite{FZV13}, amongst other
examples. If the parameters in these models are chosen appropriately, a
condensation transition may occur upon increasing a control parameter
such as the density of particles. Beyond a critical density, typically
a single condensate forms in real space that contains a finite fraction of the total number
of particles.
For example,
in the context of traffic models the analogue of the condensate is a
traffic jam, while in the context of random network models,
the condensate is a single node that
captures macroscopic number of connections.
In the RTP model studied here, the condensate is a single large run
whose duration is a finite fraction of the total run time.
Thus the RTP condensation 
provides yet another example of this phenomenon
of real-space condensation.

The condensation transition that we demonstrate in the RTP model here
also has implication in a broader context, namely in the classical
problem in the probability theory concerning the distribution
of the sum of a large number of independent and identically
distributed (i.i.d.) random variables~\cite{Feller_book}.
To establish this connection, consider the \emph{fixed-$N$ ensemble}
RTP model in $d$-dimensions
defined above with a given tumbling rate $\gamma$ and a speed distribution
$W(v)$. Since the direction after each tumbling is chosen isotropically, the position distribution
$P(\vec R, N)\equiv P(R,N)$ is clearly isotropic, i.e., it depends only
on the total distance $R$ of the particle after $N$ steps, but not
on its direction. Note that, for simplicity, we use the same notation for $P(R,N)$, in the fixed-$N$ ensemble, and $P(R,T)$, in the fixed-$T$ ensemble. It is then convenient to
study the probability distribution $Z(X,N)$ of the
total displacement $X$ in any one of the directions 
(say for instance the $x$-direction). 
Since $X$ is the $x$-component of $\vec{R}$, 
it is easy to show that $Z(X,N)$ and $P(R,N)$ are 
simply related (see Appendix \ref{app:map}).
Let $f_d(z)$ denote the probability distribution of the $x$-component
of a random unit vector in $d$-dimensions. This can
be very simply computed (see Eq. \eqref{fd}).
Consequently, given a random vector of fixed magnitude $R=|\vec R|$,
its $X$ component has the distribution $(1/R)\, f_d(X/R)$.
Finally, if $R$ itself is distributed isotropically according 
to $P(\vec R,N)\equiv P(R,N)$, it follows that
\begin{equation}
Z(X,N) = \int_{\mathbb{R}^d}d\vec{R}~ \frac{1}{R} 
f_d\left(\frac{X}{R}\right) P(R, N) \,,
\label{relation}
\end{equation}
where $d \vec R= S_d\, R^{d-1}\, dR$ with 
$S_d= 2\pi^{d/2}/\Gamma(d/2)$ denoting the
surface area of a $d$-dimensional unit sphere. Note that $Z(X,N)$ is a probability distribution and is normalized to unity, 
\begin{equation}
\int_{-\infty}^{\infty}dX~Z(X,N)=1\,.
\label{Z_norm}
\end{equation}
The notation $Z(X,N)$ for a probability distribution may seem a bit strange at first sight. 
The reason for this choice comes from the analogy to the mass-transport models 
(see the discussion later), where $Z(X,N)$ also plays the role of a partition function. 
Hence, we stick to this somewhat unfamilar notation $Z(X,N)$.

In the limit of large $N$, we expect 
that the position distribution will exhibit the
large deviation behavior,
$P(R,N)\sim \exp\left[- N \psi(R/N)\right]$ where 
$\psi(z)$ is the associated rate function. 
Then, using Eq. \eqref{relation}, it is easy to show that
$Z(X,N)\sim \exp\left[- N \psi(X/N)\right]$, i.e., 
both $P(R,N)$ and $Z(X,N)$ share the same rate function 
$\psi(z)$ (see Appendix \ref{app:map}). 
To compute the rate function $\psi(z)$ it is more convenient 
to consider the large deviation behavior of $Z(X,N)$ and 
in this paper we will follow this route.
Now, denoting by $x_i$ the $x$-component displacement of the particle 
during the $i$-th run, one sees that
\begin{equation}
Z(X,N)=\int_{-\infty}^{\infty}dx_1\,\ldots\int_{-\infty}^{\infty}dx_N 
~\left[\prod_{i=1}^{N}p(x_i)\right]\delta\left(X-\sum_{i=1}^{N}x_i\right)\,,
\label{ZXN}
\end{equation}
where $p(x)$ denotes the PDF of the $x$-component of a single
run-vector and we have used the fact that the run-vectors are
statistically independent. The
delta function in Eq. (\ref{ZXN}) just enforces the total $x$-displacement
after $N$ steps to be $X$. Clearly $p(x)$ is symmetric around $x=0$.
The dependence on the parameters $d$, $\gamma$ and $W(v)$ is encoded in $p(x)$ (see Eq. (\ref{px})). Since $p(x)$ is normalized to unity, $Z(X,N)$ in Eq. \eqref{ZXN} manifestly satisfies the normalization condition in Eq. \eqref{Z_norm}. Thus, $Z(X,N)$ in
Eq. \eqref{ZXN} can be interpreted as the distribution of the sum of $N$ i.i.d.
random variables each drawn from a symmetric $p(x)$. 
This classical problem is well studied in the probability
literature~\cite{Feller_book}.
In particular, it is well known that, when the 
second moment of $p(x)$ is finite, $Z(X,N)$ has a Gaussian shape for 
$|X|\sim O(\sqrt{N})$ (typical fluctuation), as a consequence of the CLT. 
On the other hand, 
when $|X|\gg N$ (atypically large fluctuation), 
one obtains $Z(X,N)\sim N p(X)$, corresponding to a randomly 
chosen variable that dominates the sum~\cite{Feller_book}. 
However, it is not completely 
understood how the crossover between these `typical'
and `atypical' regimes occurs in $Z(X,N)$, 
as the `control parameter' $X$ varies. Given a $p(x)$,
is there a `sharp' phase transition at some critical value $X_c$,
or is this just a smooth crossover? 
While for a few specific examples of $p(x)$
this crossover between the typical and the atypical regimes
have been studied~\cite{nagaev}, a general criterion on $p(x)$ to
determine whether a sharp phase transition occurs is still missing.
Our analysis of the large deviation properties of the RTP
model with a general speed distribution $W(v)$ (and hence that
of $p(x)$) thus sheds light on this general question as well.

In this context, let us remark that such a criterion 
is well established when the i.i.d. random variables are
all positive, i.e., $p(x)$ has only positive support.
This situation arises in a class of mass transport models
defined on a lattice of $N$ sites with some
prescribed rates of mass transfer between neighbouring 
sites~\cite{EH05,M2008,MEZ2005,EMZ06,EHM06}.
Here $x_i\ge 0$ denotes the mass at site $i$ and the dynamics
conserves the total mass $X=\sum_{i=1}^N x_i$.
For a large class of mass transfer rates, the system
reaches at long times a stationary state where the
joint distribution of masses $\{x_i\}$ factorise, with
$p(x)$ denoting each factor that depends on the mass transfer rates~\cite{EMZ2004}.
Then,
$Z(X,N)$ in Eq. (\ref{ZXN}) just denotes the
partition function in the stationary state.
In this case where $p(x)$ has only positive support ($x$ being 
a mass), it has been shown that the criterion
for condensation depends on the tail of $p(x)$ 
for large $x$~\cite{EH05,M2008,MEZ2005,EMZ06,EM08,FZV13}.
As one varies the sum $X$, the
condensation occurs 
at some critical value $X_c\sim O(N)$, if and only if
$e^{-c\, x}<p(x)<1/x^2$ as $x\to \infty$, where $c$ is any
positive constant. For example, if $p(x)$ has a fat tail,
$p(x)\sim x^{-\gamma}$
for large $x$ with $\gamma>2$, a condensation will occur.
Similarly, if $p(x) \sim \exp[-a\, x^{\alpha}]$ for large $x$ with $a>0$
and $0<\alpha<1$ (stretched-exponential), again condensation will
occur~\cite{EMZ06}. However, if $p(x)\sim \exp[-a\, x^\alpha]$ for large $x$
with $a>0$ and $\alpha>1$, there is no condensation transition but only
a smooth crossover as $X$ varies.
In our problem, the variable $x_i$'s can be both positive and negative
with $p(x)$ symmetric, and unfortunately
we can not simply apply the same criterion
that is valid only for positive random variables. However, by generalising
the method used in Ref.~\cite{EMZ06}, we show
that it is possible to find a similar criterion
for symmetric random variables as well.

\vskip 0.3cm

\noindent Our main results in this paper are threefold:
\begin{enumerate}[(I)]
\item We identify a general criterion for condensation, 
valid for the sum of random variables with symmetric distribution $p(x)$. 
In the context of the RTP model, we show that, by properly tuning the speed 
distribution $W(v)$, one can observe a condensation transition also 
in a physically accessible dimension $d\leq 3$.

\item We focus on a family of speed distributions 
$W(v)=(\alpha/v_0)(1-v/v_0)^{\alpha}\, \theta(v_0-v)$ supported
over $v\in [0,v_0]$ and 
parametrized by $\alpha$, that allows a condensation transition
according to the general criterion mentioned above. 
For this family of $W(v)$,
we compute exactly the position distribution 
$Z(X,N)$ for large $N$ (see Fig. \ref{fig:Z}).  
In the regime where $X\sim O(N)$, we show that $Z(X,N)$ exhibits a 
large-deviation form $Z(X,N)\sim \exp[-N\, \psi(X/N)]$ 
and we compute the associated rate function $\psi(z)$. 
As the control parameter $X$ exceeds a critical value $X_c=z_c\, N$,
we show that a condensation transition occurs. The signature
of this transition is manifest in the rate function $\psi(z)$: it
develops a singularity at $z=z_c$.

\item For any $\alpha$ and $d$, 
we also compute the marginal distribution 
$p(x|X)$ of a single-run displacement, 
conditioned on the total displacement $X$ (see Fig. \ref{fig:marginal}). 
This marginal distribution $p(x|X)$ can be taken as a diagnostic
of the condensation transition, as it behaves very differently
in the `subcritical' ($X<X_c$) and the `supercritical' ($X>X_c$) phases.
We show that in the supercritical phase where $X>X_c$, 
a distinct bump appears in the tail of 
$p(x|X)$, similar to what has been observed in mass transport 
models~\cite{MEZ2005,EMZ06}.
\end{enumerate}

The rest of the paper is organized as follows. In Section 
\ref{sec:model} we present the details of the RTP model and provide a 
summary of our main results. In Section \ref{sec:criterion}, using a 
grand canonical description of the system, we present a general 
criterion for condensation, valid for a large class of RTP models. In 
Section \ref{sec:position}, we study the late-time position 
distribution, both in the typical and large-deviation regimes for the 
fixed-$N$ ensemble. We show that the phase transition manifests itself 
as a singularity of the rate function and we compute its order. To 
clarify the nature of the transition, in Section \ref{sec:marginal} we 
study the marginal probability of a single-run displacement. In Section 
\ref{sec:T}, we investigate the position distribution for the fixed-$T$ 
ensemble. In Section \ref{sec:numerics}, we present the details of the 
numerical simulations. Finally, in Section \ref{sec:conclusion} we 
conclude with a summary and some open questions. Some details of the 
computations are presented in the appendices.

\section{The model and the summary of the main results}
\label{sec:model}

Since the paper is long, it is useful to provide
a description of the model and a summary of the salient features
of the main results, so that the reader is not lost in the
details given in later sections. This is precisely the purpose
of this section, where we also direct the reader to specific
equations in later sections. 

We consider a single RTP, starting from the origin and moving in $d$ 
dimensions. At each tumbling the speed of the particle is independently 
drawn from the distribution $W(v)$. As anticipated in the introduction, 
there are two possible set-ups: the fixed-$N$ and the fixed-$T$ 
ensemble. Note that if the number $N$ of running phases is fixed, then 
the total time $T$ can fluctuate. Alternatively, in the fixed-$T$ 
ensemble one fixes the total time $T$, letting $N$ fluctuate. One 
important difference between the two models is that in the fixed-$T$ 
ensemble the last running phase is yet to be completed. Therefore, the 
displacement of the particle during the last running phases has a 
different distribution with respect to the previous displacements 
\cite{MLDM20,MLDM20a}. On the other hand, in the fixed-$N$ case, all 
displacements have the same distribution. For this reason, the analytic 
study of the fixed-$N$ ensemble is usually simpler. Since, as we shall 
see, the late-time properties of the two ensembles are very similar, we 
will focus on the fixed-$N$ ensemble for most of this paper. We will 
consider the fixed-$T$ ensemble in Section \ref{sec:T}, where we show 
that the behavior of the system is qualitatively similar for the two 
models.

Denoting by $x_1\,,\ldots\,,x_N$ the displacements in the $x$-direction of the RTP during the $N$ running phases, we have
\begin{equation}
X=\sum_{i=1}^{N}x_i\,.
\label{XN_sum}
\end{equation}
These increments $x_i$, that can be positive or negative, are i.i.d. random variables, 
drawn from the symmetric probability distribution \cite{MLDM20,MLDM20a} 
(reproduced, for convenience, in Appendix \ref{app:p(x)} in this paper)
\begin{equation}
p(x)=\int_{0}^{\infty}dv~\frac{1}{v}W(v)\int_{0}^{\infty}d\ell~\frac{1}{\ell}f_d\left(\frac{x_i}{\ell}\right)~\gamma e^{-\gamma \ell/v}\,,
\label{px}
\end{equation}
where
\begin{equation}
f_d(z)=\frac{\Gamma\left(d/2\right)}{\sqrt{\pi}\Gamma\left((d-1)/2\right)}(1-z^2)^{(d-3)/2} \theta(1-|z|)\,,
\label{fd}
\end{equation}
$\Gamma(y)$ is the Gamma function. 
It is easy to check that $p(x)$ is symmetric around $x=0$. 
The behavior of $Z(X,N)$ depends on the dimension $d$ and
the speed distribution $W(v)$ through $p(x)$ in Eq. (\ref{px}).
Since $p(x)$ is symmetric, $Z(X,N)$ is also symmetric,
and hence it is sufficient to focus on the positive side, i.e., for $X>0$. 
This PDF $Z(X,N)$ can be expressed explicitly as an $N$-fold
integral in terms of $p(x)$'s, as shown in  
Eq. \eqref{ZXN}. In this paper, we 
show that under specific conditions on $W(v)$ and $d$ the system 
undergoes a condensation phase transition at a critical value $X_c$ of the 
position $X$. For $X<X_c$ (subcritical phase), all the different runs 
$x_1\,,\ldots\,,x_N$ contribute to the total displacement by roughly the 
same amount. On the other hand, for $X>X_c$ (supercritical phase), 
a single run, which is 
referred to as the condensate, contributes to a macroscopic fraction of 
the displacement (see the right panel of Fig. \ref{fig:RTP}). Our goal 
is (I) to determine the criterion on $p(x)$ for the condensation
transition in $Z(X,N)$ as $X$ varies (II) when this criterion is satisfied,
to determine the specific value $X_c$ at 
which the system forms a condensate, and, (III) to study, for $X>X_c$, 
the nature of 
this condensate, e.g., what is the distribution of run lengths
carried by the condensate. 
The salient features of our results are highlighted below.

\vspace*{0.3cm}
\noindent{\bf (I) Criterion for condensation:} 
As in mass transport models where $p(x)$ only has positive support, 
we formulate a criterion 
for condensation in the case of symmetric $p(x)$. We show that this
criterion only depends on the large-$|x|$ behavior of 
$p(x)$ (see Section \ref{sec:criterion}). By choosing the speed 
distribution $W(v)$ appropriately, one can find $p(x)$'s that allow for 
condensation. In particular, we focus on the family of speed 
distributions

\begin{equation}
W(v)=\frac{\alpha}{v_0}\left(1-\frac{v}{v_0}\right)^{\alpha-1}\, \quad
{\rm where}\quad 0\le v\le v_0\, ,
\label{W}
\end{equation}
parametrized by $\alpha>0$. The constant $v_0>0$ represents the 
maximal speed that the particle can reach. Note that this family includes,
as a special case, the canonical RTP model where the speed
is constant from run to run. 
Indeed, by taking the limit $\alpha\to 0$ in Eq. \eqref{W}, one finds
\begin{equation}
W(v)=\delta(v-v_0).
\end{equation}

Moreover, many other relevant speed distributions belong to this class. 
For instance, choosing $\alpha=1$, one obtains the uniform speed 
distribution. Since one can always rescale space and time, 
without any loss of generality we set 
\begin{equation}
v_0=\gamma=1 
\label{vgamma_def}
\end{equation}
in the rest of the paper. Thus, our system is 
parametrized by the two scalars $d$ and $\alpha$. It turns out that 
several (but not all) properties of the condensation transition
depend only on the single parameter
\begin{equation}
\nu=\frac{(d+2\alpha-1)}{2}\, .
\label{nu_def}
\end{equation}
Indeed, applying the criterion for 
condensation to the speed distribution in Eq. \eqref{W}, we find that 
condensation occurs only for $\nu>2$.

\begin{figure}
\centering
\includegraphics[width=0.7\textwidth]{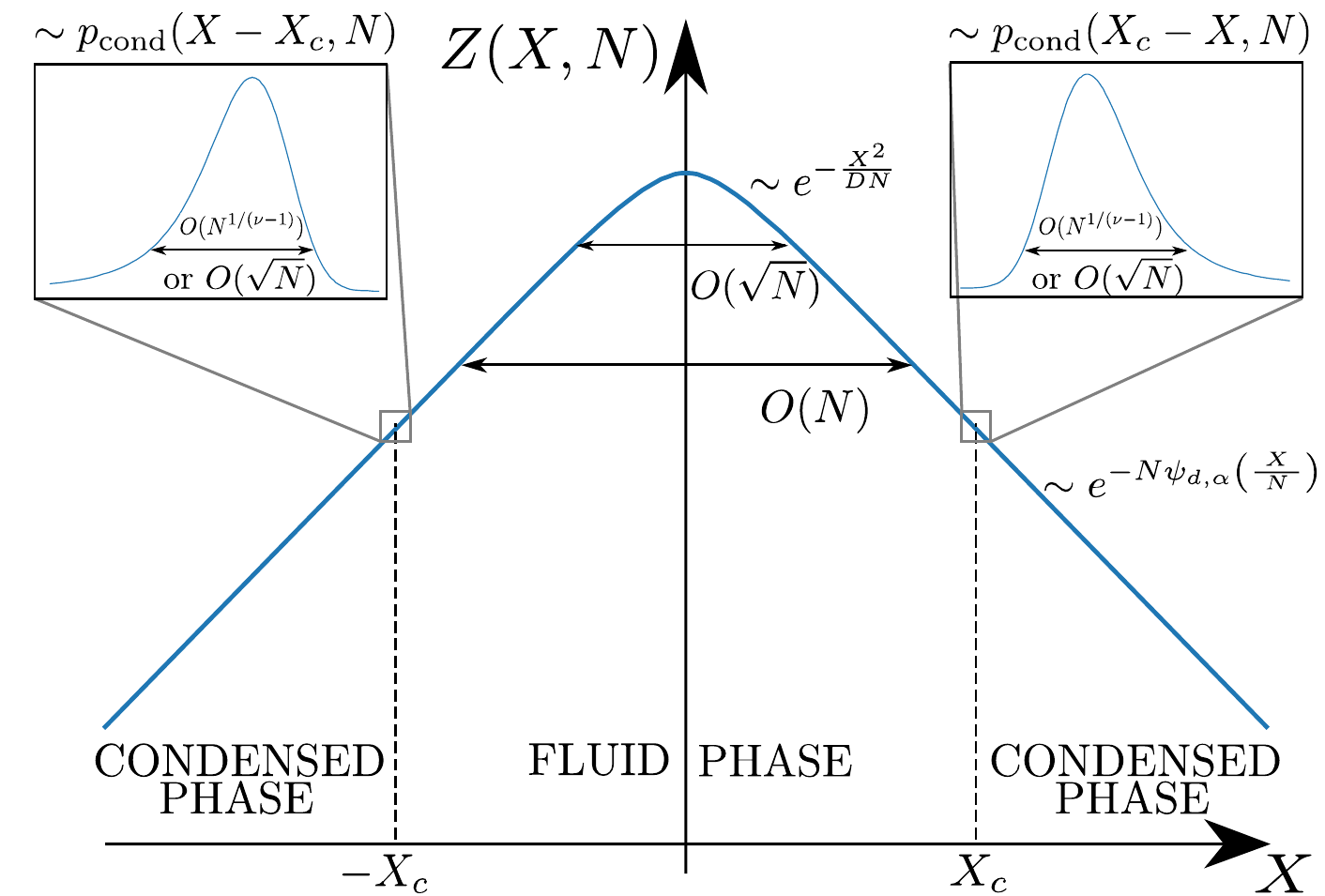} 
\caption{\label{fig:Z} Schematic representation of the PDF $Z(X,N)$, for $\nu=(d+2\alpha-1)/2>2$. For $X\sim O( \sqrt{N})$ the PDF $Z(X,N)$ is Gaussian, while for $X\sim O(N)$ it assumes the large-deviation form $Z(X,N)\sim e^{-N\psi_{d,\alpha}(X/N)}$, where the rate function $\psi_{d,\alpha}(z)$ is given in Eq. \eqref{psi_d_cond}. A dynamical phase transition occurs at $X_c\sim O(N)$, where $\psi_{d,\alpha}(z=X/N)$ is singular. In a small region around the critical point $X_c$, $Z(X,N)$ is described by the function $p_{\rm cond}(y,N)$ (see insets). For $2<\nu<3$, $p_{\rm cond}(y,N)$ has an anomalous shape, given in Eq. \eqref{p_cond_anomalous}, and it varies on a scale $O(N^{1/(\nu-1)})$. For $\nu>3$, $p_{\rm cond}(y,N)$ is Gaussian with fluctuations $\sim O(\sqrt{N})$. As a consequence of the symmetry $Z(X,N)=Z(-X,N)$, an analogous transition occurs also at $-X_c$. For $|X|<X_c$ the system is in the fluid phase, while for $|X|>X_c$ the system is in the condensed phase.}
\end{figure}

\vspace*{0.3cm}
\noindent{\bf (II) Position distribution $Z(X,N)$:} 
Thanks to the symmetry $Z(X,N)=Z(-X,N)$, it is sufficient to focus on 
the case $X>0$. In the late-time limit $N\gg 1$, we consider two 
distinct regimes. In the typical regime $X\sim O(\sqrt{N})$, we find 
the central limit behavior as expected
\begin{equation}
Z(X,N)\simeq \frac{1}{\sqrt{4\pi DN}}e^{-X^2/(4DN)}\,,
\label{typical}
\end{equation}
with
\begin{equation}
D=\frac{2}{d(\alpha+1)(\alpha+2)}\,.
\label{D}
\end{equation}

At this scale, no sign of activity is present. However, the signatures 
of the active nature of the particle can be observed in the tails of 
$Z(X,N)$, outside the typical Gaussian region. Indeed, in the 
atypical regime $X\sim O(N)$, we show that the PDF of $X$ admits the 
large-deviation form
\begin{equation}
Z(X,N)\sim \exp\left[-N ~\psi_{d,\alpha}\left(\frac{X}{ N}\right)\right]\,.
\label{large_deviation}
\end{equation}
The rate function $\psi_{d,\alpha}(z)$ depends on both
parameters $d$ and $\alpha$, and its exact expression
for any $d$ and $\alpha>0$ is given in Eq. \eqref{psi_d_z} for $\nu<2$, 
and in Eq. \eqref{psi_d_cond} for $\nu>2$. 
We will consider the scaled displacement $z=X/N$ as our control parameter. 
In particular, for $\nu<2$, $\psi_{d,\alpha}(z)$ is analytic for any $z>0$, 
while for $\nu>2$ it becomes singular at the critical point $z=z_c$.
The critical value $z_c$ also depends on both parameters $d$
and $\alpha$ and is given explicitly 
in Eq. \eqref{zc}. 
For $z>z_c$, the rate function becomes exactly linear. 
The non-analyticity of the rate function signals the presence of a 
dynamical phase transition at the critical position $X_c=z_c N$. 
This is equivalent to the non-analyticity of the free energy in the case of 
equilibrium phase transitions, with the rate function playing the
role of free energy. The free energy in equilibrium systems at the
critical point is characterized by the order of its non-analyticity.
The transition is of order $n$ 
if the $n$-th derivative of $\psi_{d,\alpha}(z)$ is discontinuous, 
while all the lower-order derivatives are continuous. 
In our model, we find that the order of the non-analyticity $n$ at the condensation
transition is given by
\begin{equation}
n=
\begin{cases}
\left\lceil\frac{\nu-1}{\nu-2}\right\rceil\,\,\, & \text{for } 2<\nu<3,\\
\\
 2 \,\,\,& \text{for } \nu>3\,,\\
\end{cases}
\label{order_intro}
\end{equation}
where $\left\lceil y\right\rceil$ denotes the smallest integer larger than or equal to $y$.
As a consequence of the $X\to-X$ symmetry of the process, 
an analogous transition occurs also at $-X_c$.

For $\nu>2$, we next zoom in
the region around the critical point $z=z_c$ and
investigate $Z(X,N)$ on a finer scale around $X=X_c$ (see Fig. \ref{fig:Z}).
By computing $Z(X,N)$ in the vicinity of $X_c=z_c\,N$,
we find that
\begin{equation}
\label{ZXN_critical}
Z(X,N)\simeq C_N~p_{\rm cond}\left(X_c-X,N\right)\,,
\end{equation}
where $C_N$ is a positive constant and the function
$p_{\rm cond}(y,N)$ depends on $N$, $\alpha$, and $d$.  
For $2<\nu<3$, we find that
\begin{equation}
p_{\rm cond}(y,N)\simeq  \frac{1}{N^{1/(\nu-1)}} V_{\nu}\left( \frac{y}{N^{1/(\nu-1)}}\right)\,,
\label{p_cond_anomalous}
\end{equation}
where the function $V_{\nu}(y)$ is given in Eq. \eqref{V} (see also
Fig. (\ref{fig:V}) for a plot of this function). 
On the other hand, for $\nu>3$, we obtain that, for $|y|\ll\sqrt{N\log(N)}$,
\begin{equation}
p_{\rm cond}(y,N)\simeq \frac{1}{\sqrt{4\pi a_{d,\alpha}N}} ~e^{-y^2/(4a_{d,\alpha}N)}\,,
\label{p_cond_normal}
\end{equation}
where $a_{d,\alpha}$ is a positive constant given in Eq. 
\eqref{a_d_alpha}. For $\nu>3$, the Gaussian shape in Eq. 
\eqref{p_cond_normal} is only valid for $|y|\ll\sqrt{N\log{N}}$. Outside 
this region, $p_{\rm cond}(y,N)$ has a power-law tail (see Eq. 
\eqref{def_p_cond_norm}). Adapting the same terminology as in mass
transport models~\cite{MEZ2005,EMZ06}, we will call
the condensate `anomalous' for $2<\nu<3$ and `normal' for $\nu>3$. 
Interestingly, as we will see later, the behavior of 
$Z(X,N)$ close to the critical point in Eq. (\ref{ZXN_critical})
also determines the size and the nature of the condensate that forms when
$X>X_c$.
More precisely, we show that the same function $p_{\rm cond}(y,N)$ that characterizes
$Z(X,N)$ near the critical point in Eq. (\ref{ZXN_critical}) and  which is 
positive and normalized to one, indeed also describes the size distribution
of the condensate, i.e., the probability distribution
of the run length carried by the condensate (hence the subscript in $p_{\rm cond}(y)$)
when the condensate forms.
A schematic representation of the different regimes of 
$Z(X,N)$ as a function of $X$ is shown in Fig. \ref{fig:Z}.

\begin{figure}
\centering
\includegraphics[width=0.7\textwidth]{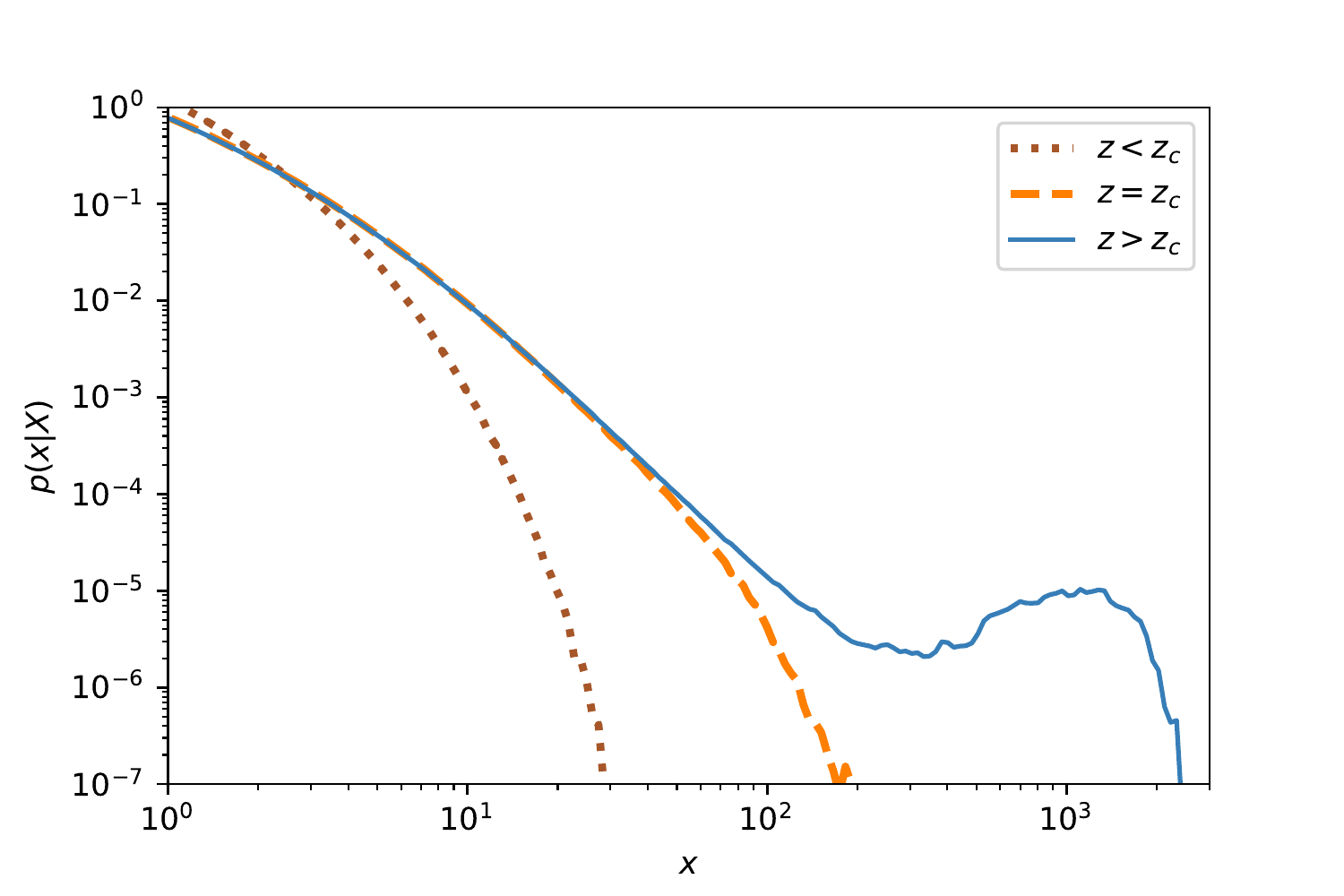} 
\caption{\label{fig:marginal} Numerical curves of the marginal probability $p(x|X)$ of a single-run displacement, for $\alpha=0$ and $d=8$. For $z<z_c$ (dotted brown line), the system is in the fluid phase and $p(x|X)$ decays exponentially fast for large $x$. At the critical point $z=z_c$ (orange dashed line), $p(x|X)$ develops a power-law tail. For $z>z_c$ (blue continuous line), the system is in the condensed phase and $p(x|X)$ still has a power-law tail and a condensate bump appears at $x=X_{\rm ex}$.
}
\end{figure}

\vspace*{0.3cm}

\noindent{\bf (III) Single-run marginal distribution $p(x|X)$:} 
To understand better the nature of the dynamical phase transition described 
above, it is useful to study the PDF of the single-run marginal distribution 
$p(x|X)$, conditioned on the total displacement $X$. This
is obtained by integrating the joint distribution of $\{x_i\}$'s
over $(N-1)$ variables, while keeping fixed the sum $X=\sum_{i=1}^N x_i$
and the value of one of them, say the first one, at $x_1=x$.
We consider only the case $\nu>2$, where the transition surely occurs.
This conditional distribution $p(x|X)$ can be taken as a clear
diagnostic for the condensation transition, since
it behaves very differently in the subcritical ($0<z<z_c$) and
supercritical ($z>z_c$) phases (see Fig. \ref{fig:marginal}).

\begin{figure}
\centering
\includegraphics[width=0.7\textwidth]{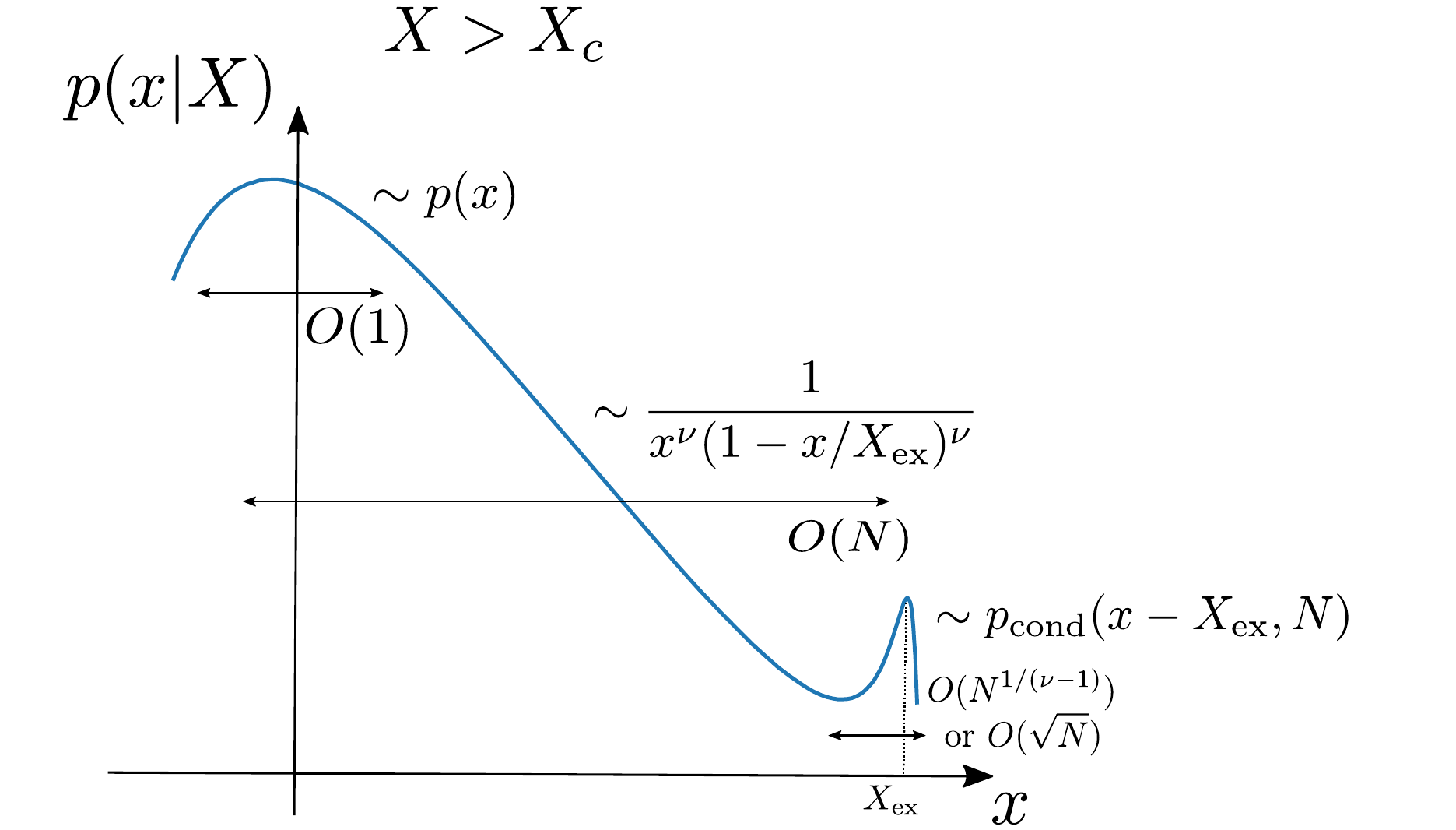}
\caption{\label{fig:scales} Qualitative behavior of the marginal
probability $p(x|X)$ versus $x$ in the condensate phase.
When $x\sim O(1)$, we find that $p(x|X)\sim p(x)$, 
where $p(x)$ is the distribution of a single displacement, 
given in Eq. \eqref{px}. 
For $1\ll x\ll N$, we find that
$p(x|X)\sim (x(1-x/X_{\rm ex}))^{-\nu}$, where $X_{\rm ex}=X-X_c$ and
$\nu=(d+2\alpha-1)/2$. At $x\simeq X_{\rm ex}\sim O(N)$,
a condensate bump appears in the tail of $p(x|X)$.
The shape $p_{\rm cond}(y,N)$ of the bump depends continuously on $\nu$.
For $2<\nu<3$, the condensate bump has an anomalous shape,
with fluctuations of order
$N^{1/(\nu-1)}$ (see Eq. \eqref{p_cond_anomalous}).
For $\nu>3$, the bump has a Gaussian shape with fluctuations of
order $\sqrt{N}$.
}
\end{figure}

\vskip 0.3cm

\noindent {\bf {Subcritical phase ($0<z<z_c$):}}
In this case, we show that $p(x|X)$ decreases monotonically
with increasing $x$ and for $x\gg 1$
\begin{equation}
p(x|X)\sim \frac{1}{x^{\nu}}e^{-x/\xi}\,,
\end{equation}
where $\xi>0$ depends on $z=X/N$. Thus, below the transition
$z<z_c$, the
marginal distribution $p(x|X)$ 
decays exponentially fast over a scale $\xi$. 
For this reason, all the displacements $x_1\,,\ldots x_N$ contribute 
``democratically'' to the total displacement $X$ and thus this
subcritical regime behaves like a 
\emph{fluid}. Notably, when $z\to z_c$ from below the 
typical length $\xi$ diverges.

\vskip 0.3cm

\noindent {\bf {Critical phase ($z=z_c$):}} 
Exactly at the critical point 
$z=z_c$, the conditional distribution still decays
monotonically with increasing $x$, but develops
a power-law tail for large $x$
\begin{equation}
p(x|X)\sim \frac{1}{x^{\nu}}\,, 
\label{crit.1}
\end{equation}
where we recall $\nu=(d+2\alpha-1)/2$.

\vskip 0.3cm

\noindent {\bf {Supercritical phase ($z>z_c$):}}
For $z>z_c$, the distribution $p(x|X)$ becomes a non-monotonic
function of $x$ (see Fig. \ref{fig:marginal}).
When $x\sim O(1)$, we show that $p(x|X)\sim p(x)$, i.e,
the conditioned distribution is insensitive to the constraint,
and behaves like a constraint-free system.
For $x\gg 1$, the function decays with increasing $x$ as a power law,
as at the critical point in Eq. (\ref{crit.1}). However, this
power law behavior ceases to hold when $x$ approaches
$X_{\rm ex}= X-z_c N>0$ (analogously to the excess mass
$M_{\rm ex}$ in the mass transport models~\cite{MEZ2005,EMZ06}).
For $1\ll x\ll X_{\rm ex}$ we get 
\begin{equation}
p(x|X)\simeq \frac{A_{d,\alpha}}{x^{\nu}}\frac{1}{(1-x/X_{\rm ex})^{\nu}}\,,
\label{regime2}
\end{equation}
where $A_{d,\alpha}>0$ is given in Eq. \eqref{A_d_alpha}.
This describes the shoulder region before the bump in 
Fig. \ref{fig:marginal} in the supercritical phase.
The approximate expression in Eq. \eqref{regime2} breaks down 
when $x\to X_{\rm ex}$. Indeed, at $x\sim X_{\rm ex}$, 
a bump appears in the tail of $p(x|X)$, where
\begin{equation}
p(x|X)=\frac1N p_{\rm cond}(x-X_{\rm ex},N)\,.
\end{equation}
Thus, the function $p_{\rm cond}(x-X_{\rm ex},N)$ describes the shape of the condensate. 
The bump is centered at $X_{\rm ex}$ and its width vanishes relative to its location
for large $N$ (see Fig. (\ref{fig:scales})). The area under this bump is the probability that a condensate appears in a particular single-run displacement. We find that
\begin{equation}
\int_{-\infty}^{\infty}dy~\frac{1}{N}\, p_{\rm cond}(y,N)=\frac{1}{N}\,,
\end{equation}
meaning that only one condensate appears in the system. 
We recall that, for $2<\nu<3$, $p_{\rm cond}(y,N)$ is given in Eq. 
\eqref{p_cond_anomalous} and the condensate has anomalous fluctuations 
of order $N^{1/(\nu-1)}$, where $1/2<1/(\nu-1)<1$. For this reason, we 
denote the phase $2<\nu<3$ as the anomalous condensate phase. On the 
other hand, for $\nu>3$, $p_{\rm cond}(y,N)$ is given in Eq. 
\eqref{p_cond_normal} and the bump has a normal shape around its peak, 
with fluctuations of order $\sqrt{N}$. Hence, we call this 
region the normal condensate phase. 
Note however that the Gaussian shape is valid only for 
$|y|\ll \sqrt{N \log{N}}$ and that outside this region, 
the bump has a power-law tail.

Finally, for $x\gg X_{\rm ex}$ and for any $\nu>2$, we observe that 
$p(x|X)$ gets cut-off around $X\sim O(N)$ (finite-size effect)
and this cut-off behavior can be described
by a large deviation form
\begin{equation}
p(x|X)\sim\exp\left[-N \chi\left(\frac{ x}{ N},\frac{ X}{ N}\right)\right]\,,
\end{equation}
where the rate function $\chi(y,z)>0$ is given in Eq. \eqref{Azy}. 
Thus, configurations where a single-run displacement is larger than 
$X_{\rm ex}$ become exponentially rare for large $N$. 

The qualitative behavior of $p(x|X)$ in the three phases (subcritical, 
critical and supercritical) is shown in Fig. \ref{fig:marginal}. In Fig. 
\ref{fig:scales}, we focus on the condensed phase $X>X_c$ and we present 
a schematic representation of the different regimes of $p(x|X)$ as a 
function of $x$.

\begin{figure}
\includegraphics[scale=1]{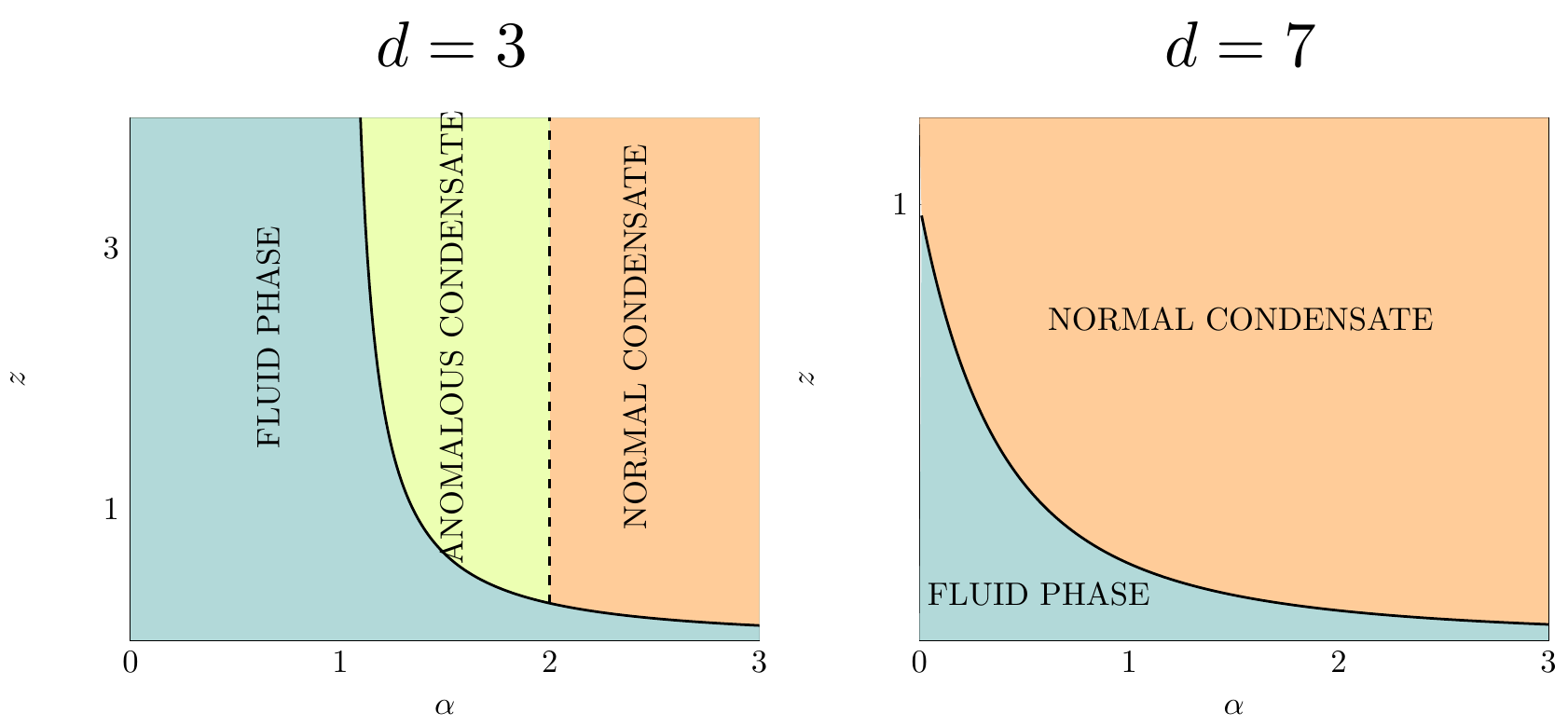} 
\caption{{\bf Left panel.}  Phase diagram in the $(\alpha,z)$ plane, for $d=3$. For $\alpha<1$ the system is always in the fluid phase, where all runs contribute by roughly the same amount to the total displacement of the particle. For $\alpha>1$, the system undergoes a dynamical condensation transition at a critical value $z_c$ of the control parameter $z$ (continuous black line). The exact expression of $z_c$ is given in Eq. \eqref{zc}. For $z<z_c$, the system is in the fluid phase, while above the transition the system is in the condensate phase. In particular, for $1<\alpha<2$, the system is in the anomalous condensate phase, while for $\alpha>2$ the system is in the normal condensate phase. {\bf Right panel.} Phase diagram in the $(\alpha,z)$ plane, for $d=7$. For $z<z_c$, the system is in the fluid phase, while for $z>z_c$ it is in the normal condensate phase. For $d\geq 7$, the system is never in the anomalous condensed phase.}
\label{fig:phase}
\end{figure}

\vskip 0.3cm

To sum up, we find that
\begin{itemize}
\item for $\nu<2$, the system is always in the fluid phase;
\item for $2<\nu<3$, the system is in the anomalous condensate phase for $X>X_c$ and the order of the transition depends continuously on $\nu$;
\item for $\nu>3$, the system is in the normal condensate phase for $X>X_c$ 
and the transition is of second order.
\end{itemize}
For the RTP model, we thus also find the two different 
types of condensed phases `anomalous' and `normal', 
as in the case of mass transport models~\cite{MEZ2005,EMZ06}. 
The behavior of the system is determined by three parameters: 
the two system parameters $(\alpha,d)$ and the control parameter 
$z=X/N$. This would correspond to a three-dimensional phase diagram, 
which is of course complicated to display.
For this reason, we present two different slices of the phase diagram. 
In the left panel of Fig. \ref{fig:phase}, we focus on the physical 
dimension $d=3$ and we show the $(\alpha,z)$ phase space. In three 
dimensions and for $\alpha<1$, condensation can not occur. Conversely, for 
$\alpha>1$, above some critical value $z_c$ of the parameter $z$ (given 
in Eq. \eqref{zc}), the system undergoes a condensation transition. In 
particular, for $1<\alpha<2$ and $z>z_c$, the condensate is anomalous.
In contrast, for $\alpha>2$ and $z>z_c$, the condensate is normal. 
Increasing the 
dimension $d$, the region of the phase space corresponding to the anomalous 
condensate phase shrinks, until, at $d=7$, it disappears. Indeed, for 
$d\geq 7$, the system can either be in the fluid ($z<z_c$) or in the normal 
condensate phase ($z>z_c$). In the right panel of Fig. \ref{fig:phase}, 
we present 
the $(z,\alpha)$ phase diagram for $d=7$ which shows that
only two phases `fluid' and `normal condensate' can occur.

\section{Grand canonical criterion for condensation}

\label{sec:criterion}

In this section, we provide a general argument that allows us to 
determine the conditions that are necessary for condensation. This 
approach is based on a grand canonical description of the system and 
will also allow us to determine the critical value $X_c$ of the total 
displacement $X$ at which the phase transition occurs. Note however that 
the method presented below does not give any information about the 
nature of the condensed phase, which will be analyzed in detail in the next sections.

In order to investigate the condensation transition, we will focus on 
the large-deviation regime where $X$ scales linearly with the number $N$ of 
runs, in the limit $N\to\infty$. We define the scaled distance $z=X/N$, 
which will be the control parameter of our system. 
To establish when condensation occurs, we adopt a grand canonical 
description, as was done for positive-only
i.i.d. random variables in the context of mass models~\cite{MEZ2005,EMZ06}. 
There will be important differences however from the mass models.
In the grand canonical approach we assume that the variables 
$x_i$ in Eq. \eqref{ZXN} become decoupled from each other. 
To do this, we remove the hard delta-function constraint in Eq. (\ref{ZXN}) and replace
it by a factor $e^{-\mu\, \sum_{i=1}^N x_i}$ 
where $\mu$ plays the role of the negative chemical potential or equivalently 
a Lagrange multiplier.
We fix the value of $\mu$ by fixing the average $\langle X\rangle$.
In other words, we let the total displacement $X$ free to fluctuate 
in the grand canonical description, but with its average $\langle X\rangle$ kept fixed.
Provided this approach works, the canonical partition function given
by the $N$-fold integral in Eq. (\ref{ZXN}) is replaced by the grand-canonical partition function
defined as
\begin{equation}
Z_{\rm GC}(\mu,N) = \int_{-\infty}^{\infty} \prod_{i=1}^N p(x_i)\, e^{-\mu\, x_i}\, dx_i =
\left[\int_{-\infty}^{\infty} p(x)\, e^{-\mu x}\, dx\right]^N\, ,
\label{ZGC.1}
\end{equation}
with $p(x)$ given in Eq. (\ref{px}).
Thus, in the grand canonical ensemble, the $N$ runs are completely independent, each drawn
from the normalized PDF
\begin{equation}
p_{\mu}(x)=\frac{e^{-\mu x}\, p(x)}{\int_{-\infty}^{\infty}dx\,e^{-\mu x}\,p(x)}\, .
\label{P_mu}
\end{equation}
We recall that the PDF $p(x)$ is symmetric around $x=0$. 
The parameter $\mu$ can be determined from the following condition on the average displacement
\begin{equation}
\langle X\rangle= \sum_{i=1}^N \langle x_i\rangle =z\, N\,,
\end{equation}
where the average is with respect to the distribution $p_\mu(x)$ in Eq. (\ref{P_mu}). This gives
\begin{equation}
z=f(\mu)\equiv \frac{\int_{-\infty}^{\infty}dx\,x~e^{-\mu x}p(x)}{\int_{-\infty}^{\infty}dx~e^{-\mu x}p(x)}\,.
\label{equation_mu}
\end{equation}
The main idea behind the condensation criterion that we are going to present is that, 
when Eq. \eqref{equation_mu} admits a solution, the canonical and grand canonical descriptions 
are equivalent and we will call the system to be in the `fluid' phase. On the other hand, 
if for some value of $z$, Eq. \eqref{equation_mu}  ceases to have a solution for $\mu$, 
then the two ensembles are not no longer equivalent, signalling a possible phase transition. 
To proceed, it is useful to define the limiting value
\begin{equation}
c=-\lim_{x\to\infty}\frac{\log(p(x))}{x}\,.
\label{c}
\end{equation}
We distinguish different cases, depending on $c$.

\vskip 0.3cm

\noindent{\bf The case $c=\infty$:} First, we consider the case $c=\infty$, corresponding to 
a PDF $p(x)$ that decays faster than any exponential for large $|x|$. Let us first examine
the two integrals, respetively in the numerator and the denominator of Eq. (\ref{equation_mu}). 
When $p(x)$ decays faster than any exponential, clearly both integrals in
Eq. (\ref{equation_mu}) exist for any $\mu$, i.e., for all
$-\infty<\mu<\infty$. Then the function $f(\mu)$ in Eq. (\ref{equation_mu}) is a monotonically
decreasing function of $\mu$ in the range $\mu\in [-\infty,\infty]$,
going from $\infty$ (as $\mu\to-\infty$) to $-\infty$ (as $\mu\to 
\infty$). Then, for any value of $z$, there is a unique solution of the equation 
\eqref{equation_mu} for $\mu$. This means that the canonical and grand canonical
descriptions are equivalent, the system remains a fluid for all $z$, and
never develops a condensate.

\vskip 0.3cm

\begin{figure*}[t]
\centering
\includegraphics[width=0.6\textwidth]{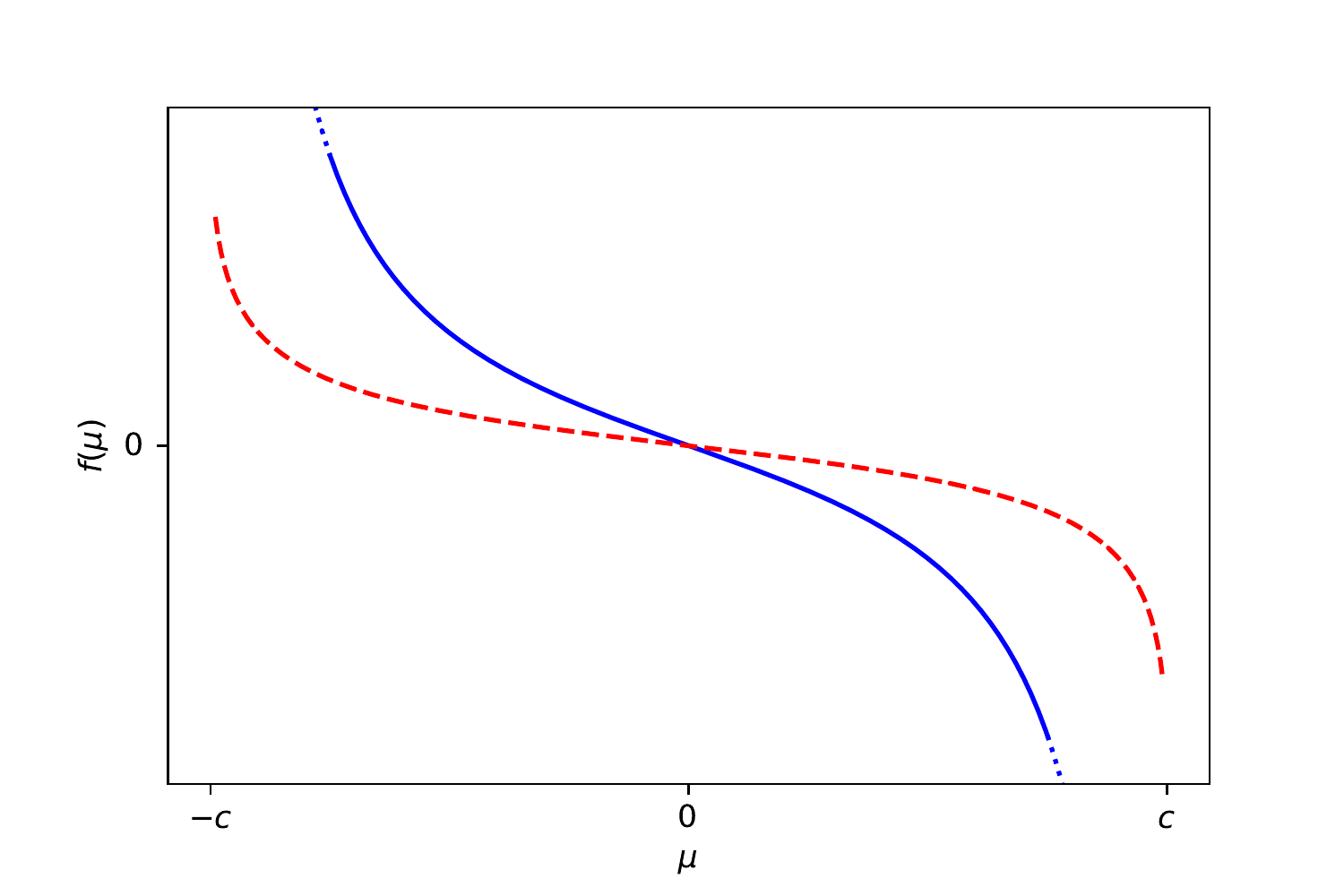} 
\caption{\label{fig:f} The function $f(\mu)$ versus $\mu$. If $f(\mu)$ diverges when $\mu\to -c$ (continuous blue curve), Eq. \eqref{equation_mu} will admit a solution for $\mu$ and no transition occurs. If $f(\mu)$ goes to a finite value $f(-c)$ when $\mu\to -c$ (dashed red curve), Eq. \eqref{equation_mu} will always admit a solution only for $z<f(-c)$ and at $z=f(-c)$ a condensation transition occurs. 
}
\end{figure*}

\noindent{\bf The case $0<c<\infty$:} This corresponds to a distribution 
$p(x)$ that decays exponentially fast as 
$p(x)\sim e^{-c|x|}$ for large $|x|$. 
Then, the parameter $\mu$ can only take values only in the interval 
$(-c,c)$ in order that both integrals in Eq. (\ref{equation_mu}) converge. 
It is useful to define the auxiliary function 
\begin{equation}
\tilde{p}(x)=p(x)e^{c|x|}\,.
\end{equation}
The function $f(\mu)$ in Eq. (\ref{equation_mu})
is again a decreasing odd function of $\mu$, but now only in
the bounded range $\mu\in[-c,c]$. 
It is then easy to see from Eq. (\ref{equation_mu})
that if $\tilde{p}(x)$ decays slower than $1/|x|^2$ for large $|x|$, 
then $f(\mu)$ diverges at the two edges:
$f(\mu)\to +\infty$ as $\mu\to -c$ and $f(\mu)\to -\infty$
as $\mu\to c$ (see Fig. \ref{fig:f}). Hence, for a given $z$, one can always find a
solution to the equation $f(\mu)=z$ in Eq. (\ref{equation_mu}).
Consequently, there is no transition. 

On the other hand, if $\tilde{p}(x)$ decays faster than $1/|x|^2$ then
the integrals in Eq. (\ref{equation_mu}) are convergent
for all $\mu\in [-c,c]$. In particular, at the left edge, the
function $f(\mu)$ approaches 
\begin{equation}
z_c=f(-c)=\frac{\int_{-\infty}^{\infty}dx\,x~e^{c x}p(x)}{\int_{-\infty}^{\infty}dx~e^{c x}p(x)}<\infty\,.
\label{f-c}
\end{equation}
Thus, for $z<f(-c)$, a solution of Eq. \eqref{equation_mu} always exists. 
On the other hand, for $z>f(-c)$, there is no solution to 
Eq. \eqref{equation_mu}, signalling a 
condensation transition. Thus, the phase transition occurs at the 
critical value $z_c=f(-c)$. Using the symmetry, a similar condensation
will also occur for $z<-z_c=f(c)$. 
Note however that this grand canonical 
description does not shed light on the precise nature of the 
condensed phase. Indeed, to understand the behavior of the system above the 
transition, a detailed analysis of the canonical PDF 
in Eq. \eqref{ZXN} is required as in the case of
mass transport models~\cite{EMZ06}.

\vskip 0.3cm

\noindent{\bf The case $c=0$:} In this case, $p(x)$ decays slower than an exponential
as $|x|\to \infty$. Thus, the integrals in the denominator and numerator 
of Eq. (\ref{equation_mu}) exist for $\mu=0$. For any nonzero $\mu$, the integrals
diverge, either as $x\to -\infty$ (if $\mu>0$), or as $x\to \infty$ (if $\mu<0$).  
Thus, the grand canonical description fails completely here.
However, we believe that the system still 
undergoes a condensation transition if $p(x)$ decays, for large $|x|$ faster than $1/|x|^3$. 
The reason behind this conjecture is the following. If $p(x)$ decays faster than $1/|x|^3$, 
then its second moment is finite and the CLT applies. Therefore, for $X\sim \sqrt{N}$, 
$Z(X,N)$ assumes a Gaussian shape. On the other hand, for $X\sim N$, we expect
\begin{equation}
Z(X,N)\sim N p(X)\,,
\label{tail}
\end{equation}
where the right-hand side corresponds to a configuration where one of the runs absorbs the 
whole displacement $X$. 
Thus, $Z(X,N)$ is described by two regimes: 
the typical Gaussian regime for $X\sim \sqrt{N}$ and the fat-tailed regime for $X\sim N$.
In this case, for $X\sim O(N)$, the distribution $Z(X,N)$ does not have
a large deviation behavior of the type, $Z(X,N)\sim \exp[-N\, \psi(X/N)]$,
and thus condensation can not happen on a scale $X\sim O(N)$. However, if the central
CLT region has to match the tail behvaior in Eq. (\ref{tail}), 
we believe that a condensation should occur at a shorter scale $X\sim N^{\gamma}$,
where $1/2<\gamma<1$.
This has already been hinted in Ref.~\cite{GM2019} which studied a particular example,
though there $p(x)$ was asymmetric.

In the complementary case when $p(x)$ decays slower than $1/|x|^3$, 
the CLT does not hold and, already in the typical regime, $X$ is dominated by the maximum 
of $x_1\,,\ldots x_N$ \cite{BG90}. Thus, in this case, condensation spontaneously occurs at 
any scale and no dynamical phase transition takes place.

In the rest of this section, we will show that one can obtain several RTP models that satisfy the condensation criterion. This can be achieved by properly tuning the speed distribution $W(v)$. Below, we provide few examples of $W(v)$ that lead to condensation.

\begin{itemize}

\item $W(v)=\alpha(1-v)^{\alpha-1}$ with $0<v<1$, as mentioned
in Eq. (\ref{W}) with $v_0=1$. In this case, for arbitrary $d$, one can show that for large $|x|$ (see Appendix \ref{app:asymp})
\begin{equation}
p(x)\simeq A_{d,\alpha}  e^{- |x|} \frac{1}{|x|^{\nu}}~ \,,
\label{px_asym}
\end{equation}
where 
\begin{equation}
A_{d,\alpha}= \frac{\Gamma\left(d/2\right)\alpha 
\Gamma\left(\alpha\right) }{\sqrt{\pi}}2^{(d-3)/2}\, \quad {\rm and}
\quad \nu= \frac{(d+2\alpha-1)}{2}\, .
\label{A_d_alpha}
\end{equation}
In this case $c=1$ from Eq. (\ref{c})
and, applying the criterion described above, 
we find that the transition is possible only for $\nu>2$, 
i.e., for $d+2\alpha>5$. 
Recalling that the limit $\alpha\to 0$ corresponds to the canonical RTP 
model with fixed velocity (i.e., $W(v)=\delta(v-1)$), 
we recovered that for the fixed-velocity RTP model condensation is 
possible only for $d>5$. This was first observed in \cite{PTV2020}. 
Plugging the expression for $p(x)$, given in Eq. \eqref{px}, 
into Eq. \eqref{f-c}, we obtain the critical value $z_c$
explicitly, valid for arbitrary $d$ and $\alpha$,
\begin{equation}
z_c=\frac{4}{d(1+\alpha)(2+\alpha)}\frac{{}_4  F_3\left[3/2,~3/2,~2,~2;(2+d)/2,~(3+\alpha)/2,~(4+\alpha)/2;1\right]}{{}_4  F_3\left[1/2,~1/2,~1,~1; d/2,~(1+\alpha)/2,~(2+\alpha)/2;1\right]}\,,
\label{zc_grand}
\end{equation}
where ${}_4  F_3$ is the standard hypergeometric function
defined more precisely in Eq. (\ref{definition_pFq}).
In the next sections, we will focus on this family of speed distributions, 
parametrized by $\alpha$.

\item $W(v)=\sqrt{\frac{2}{\pi}}\,e^{-v^2/2}$ with $v>0$. 
Considering $d=1$ and using Eq. \eqref{px}, one can show that, for $|x|\gg 1$,
\begin{equation}
p(x)\sim {|x|}^{-1/3}e^{-3| x|^{2/3}/2}\,.
\end{equation}
In this case $p(x)$ decays slower than any exponential, thus $c=0$. 
Moreover, $p(x)$ decays faster than $1/|x|^3$ and thus according
to our conjecture, a condensation transition should occur.
The condensation transition in the RTP
model with this particular half-Gaussian speed distribution 
was studied in Ref.~\cite{GM2019}, but in the
presence of an additional constant force.

\item $W(v)\sim 1/v^\beta$ for large $v$ with $\beta>1$. 
In this example, for $d=1$, it is easy to show 
from Eq. (\ref{px}) that 
$p(x)\sim 1/|x|^\beta$ for large $|x|$. 
Thus, we find $c=0$ and one has condensation only if $\beta>3$, according to our conjecure.

\end{itemize}

Let us recall that in the case of the sum of 
positive-only i.i.d. random variables, 
a similar condensation criterion was established~\cite{EMZ06}. 
In that case, one can still define the limiting value $c$ in Eq. \eqref{c}. 
Then, if $c=\infty$, no condensation happens. 
For $0\leq c<\infty$, condensation happens only if $\tilde{p}(x)=e^{cx}p(x)$ decays 
faster than $1/x^2$.

\section{Position distribution}

\label{sec:position}

In this section, we want to investigate the PDF $Z(X,N)$ of the total $x$-component 
displacement $X$, where $X=\sum_{i=1}^{N}x_i$, by analysing fully the $N$-fold
integral in Eq. (\ref{ZXN}), thus going beyond the grand canonical description
discussed in the previous section. 
We will first derive an exact expression for $Z(X,N)$, valid for any $X$ and $N$. 
Then, focusing on large $N$, we study both the typical regime $X\sim O(\sqrt{N})$, 
where $Z(X,N)$ is Gaussian, and the large-deviation regime $X\sim O(N)$, where $Z(X,N)$ 
assumes a large deviation form, $Z(X,N)\sim \exp\left[-N \psi_{d,\alpha}(X/N)\right]$, 
with a rate function $\psi_{d,\alpha}(z)$ that we compute exactly. Under specific conditions
on $d$ and $\alpha$, we show that $\psi_{d,\alpha}(z)$ becomes singular at a critical value 
$z_c$ of the scaled displacement $z=X/N$. This singularity corresponds to a condensation 
phase transition.

We recall that the PDF $Z(X,N)$ can be written as (see Eq. \eqref{ZXN})
\begin{equation}
Z(X,N)=\int_{-\infty}^{\infty}dx_1\,
\ldots\int_{-\infty}^{\infty}dx_N\,\prod_{i=1}^{N}p(x_i)~\delta\left(X-\sum_{i=1}^N x_i\right)\,,
\label{eq:PXN_1}
\end{equation}
where the delta function constraints the final position to be $X$ and 
$p(x)$ is given in Eq. \eqref{px}, with 
$W(v)= \alpha(1-v)^{\alpha-1}$ for $0\le v\le 1$ and $W(v)=0$ otherwise. 
To proceed, we recall the integral representation of the delta function
\begin{equation}
\delta(X)=\frac{1}{2\pi i}\int_{\Gamma}dq~e^{-qX}\,,
\end{equation}
where the integral is performed over the imaginary-axis Bromwich contour $\Gamma$ in the complex $q$ plane. Plugging this integral expression into Eq. \eqref{eq:PXN_1}, we find
\begin{equation}
Z(X,N)=\frac{1}{2\pi i}\int_{\Gamma}dq~e^{qX}\left[\hat{p}(q)\right]^N\,,
\label{eq:PXN_2}
\end{equation}
where 
\begin{equation}
\hat{p}(q)=\int_{-\infty}^{\infty}dx~e^{-qx}p(x)\,.
\label{FT.1}
\end{equation}

Substituting $W(v)=\alpha\, (1-v)^{\alpha-1}$ over $v\in [0,1]$
in Eq. (\ref{px}), we first evaluate $p(x)$ and then compute
$\hat{p}(q)$ using Eq. (\ref{FT.1}). Using Mathematica, we get
\begin{equation}
\hat{p}(q)={}_4 F_3\left(\frac12,\frac12,1,1;\frac{d}{2},\frac{1+\alpha}{2},\frac{2+\alpha}{2};q^2\right)\,,
\label{hat-p}
\end{equation}
where ${}_p F_q(\alpha_1\,,\ldots\alpha_p;\beta_1\ldots\beta_q;q)$ denotes the generalized hypergeometric function, defined as
\begin{equation}
{}_p F_q(\alpha_1\,,\ldots\alpha_p;\beta_1\ldots\beta_q;z)=\sum_{n=0}^{\infty}\frac{(\alpha_1)_n\ldots (\alpha_p)_n}{(\beta_1)_n\ldots(\beta_q)_n}\frac{z^n}{n!}\,,
\label{definition_pFq}
\end{equation}
where $(a)_n$ is the rising factorial (or Pochhammer symbol), defined as
\begin{equation}
(a)_n=\begin{cases}
1~~~&\text{  if  }\,\,n=0\\
\\
a(a+1)(a+2)\ldots(a+n-1)~~~&\text{  if  }\,\,n\geq 1~.\\
\end{cases}
\end{equation} Thus, we find
\begin{equation}
Z(X,N)=\frac{1}{2\pi i}\int_{\Gamma}dq~\exp\left[qX+NS_{d,\alpha}(q)\right]\,,
\label{eq:PXN_3}
\end{equation}
where
\begin{equation}
S_{d,\alpha}(q)=\log\left[{}_4 F_3\left(\frac12,\frac12,1,1;\frac{d}{2},\frac{1+\alpha}{2},\frac{2+\alpha}{2};q^2\right)\right]\,.
\label{S}
\end{equation}
Note that this result is exact for any $X$ and $N$. We are now interested in extracting the behavior of $Z(X,N)$ in the limit of large $N$.

\subsection{Typical regime}
\begin{figure}
\centering

\includegraphics[scale=1]{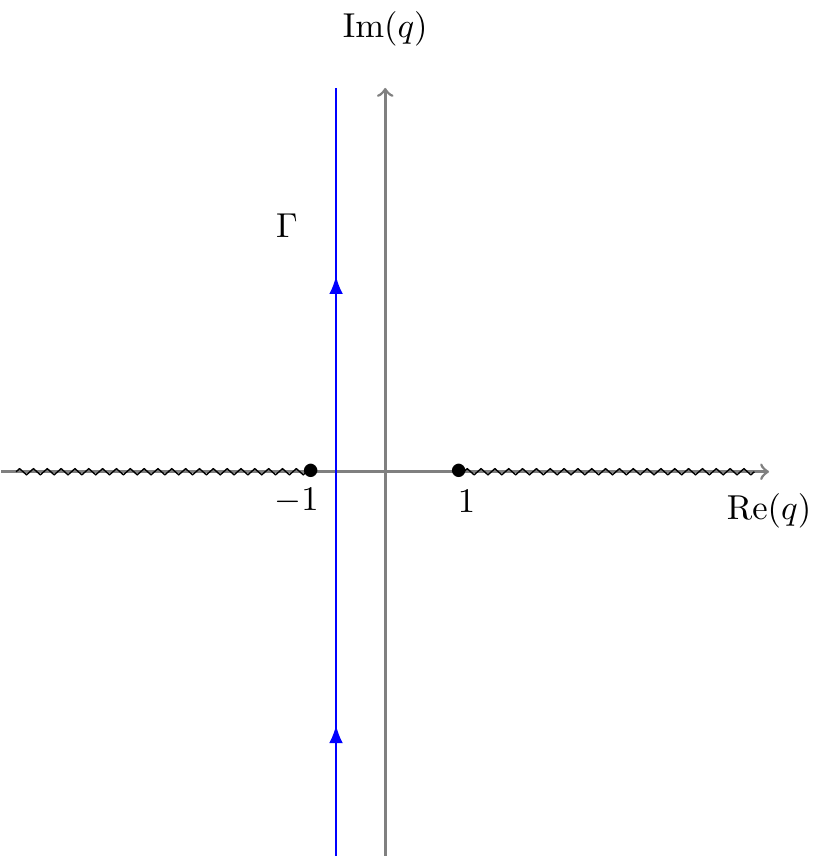} 

\caption{Analytic structure of the function $S_{d,\alpha}(q)$, given in Eq. \eqref{S}. For any $d$ and $\alpha$, $S_{d,\alpha}(q)$ has two branch cuts (the grey wiggly lines in figure) in the real-$q$ axis, for $q<-1$ and for $q>1$. The continuous blue line represents the Bromwich contour $\Gamma$, defined in the text.}
\label{fig:S}
\end{figure}

First, we investigate the typical regime where $X\sim \sqrt{N}$. 
Substituting $X=\sqrt{N}\,y$ in Eq. \eqref{eq:PXN_3}, where the variable 
$y$ is assumed to be of order one, we obtain
\begin{equation}
Z(X=\sqrt{N}\,y,N)=\frac{1}{2\pi i}\int_{\Gamma}dq~\exp\left[q\sqrt{N}\,y+
N\,S_{d,\alpha}(q)\right]\,.
\label{eq:PXN_4}
\end{equation}
We now perform the change of variable $q\to q \sqrt{N}$ and we obtain
\begin{equation}
Z(X=\sqrt{N}y,N)=\frac{1}{2\pi i\sqrt{N}}\int_{\Gamma}dq~\exp\left[q~y+NS_{d,\alpha}\left(\frac{q}{\sqrt{N}}\right)\right]\,.
\label{eq:PXN_5}
\end{equation}
We expand the right-hand side of Eq. \eqref{eq:PXN_5} for large $N$, using Eq. \eqref{S} and the small-argument expansion of the generalized hypergeometric function \cite{Gradshteyn_book}, and we find
\begin{equation}
Z(X=\sqrt{N}y,N)\simeq\frac{1}{2\pi i\sqrt{N}}\int_{\Gamma}dq~\exp\left(q~y+\frac{2 q^2}{d (\alpha+1)(\alpha+2)}\right)\,.
\label{eq:PXN_6}
\end{equation}
Finally, performing the Gaussian integral over $q$ we obtain the
results announced in Eqs. (\ref{typical}) and (\ref{D}). Thus, in this regime, the distribution of the final position of the 
particle is Gaussian. This is a consequence of the CLT, since $X$ is the 
sum of $N$ i.i.d. random variables with finite variance. This is consistent with the fact that, for any $N$, the variance of $X$ is simply given by
\begin{equation}
\langle X^2\rangle=N\int_{-\infty}^{\infty}dx~x^2~p(x)= \frac{4N}{d(\alpha+1)(\alpha+2)}\,.
\end{equation}
The result in Eq. \eqref{typical} tells us that, for late times, the RTP has typically 
a diffusive behavior, similar to the one of a passive Brownian motion. 
In other words, at the scale $X\sim \sqrt{N}$ the position distribution 
of the RTP does not show any signs of activity. In order to observe the 
signatures of the active nature of the particle, it is necessary to 
investigate the large-deviation regime, where $X\sim N$. It is possible 
to show that the result in Eq. \eqref{typical} is valid on a larger 
region than the one predicted by the CLT, for any $|X|\ll N^{3/4}$ (see 
Appendix \ref{app:CLT}).

\subsection{Large-deviation regime}

\begin{figure*}[t]
\centering
\includegraphics[width=0.6\textwidth]{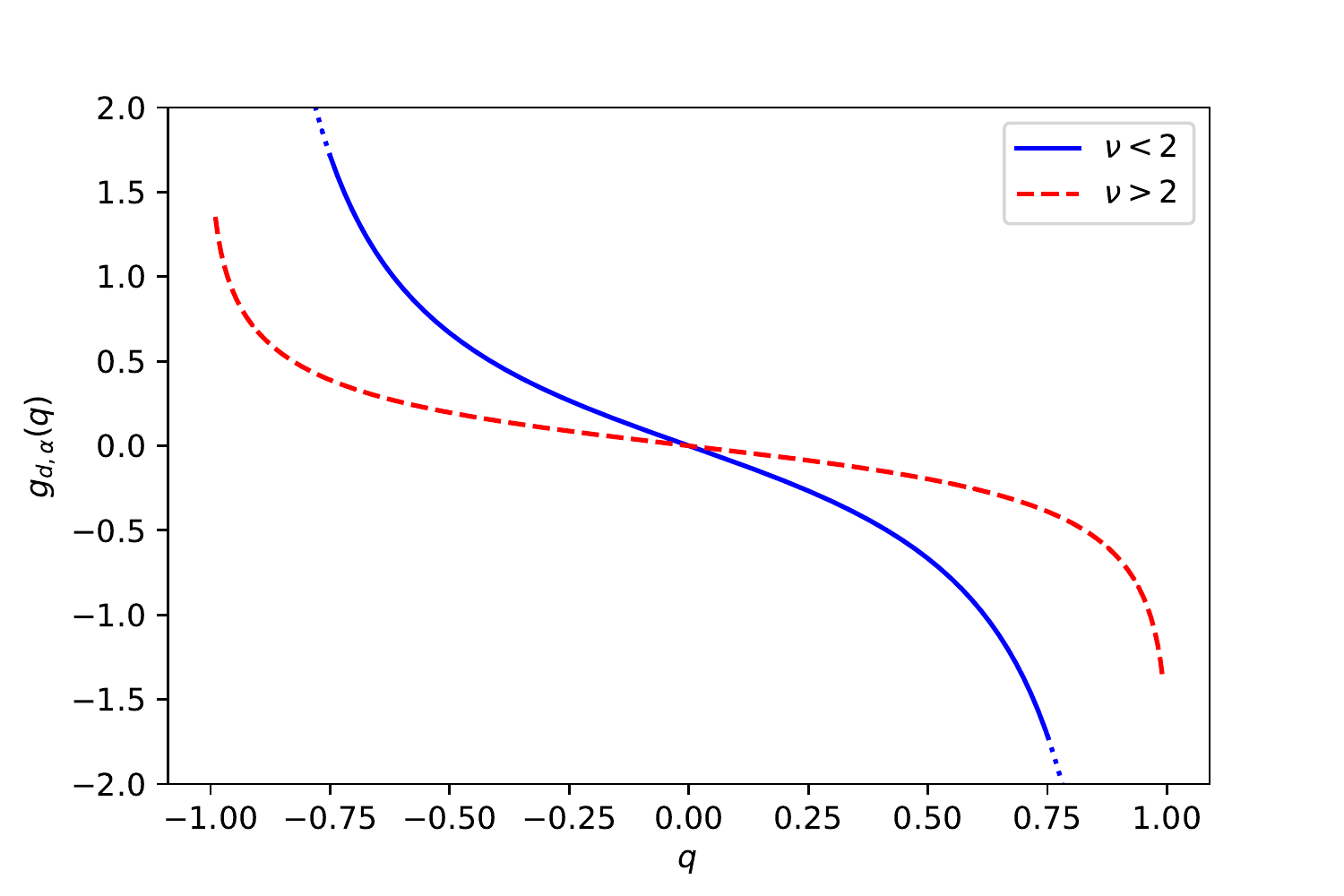} 
\caption{\label{fig:g} The function $g_{d,\alpha}(q)$ versus $q$ for different values of $\nu=(d+2\alpha-1)/2$. For any $\nu$, $g_{d,\alpha}(q)$ is a decreasing odd function of $q$. For $\nu<2$, $g_{d,\alpha}(q)$ diverges when $q\to -1$, while for $\nu>2$ it goes to the finite value $g_{d,\alpha}(-1)$.
}
\end{figure*}

To proceed, we define the rescaled variable $z=X/N$. 
From Eq. \eqref{eq:PXN_3}, we obtain
\begin{equation}
Z(X=N z,N)=\frac{1}{2\pi i }\int_{\Gamma}dq~e^{N \left[qz+S_{d,\alpha}(q)\right]}\,.
\label{eq:PXN_7}
\end{equation}
where $S_{d,\alpha}(q)$ is given in Eq. \eqref{S}. We recall that the integral in Eq. \eqref{eq:PXN_7} is performed over the imaginary-axis Bromwich contour $\Gamma$ in the complex-$q$ plane. For any $d$ and $\alpha$, the complex function $S_{d,\alpha}(q)$ has two branch cuts running in the real-$q$ axis for $q<-1$ and $q>1$ (see Fig. \ref{fig:S}). 

\begin{figure*}[t]
\centering
\includegraphics[width=1\textwidth]{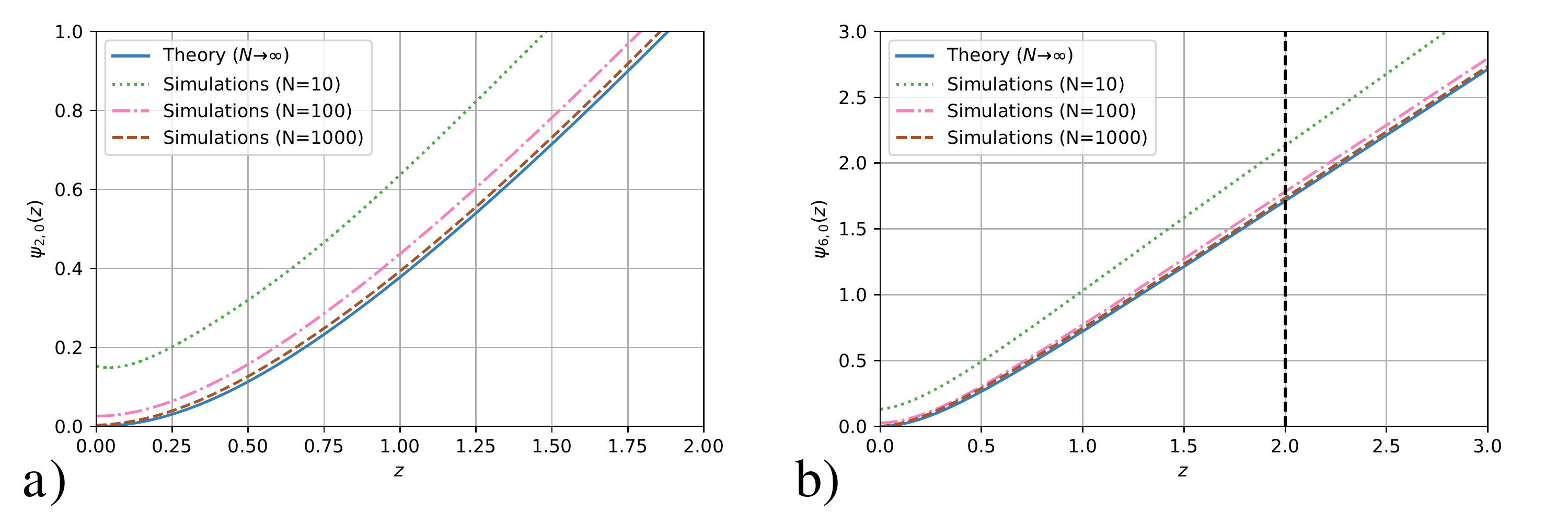} 
\caption{\label{fig:psi_int} {\bf a)} Rate function $\psi_{d,\alpha}(z)$ versus $z$, for $d=2$ and $\alpha=0$. The continuous blue line corresponds to the exact result in Eq. \eqref{psi_2_0}, valid in the limit $N\to\infty$. For this choice of the parameters $d$ and $\alpha$, no transition occurs.
{\bf b)} Rate function $\psi_{d,\alpha}(z)$ versus $z$, for $d=6$ and $\alpha=0$. The continuous blue line corresponds to the exact result in Eq. \eqref{psi_d_cond}. The vertical dashed line signals the critical point $z_c$ at which the phase transition occurs, for $z>z_c$ the rate function becomes exactly linear. In both panels, the colored dashed lines are the results of numerical simulations obtained at finite $N$, as described in Section \ref{sec:numerics}.
}
\end{figure*}

We first try to compute the integral in Eq. \eqref{eq:PXN_7} by 
saddle-point approximation. Assuming a saddle-point exists, it
must satisfy $\frac{d}{dq}[qz+S_{d,\alpha}(q)]=0$. This gives the
saddle-point equation 
\begin{equation}
z=g_{d,\alpha}(q)\equiv -S'_{d,\alpha}(q)\,.
\label{SPE}
\end{equation}
Using the expression of $S_{d,\alpha}(q)$ in Eq. \eqref{S}, we find
\begin{equation}
g_{d,\alpha}(q)=-\frac{4q}{d(\alpha+1)(\alpha+2)}\frac{{}_4 F_3\left(\frac32,\frac32,2,2;\frac{2+d}{2},\frac{3+\alpha}{2},\frac{4+\alpha}{2};q^2\right)}{{}_4 F_3\left(\frac12,\frac12,1,1;\frac{d}{2},\frac{1+\alpha}{2},\frac{2+\alpha}{2};q^2\right)}\,.
 \label{gd}
 \end{equation}
Note that, identifying $\mu= q$, the saddle point equation \eqref{SPE} 
is the same condition as the one that fixes the chemical potential $\mu$ 
in Eq. \eqref{equation_mu} in the grand canonical argument for 
condensation. One can check that, since $z$ is real, the solution $q^*(z)$ of the 
saddle-point equation in \eqref{SPE} has to be real. Moreover, due to the branch cuts of the function 
$S_{d,\alpha}(q)$ (see Fig. \ref{fig:S}), $q^*(z)$ has to belong to the real interval 
$(-1,1)$. Therefore, it is instructive to analyze the behavior of 
$g_{d,\alpha}(q)$ for $q\in(-1,1)$. First of all, for any $d$ and 
$\alpha$, it is easy to show that $g_{d,\alpha}(q)$ is a decreasing odd 
function of $q$ along the real interval $q\in [-1,1]$, such that 
$g_{d,\alpha}(q)>0$ for $q<0$ and 
$g_{d,\alpha}(q)<0$ for $q>0$ (see Fig. \ref{fig:g}). To proceed, we 
need the following asymptotic expansion for the generalized 
hypergeometric function, valid for $q\to 1$ from below~\cite{buhring}
\begin{equation}
{}_4 F_3\left(\alpha_1,~\alpha_2,~\alpha_3,~\alpha_4 ;\beta_1,~\beta_2,~\beta_3;q\right)=\sum_{n=0}^{\infty}a_n(1-q)^n+(1-q)^\varphi\,\sum_{n=0}^{\infty}b_n(1-q)^n\,,
\label{eq:hypergeom1_N}
\end{equation}
where $a_n$ and $b_n$ are constants that depend on the parameters of the function (for the precise expressions of $a_n$ and $b_n$ see \cite{buhring}) and
\begin{equation}
\varphi=\sum_{j=1}^{3}\beta_j-\sum_{j=1}^{4}\alpha_j\,.
\end{equation}
Note that the formula in Eq. \eqref{eq:hypergeom1_N} is only valid if $\varphi$ is not an integer. In the case of integer $\varphi$, logarithmic corrections are present in the asymptotic expansion in Eq. \eqref{eq:hypergeom1_N} \cite{buhring}. Using Eq. \eqref{eq:hypergeom1_N}, it is easy to show that, for $\nu< 2$ (where we recall that $\nu=(d+2\alpha-1)/2$), $g_{d,\alpha}(q)$ diverges when $q\to - 1$. Thus, for $\nu<2$ the saddle-point equation \eqref{SPE} admits a unique solution for any $z$ and we obtain 
\begin{equation}
Z(X,N)\simeq \frac{1}{\sqrt{2\pi  |S''_{d,\alpha}\left[q^*(X/N)\right]|N }}\exp\left[-N \psi_{d,\alpha}\left(\frac{ X}{N}\right)\right]\,,
\label{eq:d<5}
\end{equation}
where 
\begin{equation}
\psi_{d,\alpha}(z)=- z~q^*(z)-S_{d,\alpha}(q^*(z))\,,
\label{psi_d_z}
\end{equation}
$q^*(z)$ is the unique solution of Eq. \eqref{SPE} and $S''_{d,\alpha}(q)$ is the second derivative of $S_{d,\alpha}(q)$ with respect to $q$. For special values of $d$ and $\alpha$, it is possible to find an explicit expression for $\psi_{d,\alpha}(z)$. For instance, in the special case $d=2$ and $\alpha=0$, we find
\begin{equation}
\psi_{2,0}(z)=\frac{1}{2}\left[\sqrt{1+4 z^2}-1+\log\left(\frac{\sqrt{1+4z^2}-1}{2z^2}\right)\right]\,.
\label{psi_2_0}
\end{equation}
The rate function $\psi_{2,0}(z)$ is shown in Fig. \ref{fig:psi_int} and 
it is in good agreement with numerical simulations performed for 
$N=10^4$.

On the other hand, for $\nu>2$ one has
\begin{equation}
0<g_{d,\alpha}(-1)<+\infty\,.
\end{equation}
Thus, for small positive $z$ the saddle-point equation \eqref{SPE} admits a unique solution $-1<q^*(z)<0$ and the position distribution $Z(X,N)$ is still given by the expression in Eq. \eqref{eq:d<5}. However, increasing $z$, the solution $q^*(z)$ decreases until, at the critical value $z_c=g_{d,\alpha}(-1)$, it encounters the branch cut at $q=-1$ (see Fig. \ref{fig:S}). For $z>z_c$, $q^*(z)$ freezes at the value $-1$. Indeed, increasing $z$ above $z_c$, Eq. \eqref{SPE} has no solution and the integral in Eq. \eqref{eq:PXN_7} cannot be computed via the saddle-point approximation. Nevertheless, this integral is dominated by values of $q$ close to $q=-1$ and therefore one can approximate
\begin{equation}
Z(X=N\, z,N)\sim \exp\left[-N \left(z-S_{d,\alpha}(-1)\right)\right]\,.
\label{eq:d>5}
\end{equation}
Hence, the rate function can be written for $\nu>2$, as
\begin{equation}
\psi_{d,\alpha}(z)=
\begin{cases}
- z~q^*(z)-S_{d,\alpha}(q^*(z))~~~~&\text{  for  }z<z_c\\
\\
z-S_{d,\alpha}(-1)~~~~&\text{  for  }z>z_c\\
\end{cases}
\label{psi_d_cond}
\end{equation}
where $q^*(z)$ is the unique solution of Eq. \eqref{SPE} and
\begin{equation}
z_c=\frac{4}{d(\alpha+1)(\alpha+2)}\frac{{}_4 F_3\left(\frac32,\frac32,2,2;\frac{2+d}{2},\frac{3+\alpha}{2},\frac{4+\alpha}{2};1\right)}{{}_4 F_3\left(\frac12,\frac12,1,1;\frac{d}{2},\frac{1+\alpha}{2},\frac{2+\alpha}{2};1\right)}\,.
\label{zc}
\end{equation}
Note that, as expected, the critical value $z_c$ is the same as the one predicted by the grand canonical argument in Section \ref{sec:criterion} (see Eq. \eqref{zc_grand}).

For $z>z_c$, to find the prefactor of the expression 
in Eq. \eqref{eq:d>5}, one has to compute the contour integral 
in Eq. \eqref{eq:PXN_7}. We perform this calculation in the case 
where $\nu=(d+2\alpha-1)/2$ is not an integer for simplicity. However, this
calculation can be extended easily to arbitrary $\nu$. 
Since we expect the integral to be dominated by values of $q$ close to $-1$, it is useful to perform the change of variable $q\to s=(q+1)N$ in Eq. \eqref{eq:PXN_7}, which yields
\begin{equation}
Z(X=N\, z,\,N)=\frac{1}{2\pi i }N\int_{\Gamma}ds~e^{N \left[(s/N-1)z+S_{d,\alpha}(-1+s/N)\right]}\,.
\label{eq:PXN_limit1}
\end{equation}
Using the asymptotic expression of the hypergeometric function close to unit argument in Eq. \eqref{eq:hypergeom1_N}, we expand the exponent for large $N$ and we find
\begin{equation}
Z(X=N\,z\,N)=B_N e^{-Nz}\frac{1}{2\pi i }N\int_{\Gamma}ds~\exp\left[(z-z_c)~s+a_{d,\alpha}~\frac{s^2}{N}\ldots+b_{d,\alpha}~
\frac{s^{\nu-1}}{N^{\nu-2}}+\ldots\right]\,,
\label{eq:PXN_limit2}
\end{equation}
where $s^{\nu-1}$ is the leading singular term and
\begin{equation}
B_N=e^{N S_{d,\alpha}(-1)}\,.
\label{BN}
\end{equation}
The constants $a_{d,\alpha}$ and $b_{d,\alpha}$ can be exactly computed \cite{buhring}. In particular, we find that  $a_{d,\alpha}=0$ for $\nu<3$ and
\begin{eqnarray}
\label{a_d_alpha}
 a_{d,\alpha}&=&\frac{\Gamma\left(\frac{d}{2}\right)\Gamma\left(\frac{1+\alpha}{2}\right)\Gamma\left(\frac{2+\alpha}{2}\right)}{8 {}_4 F_3\left(\frac12,\frac12,1,1;\frac{d}{2},\frac{1+\alpha}{2},\frac{2+\alpha}{2};1\right)^2} \left[
\frac{2~ {}_4 F_3\left(\frac12,\frac12,1,1; \frac{d}{2},\frac{1+\alpha}{2},\frac{2+\alpha}{2};1\right)}{\Gamma\left(\frac{d}{2}\right)\Gamma\left(\frac{1+\alpha}{2}\right)\Gamma\left(\frac{2+\alpha}{2}\right)}\right. \left(\frac{ {}_4 F_3\left(\frac32,\frac32,2,2; \frac{2+d}{2},\frac{3+\alpha}{2},\frac{4+\alpha}{2};1\right)}{\Gamma\left(\frac{2+d}{2}\right)\Gamma\left(\frac{3+\alpha}{2}\right)\Gamma\left(\frac{4+\alpha}{2}\right)}\right.\nonumber \\
&+&\left. 18\frac{ {}_4 F_3\left[\frac52,\frac52,3,3; \frac{4+d}{2},\frac{5+\alpha}{2},\frac{6+\alpha}{2};1\right]}{\Gamma\left(\frac{4+d}{2}\right)\Gamma\left(\frac{5+\alpha}{2}\right)\Gamma\left(\frac{6+\alpha}{2}\right)} \right) -\left. \left(\frac{ {}_4 F_3\left(\frac32,\frac32,2,2; \frac{2+d}{2},\frac{3+\alpha}{2},\frac{4+\alpha}{2};1\right)}{\Gamma\left(\frac{2+d}{2}\right)\Gamma\left(\frac{3+\alpha}{2}\right)\Gamma\left(\frac{4+\alpha}{2}\right)}\right)^2\right]
\end{eqnarray}
for $\nu>3$, while 
\begin{equation}
b_{d,\alpha}=\Gamma(1-\nu)A_{d,\alpha}\,,
\label{b_d_alpha}
\end{equation}
for any $\nu$, where $A_{d,\alpha}$ is given
in Eq. (\ref{A_d_alpha}). 
For $\nu>3$, one can check that $a_{d,\alpha}$ is positive. 
Expanding Eq. \eqref{eq:PXN_limit2} for large $N$ we find
\begin{equation}
Z(X=N z,N)=B_N e^{-Nz}\frac{1}{2\pi i }N\int_{\Gamma}ds~e^{z_{\rm ex}s}~\left[1+a_{d,\alpha}~\frac{s^2}{N}\ldots+b_{d,\alpha}
\frac{s^{\nu-1}}{N^{\nu-2}}+\ldots\right]\,,
\label{eq:PXN_limit3}
\end{equation}
where $z_{\rm ex}=z-z_c$ is assumed to be $O(1)$. It is possible to show that, for any $a\geq0$, (see Appendix A.3 of Ref. \cite{EMZ06})
\begin{equation}
\frac{1}{2\pi i}\int_{\Gamma}ds~e^{z_{\rm ex}s}s^{a-1}=\frac{\sin(\pi a)}{\pi}\Gamma(a)\,.
\label{integral_a}
\end{equation}
Thus, when $a$ is integer, the integral above vanishes. Therefore, since we are assuming that $\nu$ is not an integer, the leading term in Eq. \eqref{eq:PXN_limit3} is, using the expression for $b_{d,\alpha}$ given in Eq. \eqref{b_d_alpha},
\begin{equation}
 Z(X=N z,N)\simeq B_N e^{-Nz}\frac{1}{z_{\rm ex}^{\nu}}\frac{b_{d,\alpha}}{N^{\nu-1}}\frac{\sin(\pi \nu)}{\nu}\Gamma(\nu)\Gamma(1-\nu)A_{d,\alpha}\,.
\end{equation}
Finally, using the relation $\Gamma(\nu)\Gamma(1-\nu)=\pi/\sin(\nu\pi)$, we find
\begin{equation}
Z(X,N)\simeq B_N A_{d,\alpha}\frac{N}{ (X-X_c)^{\nu}}e^{- X}\,.
\label{eq:PXN_limit6}
\end{equation}
When $\nu$ is an integer, a similar argument can be applied. Note that the expression in Eq. \eqref{eq:PXN_limit6} can be rewritten, using the large-$x$ expansion of $p(x)$ in Eq. \eqref{px_asym}, as
\begin{equation}
Z(X,N)\simeq C_N  N p(X-X_c)\,,
\label{eq:PXN_limit7}
\end{equation}
where $p(x)$ is the PDF of a 
single-run displacement and $C_N=B_N e^{-X_c}$. 
This expression can be interpreted as follows. Above the critical value 
$X=X_c$, all the extra displacement $X_{\rm ex}=X-X_c$ is absorbed by a 
single run, the condensate. The probability weight associated to the 
condensate is therefore $p(X-X_c)$ and the factor $N$
in Eq. (\ref{eq:PXN_limit7}) arises
since the condensate can be 
any one of the $N$ runs. The factor $C_N$ in 
Eq. \eqref{eq:PXN_7} is the probability weight of the other $N-1$ sites, 
which becomes independent of $X$ above the transition.

\subsection{Order of the transition}

It is also interesting to compute the order of the phase transition described above. We recall that the system undergoes a transition of order $n$ if the $n$-th derivative of $\psi_{d,\alpha}(z)$ is discontinuous, while all lower-order derivatives are continuous. Thus, we need to investigate the asymptotic behavior of $\psi_{d,\alpha}(z)$ close to the transition. In the limit $z\to z_c$ from below, we know that a solution $q^*(z)$ of the saddle point equation always exists. Moreover, since we know that exactly at the critical point $z=z_c$ the saddle point $q^*(z)$ encounters the branch cut at $q=-1$, we expect that $q^*(z)$ is close to $-1$ near the transition. Plugging $q=-1+s$ into the exponent of the integrand in Eq. \eqref{eq:PXN_7} and using Eq. \eqref{eq:hypergeom1_N} to expand for small $s$, we obtain
\begin{equation}
z~q+S_{d,\alpha}(q)\simeq S_{d,\alpha}(-1)-z +z_{\rm ex}s+c_{\nu}\,s^{\eta}\,,
\label{eq:expansion}
\end{equation}
where $c_{\nu}=b_{d,\alpha}$ for $2<\nu<3$ and $c_{\nu}=a_{d,\alpha}$ for $\nu>3$ (where $a_{d,\alpha}$ and $b_{d,\alpha}$ are given in Eqs. \eqref{a_d_alpha} and \eqref{b_d_alpha}), and
\begin{equation}
\eta=
\begin{cases}
\nu-1 & \text{for } 2<\nu<3,\\
\\
 2 & \text{for } \nu>3.\\
\end{cases}
\label{bd}
\end{equation}
We recall that $S_{d,\alpha}(q)$ is defined in Eq. \eqref{S}. Setting to zero the first derivative with respect to $s$ of the expression in Eq. \eqref{eq:expansion}, we obtain 
\begin{equation}
s=\left(\frac{z_c-z}{c_{\nu}~\eta}\right)^{1/(\eta-1)}\,.
\end{equation}
Thus, the saddle point is located at, for $z\to z_c$, 
\begin{equation}
q^*(z)\simeq-1+\left(\frac{z_c-z}{c_{\nu}~\eta}\right)^{1/(\eta-1)}\,.
\end{equation}
Plugging this value into Eq. \eqref{psi_d_cond}, we find that when $z\to z_c$ from below
\begin{equation}
\psi_{d,\alpha}(z)\simeq z-S_{d,\alpha}(-1)+\left((c_{\nu}~\eta)^{-1/(\eta-1)}-\frac{c_{\nu}}{(c_{\nu}~\eta)^{\eta/(\eta-1)}}\right)~(z_c-z)^{\eta/(\eta-1)}\,.
\label{psi_d_expansion_below}
\end{equation}
Recalling that for $z>z_c$
\begin{equation}
\psi_{d,\alpha}(z)= z-S_{d,\alpha}(-1)\,,
\end{equation}
we find that the order $n$ of the transition is $\left\lceil \eta/(\eta-1)\right\rceil$, where $\left\lceil y\right\rceil$ denotes the smallest integer larger than or equal to $y$. Using the expression for $\eta$ given in Eq. \eqref{bd}, we find
\begin{equation}
n=
\begin{cases}
\left\lceil\frac{\nu-1}{\nu-2}\right\rceil & \text{for } 2<\nu<3,\\
\\
 2 & \text{for } \nu>3.\\
\end{cases}
\label{order}
\end{equation}
In other words, we observe a second-order phase transition for $\nu>3$, while, for $2<\nu<3$, the order of the transition depends continuously on $\nu$. For instance, when $\nu=5/2$ we find $n=3$. Notably, the order $n$ diverges when $\nu\to 2$, in agreement with the fact that for $\nu=2$ all derivatives of $\psi_{d,\alpha}(z)$ are continuous.

\subsection{Asymptotics of $\psi_{d,\alpha}(z)$}

Next, we are interested in the asymptotic behavior of $\psi_{d,\alpha}(z)$ for small and large $z$. For small enough $z$, a solution of the saddle point equation \eqref{SPE} always exists, with $q^*(z)$ small. Expanding Eq. \eqref{SPE} for small $q^*(z)$, we find
\begin{equation}
q^*(z)\simeq -\frac{d(\alpha+1)(\alpha+2)}{4}z.
\end{equation}
Plugging this solution into Eq. \eqref{psi_d_z} and expanding for small $z$, we obtain 
\begin{equation}
\psi_{d,\alpha}(z)\simeq \frac{d(1+\alpha)(2+\alpha)}{8}z^2\,.
\end{equation}
Comparing this result with Eq. \eqref{typical}, we notice that the small-argument behavior of the rate function smoothly connects with the typical Gaussian behavior in Eq. \eqref{typical}. We next consider the large-$z$ behavior of $\psi_{d,\alpha}(z)$. It is useful to distinguish different cases, depending on $\nu$.

\vskip 0.3cm

\noindent{\bf The case $\nu>2$:} For $\nu>2$, we already know that for $z>z_c$ the rate function is exactly given by just a linear function
\begin{equation}
\psi_{d,\alpha}(z)=z-S_{d,\alpha}(-1)\,,
\end{equation}
where $S_{d,\alpha}(q)$ is given in Eq. \eqref{S}. 

\vskip 0.3cm

\noindent{\bf The case $1<\nu<2$:} When $\nu<2$, for large $z$, we know that a unique solution $q^*(z)$ of the saddle point Eq. \eqref{SPE} exists for any $z$. Moreover, in the limit of large $z$, we expect $q^*(z)$ to be close to $-1$. Therefore, we plug $q=-1+s$ into the exponent of the integrand in Eq. \eqref{eq:PXN_7} and we use Eq. \eqref{eq:hypergeom1_N} to expand for small $s$. For $1<\nu<2$, we obtain
\begin{equation}
z~(-1+s)+S_{d,\alpha}(-1+s)\simeq z(-1+s)+S_{d,\alpha}(-1)+\tilde{a}s^{\nu-1}\,,
\end{equation}
where $\tilde{a}<0$ is a constant that depends on $d$ and $\alpha$. Setting to zero the first derivative of this expression with respect to $s$, we obtain
\begin{equation}
z+\tilde{a}(\nu-1)s^{\nu-2}=0\,.
\end{equation}
Thus, we find that, for large $z$, the solution of the saddle point equation can be written as 
\begin{equation}
q^*(z)\simeq-1+(\tilde{a}(1-\nu)/z)^{1/(2-\nu)}\,.
\end{equation}
Plugging this expression for $q^*(z)$ into Eq. \eqref{psi_d_z} and expanding for large $z$, we finally obtain
\begin{equation}
\psi_{d,\alpha}(z)=z-S_{d,\alpha}(-1)+O(z^{(1-\nu)/(2-\nu)})\,.
\end{equation}

\vskip 0.3cm

\noindent{\bf The case $0<\nu<1$:}  When $0<\nu<1$ the procedure above yields, to leading order in $s$,
\begin{equation}
z~(-1+s)+S_{d,\alpha}(-1+s)\simeq z(-1+s)+(\nu-1)\log\left(s\right)\,.
\end{equation}
Setting to zero the first derivative of this expression with respect to $s$, we find that the saddle point equation becomes, for $z\gg 1$,
\begin{equation}
z+\frac{\nu-1}{s}=0\,.
\end{equation}
Thus, we find that, to leading order, $q^*(z)\simeq -1+(1-\nu)/z$. Plugging this expression in Eq. \eqref{psi_d_z} and expanding for large $z$, we find that, for $0<\nu<1$,
\begin{equation}
\psi_{d,\alpha}(z)= z-(1-\nu)\log(z)+O(1)\,.
\end{equation}

To summarize, we have shown that, for $z\gg 1$,
\begin{equation}
\psi_{d,\alpha}(z)=
\begin{cases}
z-(1-\nu)\log(z)+O(1)& \text{for } 0<\nu<1\\
\\
z-S_{d,\alpha}(-1)+O(z^{(1-\nu)/(2-\nu)}) & \text{for } 1<\nu<2\\
\\
\\
z-S_{d,\alpha}(-1) & \text{for } \nu>2\,,\\
\\
\end{cases}
\end{equation}
where we recall that $S_{d,\alpha}(q)$ is given in Eq. \eqref{S}.

\subsection{Vicinity of the critical point: intermediate matching regime}
\label{sec:intermediate}

We now focus on the case $\nu>2$ where a condensation
is guaranteed to occur as $X$ exceeds a critical value $X_c$.
We want to investigate the behavior of $Z(X,N)$ in a small neighborhood of 
the critical point $X=X_c$ for large $N$ (see Fig. \ref{fig:Z}). 
In the following, we will assume that $\nu$ is not an integer number. 
The discussion below can be easily generalized to the case 
where $\nu$ is integer. We recall that $X_c=z_c N$, where the critical 
value $z_c$, given in Eq. \eqref{zc}, is of order one. 
Close to the transition, we write $X$ as
\begin{equation}
X=X_c+y N^{\lambda}\,,
\label{X_y}
\end{equation}
where $y$ is an order-one variable and $0<\lambda<1$ can be adjusted 
depending on $\nu$. Close to the transition, i.e. 
for $|X_{\rm ex}|=|X-X_c|\ll N$, we know that the contour integral in Eq. \eqref{eq:PXN_7} is dominated by values of $q$ close to $-1$. Thus, performing the change of variable $q\to s=-1+q$, we find
\begin{equation}
Z(X,N)=\frac{1}{2\pi i }\int_{\Gamma}ds~e^{(-1+s) X+N~S_{d,\alpha}(-1+s)}\,,
\end{equation}
where $S_{d,\alpha}(q)$ is given in Eq. \eqref{S}. Using Eq. \eqref{eq:hypergeom1_N}, we expand for small $s$ and we obtain
\begin{equation}
Z(X,N)\simeq B_N ~e^{-X}\frac{1}{2\pi i }\int_{\Gamma}ds~\exp\left[(X-X_c)s+N\left(a_{d,\alpha}~s^2+\ldots+b_{d,\alpha}~s^{\nu-1}+\ldots\right)\right]\,,
\end{equation}
where $s^{\nu-1}$ is the first non-analytic term of the expansion and $B_N$ is given in Eq. \eqref{BN}.
The constants $a_{d,\alpha}$ and $b_{d,\alpha}$ are given in Eqs. \eqref{a_d_alpha} and \eqref{b_d_alpha}. Using Eq. \eqref{X_y}, we obtain
\begin{equation}
Z(X,N)\simeq B_N e^{-X}\frac{1}{2\pi i }\int_{\Gamma}ds~\exp\left[ysN^{\lambda}+N\left(a_{d,\alpha}~s^2+\ldots+b_{d,\alpha}~s^{\nu-1}+\ldots\right)\right]\,.
\end{equation}
After the change of variable $s\to \tilde{s}=s N^{\lambda}$, we get
\begin{equation}
Z(X,N)\simeq B_N e^{-X}\frac{N^{-\lambda}}{2\pi i }\int_{\Gamma}d\tilde{s}~\exp\left[y\tilde{s}+a_{d,\alpha}~\tilde{s}^2 N^{1-2\lambda}+\ldots+b_{d,\alpha}~\tilde{s}^{\nu-1}N^{1-(\nu-1)\lambda}+\ldots\right]\,.
\label{PXN_lambda}
\end{equation}
Let us now consider the two cases $2<\nu<3$ and $\nu>3$ separately.

\begin{figure*}[t]
\centering
\includegraphics[width=0.8\textwidth]{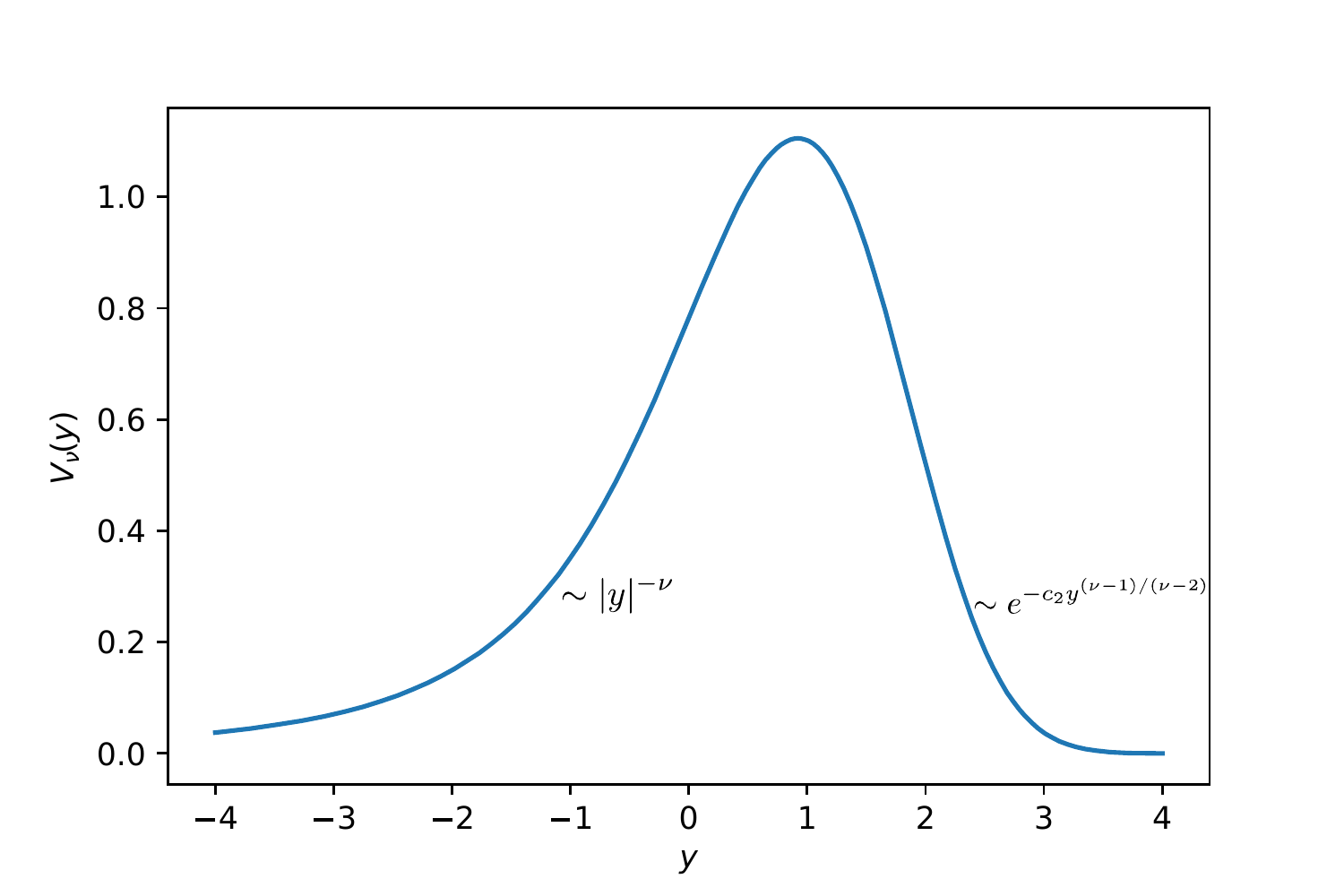} 
\caption{\label{fig:V} The function $V_{\nu}(y)$ versus $y$, for $\nu=5/2$. For $y\to\infty$, $V_{\nu}(y)$  decays exponentially fast as $e^{-c_2 y^{(\nu-1)/(\nu-2)}}$. For $y\to-\infty$, it has a power-law tail  $V_{\nu}(y)\sim |y|^{-\nu}$.
}
\end{figure*}

For $2<\nu<3$, we set $\lambda=1/(\nu-1)$ and, using the definition of $y$ in Eq. \eqref{X_y}, we obtain, to leading order,
\begin{equation}
Z(X,N)\simeq C_N \frac{1}{N^{1/(\nu-1)}}V_{\nu}\left(\frac{X_c-X}{N^{1/(\nu-1)}}\right)\,,
\label{PXN_V}
\end{equation}
where $C_N=B_N e^{-X}$ and
\begin{equation}
V_{\nu}(y)=\frac{1}{2\pi i }\int_{\Gamma}ds~\exp\left[-y~s+b_{d,\alpha}~s^{\nu-1}\right]\,.
\label{V}
\end{equation}
This same function $V_{\nu}(y)$ 
also appeared in the
analysis of the partition function
in mass transport models~\cite{EMZ06,EM08} and 
is shown in Fig. \ref{fig:V}. It has the following asymptotic 
behaviors~\cite{EMZ06} 
\begin{equation}
V_{\nu}(y)\simeq
\begin{cases}
A_{d,\alpha}|y|^{-\nu} &\text{  for  }y\to -\infty\\
\\
c_1~y^{(3-\nu)/(2(\nu-2))} e^{-c_2~ y^{(\nu-1)/(\nu-2)}}&\text{  for  }y\to \infty\,,
\end{cases}
\label{V_asymp}
\end{equation}
where $A_{d,\alpha}$ is given in Eq. \eqref{A_d_alpha},
\begin{equation}
c_1=\frac{1}{\left(2\pi(\nu-2)(b_{d,\alpha}(\nu-1))^{1/(\nu-2)}\right)^{1/2}}
\label{c1}
\end{equation}
and
\begin{equation}
c_2=\frac{\nu-2}{(\nu-1)(b_{d,\alpha}(\nu-1))^{1/(\nu-2)}}\,.
\label{c2}
\end{equation}
Performing the change of variable $s=r e^{\pm i \pi/2}$ for the upper and lower part of the imaginary-axis contour $\Gamma$, it is possible to rewrite the expression for $V_{\nu}(y)$ in Eq. \eqref{V} as
\begin{equation}
V_{\nu}(y)=\frac{1}{\pi  }\int_{0}^{\infty}dr~e^{b_{d,\alpha} \sin(\pi\nu/2)r^{\nu-1}}\cos\left[b_{d,\alpha}\cos(\pi\nu/2)r^{\nu-1}+yr\right]\,,
\label{V1}
\end{equation}
which can be easily evaluated numerically. Moreover, it is easy to check that $V_{\nu}(y)$ is positive and normalized to one. Using the asymptotic result for $y\to\infty$, one can check that the expression for $Z(X,N)$ in Eq. \eqref{PXN_V} matches smoothly to the expression obtained by saddle-point approximation for $z<z_c$ (see Eq. \eqref{psi_d_expansion_below}). Similarly, using the expansion for $y\to-\infty$, we observe that for $X_{\rm ex}=X-X_c\gg N^{1/(\nu-1)}$
\begin{equation}
Z(X,N)\simeq B_N e^{-X}\frac{N~A_{d,\alpha}}{(X-X_c)^{\nu}}\,,
\end{equation}
in agreement with the result in Eq. \eqref{eq:PXN_limit6}.

When $\nu>3$, we set $\lambda=1/2$ in Eq. \eqref{PXN_lambda} and we obtain
\begin{equation}
Z(X,N)\simeq B_N e^{-X}\frac{N^{-1/2}}{2\pi i }\int_{\Gamma}d\tilde{s}~\exp\left[y\tilde{s}+a_{d,\alpha}~\tilde{s}^2 \right]\,.
\label{ZXN_1/2}
\end{equation}
Computing the integral over $\tilde{s}$ and using the definition of $y$, we get
\begin{equation}
Z(X,N)\simeq C_N \frac{1}{\sqrt{4\pi a_{d,\alpha} N}} \exp\left[-\frac{\left(X-X_c\right)^2}{4a_{d,\alpha}N}\right]\,,
\label{gaussian_PX}
\end{equation}
where $C_N=B_N e^{-X_c}$. Thus, for $\nu>3$ the PDF of $X$ has, in the critical region where $|X-X_c|\sim \sqrt{N}$, a Gaussian shape. Actually, it is possible to check that this Gaussian form remains valid, on the left tail, on a larger region, depending on $\nu$. For instance, for $\nu>4$, it is valid up to $X_c-X\sim N^{3/4}$. Conversely, on the right tail, the expression in Eq. \eqref{gaussian_PX} remains valid up to $X-X_c\sim\sqrt{N\log(N)}$. Beyond this scale, i.e., for $X-X_c\gg \sqrt{N\log(N)}$, it is possible to show that 
\begin{equation}
Z(X,N)\simeq B_N e^{-X}\frac{N~A_{d,\alpha}}{(X-X_c)^{\nu}}\, ,
\label{gaussian_PX2}
\end{equation}
in agreement with the expression in Eq. \eqref{eq:PXN_limit6}, obtained for $X-X_c\sim O(N)$.

In order to describe the crossover between the Gaussian shape in Eq. \eqref{gaussian_PX} and the power-law tail in Eq. \eqref{gaussian_PX2}, one needs to keep the first singular term in the expansion in Eq. \eqref{ZXN_1/2}. From Eq. \eqref{PXN_lambda}, we obtain
\begin{equation}
Z(X,N)\simeq B_N  e^{-X}\frac{N^{-1/2}}{2\pi i }\int_{\Gamma}d\tilde{s}~\exp\left[y\tilde{s}+a_{d,\alpha}~\tilde{s}^2+b_{d,\alpha}\tilde{s}^{\nu-1}N^{-(\nu-3)/2} \right]\,.
\end{equation}
For large $N$, the integrand can be written as
\begin{equation}
Z(X,N)\simeq B_N e^{-X}\frac{1}{\sqrt{N}}~	g_N\left(\frac{X-X_c}{\sqrt{N}}\right)\,,
\label{ZgN}
\end{equation}
where 
\begin{equation}
g_N(y)=\frac{1}{2\pi i }\int_{\Gamma}ds~e^{ys+a_{d,\alpha}s^2}\left(1+b_{d,\alpha}s^{\nu-1}N^{-(\nu-3)/2}\right) \,.
\end{equation}
This function $g_N(y)$ can be rewritten as the sum of a Gaussian part and of power-law part
\begin{equation}
g_N(y)=\frac{1}{\sqrt{4\pi a_{d,\alpha}}}e^{-y^2/(4a_{d,\alpha})}+\frac{b_{d,\alpha}}{N^{(\nu-3)/2}}\frac{1}{2\pi i }\int_{\Gamma}ds~e^{ys+a_{d,\alpha}s^2}s^{\nu-1} \,.
\end{equation}
When $y\sim O(1)$, the Gaussian term is always leading and one obtains the result in Eq. \eqref{gaussian_PX}. On the other hand, when $y\sim O(\sqrt{N})$, the power-law part dominates, coherently with the result in Eq. \eqref{gaussian_PX2}. We now want to describe the crossover between these two regimes. When $y\gg 1$ the integral over $s$ can be approximated as 
\begin{equation}
g_N(y)\simeq\frac{1}{\sqrt{4\pi a_{d,\alpha}}}e^{-y^2/(4a_{d,\alpha})}+\frac{A_{d,\alpha}}{N^{(\nu-3)/2}}\frac{1}{y^{\nu} }\,,
\label{gN}
\end{equation}
where we have used the expression for $b_{d,\alpha}$, given in Eq. \eqref{A_d_alpha}. We want to find $c_N$ and $d_N$, such that $w=(y-c_N)/d_N$ is fixed for large $N$. Plugging
\begin{equation}
y=c_N+d_N w
\end{equation}
in Eq. \eqref{gN}, we obtain
\begin{equation}
g_N(y=c_N+d_N w)=\frac{A_{d,\alpha}}{N^{(\nu-3)/2}}\frac{1}{c_N^{\nu}(1+d_N w/c_N)^{\nu} }\left[1+\frac{N^{(\nu-3)/2}c_N^{\nu}(1+d_N w/c_N)^{\nu} }{\sqrt{4\pi a_{d,\alpha}}~A_{d,\alpha}}e^{-(c_N^2+2c_N d_N w+d_N^2 w^2)/(4a_{d,\alpha})}\right]\,.
\end{equation}
We now choose $c_N$ such that
\begin{equation}
N^{(\nu-3)/2}c_N^{\nu}e^{-c_N^2/(4a_{d,\alpha})}=1\,.
\end{equation}
Thus, to leading order
\begin{equation}
c_N\simeq \sqrt{2a_{d,\alpha}(\nu-3)\log(N)}\,.
\label{cN}
\end{equation}
Moreover, we choose $d_N=1/c_N$. Then, to leading order, we obtain
\begin{equation}
g_N(y=c_N+d_N w)=\frac{A_{d,\alpha}}{N^{(\nu-3)/2}}\frac{1}{(2a_{d,\alpha}(\nu-3)\log(N))^{\nu/2}}\left[1+\frac{	1}{\sqrt{4\pi a_{d,\alpha}}~A_{d,\alpha}}e^{- w/(2a_{d,\alpha})}\right]\,.
\end{equation}
Finally, using Eq. \eqref{ZgN}, we find that
\begin{equation}
Z(X,N)\simeq B_N \frac{A_{d,\alpha}}{c_N^{\nu} N^{(\nu-3)/2}} e^{-X}~ h\left[c_N\left(\frac{X-X_c}{\sqrt{N}}-c_N\right)\right]\,,
\end{equation}
where $c_N$ is given in Eq. \eqref{cN} and
\begin{equation}
h(w)=1+\frac{	1}{\sqrt{4\pi a_{d,\alpha}}~A_{d,\alpha}}e^{- w/(2a_{d,\alpha})}\,.
\label{h}
\end{equation}
Overall, we have shown that the crossover occurs for $X-X_c\sim\sqrt{N\log(N)}$ and that it is described by the function $h(w)$.

To summarize, we have shown that, for any $\nu>2$ and for $X\sim X_c$, the PDF $Z(X,N)$ can be always written as
\begin{equation}
Z(X,N)\simeq C_N  p_{\rm cond}(X_c-X,N)\,,
\label{Z_cond}
\end{equation}
where the function $p_{\rm cond}(y,N)$ assumes different expressions depending on $\nu$. The reason behind the choice of the subscript \emph{cond} will become clear in the next section. Using Eq. \eqref{PXN_V}, we find that for $2<\nu<3$
\begin{equation}
p_{\rm cond}(y,N)\simeq \frac{1}{N^{1/(\nu-1)}}V_{\nu}\left(\frac{y}{N^{1/(\nu-1)}}\right)\,,
\label{def_p_cond_anom}
\end{equation}
where the function $V_{\nu}(y)$ in Eq. \eqref{V}. On the other hand, for $\nu>3$, we find
\begin{equation}
p_{\rm cond}(y,N)\simeq
\begin{cases}
\frac{N A_{d,\alpha} }{|y|^{\nu}} ~~~&\text{   for   }y\ll-\sqrt{N\log(N)}~,\\
\\
\frac{A_{d,\alpha}}{c_N^{\nu} N^{(\nu-3)/2}}  h\left[\frac{c_N}{\sqrt{N}}\left(|y|-\sqrt{N}c_N\right)\right]
~~~&\text{   for   } y\simeq-\sqrt{N\log(N)}~,\\
\\
\frac{1}{\sqrt{4\pi a_{d,\alpha} N}} e^{-y^2/(4a_{d,\alpha}N)}~~~&\text{   for   }y\gg-\sqrt{N\log(N)}\,,
\end{cases}
\label{def_p_cond_norm}
\end{equation}
where $A_{d,\alpha}$ and $a_{d,\alpha}$ are given in Eqs. \eqref{A_d_alpha} and \eqref{a_d_alpha}. The function $h(w)$ is given in Eq. \eqref{h} and $c_N\sim \sqrt{\log(N)}$ is given in Eq. \eqref{cN}. In both cases, it is possible to check that the function $p_{\rm cond}(y,N)$ is positive and normalized over $y$, for large $N$. 	Using Eqs. \eqref{def_p_cond_anom} and \eqref{def_p_cond_norm}, we find that for $y\to \infty$
\begin{equation}
p_{\rm cond}(y,N)\sim
\begin{cases}
\exp\left[-c_2 \frac{y^{(\nu-1)/(\nu-2)}}{N^{1/(\nu-2)}}\right]~~~&\text{   for   }2<\nu<3~,\\
\\
\exp\left[-\frac{y^2}{4a_{d,\alpha}N}\right]
~~~&\text{   for   } \nu>3\,,
\end{cases}
\label{p_cond_asympt1}
\end{equation}
where $c_2$ is given in Eq. \eqref{c2}. On the other hand, for $y\to-\infty$, we obtain
\begin{equation}
p_{\rm cond}(y,N)\simeq \frac{NA_{d,\alpha}}{|y|^{\nu}}\,,
\label{p_cond_asympt2}
\end{equation}
for any $\nu>2$.

\section{Marginal distribution of a single jump}
\label{sec:marginal}

In this section we investigate the marginal PDF $p(x|X)$ of a single-run 
displacement $x$, conditioned on the final $x$-component displacement 
$X$. This means that if we pick at random one of the $N$ runs with
the total displacement $X$ fixed, what
is the distribution of the size of this run? 
Note that the displacements $x_1\,,\ldots\,,x_N$ are i.i.d. random 
variables, since we are considering the fixed-$N$ ensemble. Thus, $x$ 
can be identified with any of these variables, say for simplicity 
$x=x_1$. Then, the conditional PDF of $x$ is given by
\begin{equation}
p(x|X)=\frac{p(x)\int_{-\infty}^{\infty}dx_2~\ldots\int_{-\infty}^{\infty}dx_N\left[\prod_{i=2}^N p(x_i)\right]\delta\left(X-x-\sum_{i=2}^{\infty}x_i\right)}{\int_{-\infty}^{\infty}dx_1~\ldots\int_{-\infty}^{\infty}dx_N\left[\prod_{i=1}^N p(x_i)\right]\delta\left(X-\sum_{i=1}^{\infty}x_i\right)}\,,
\end{equation}
which can be rewritten as
\begin{equation}
p(x|X)=p(x)~\frac{Z(X-x,N-1)}{Z(X,N)}\,,
\label{marginal}
\end{equation}
where we have used the definition of $Z(X,N)$ in Eq. \eqref{ZXN}.
We are interested in the large-deviation regime where $X=zN$ and 
$z$ is of order one. As explained in the previous section, 
for $\nu>2$, the system undergoes a phase transition at a 
critical value $z_c$ of the parameter $z$.
In this section we show that this transition shows up very clearly
in the marginal distribution $p(x|X)$ which has very different
behavior in the subcritical ($X<z_c N$) and the supercritical ($X>z_c N$)
phases. In the subcritical `fluid' phase, $p(x|X)$ is a monotonically
decreasing function of $x$ with an exponential tail.
For $X=z_c N$ we have a critical fluid where $p(x|X)$ still
decays monotonically with increasing $x$, but now
as a power law $\sim x^{-\nu}$ for large $x$. Finally,
in the supercritical phase ($X>z_c N$), $p(x|X)$ becomes non-monotonic as a function of $x$, developing
in particular a bump centered at $x= X_{\rm ex}= X- z_c N$ (see
Fig. \ref{fig:marginal}). Let us discuss these three cases separately.

\vskip 0.3cm

\noindent{{\bf {Subcritical phase ($X<X_c$):}}} Plugging the expression for $Z(X,N)$ given in Eq. \eqref{eq:PXN_7}, into Eq. \eqref{marginal}, we obtain
\begin{equation}
p(x|X)\simeq p(x)~\frac{\int_{\Gamma}dq~e^{-qx}e^{N \left[qz+S_{d,0}(q)\right]}}{\int_{\Gamma}dq~e^{N \left[qz+S_{d,0}(q)\right]}}\,,
\label{marginal1}
\end{equation}
where the integrals are performed over the imaginary-axis Bromwich contour $\Gamma$ (see Fig. \ref{fig:S}). In Section \ref{sec:position}, we have shown that the integrals above are dominated by the solution $q^*(z)$ of the saddle point equation \eqref{SPE}. In the subcritical phase $z<z_c$ such solution $q^*(z)>-1$ always exists. Thus, from Eq. \eqref{marginal1}, we obtain
\begin{equation}
p(x|X)\simeq p(x)e^{-q^*(z)\,x}\,.
\label{marginal2}
\end{equation}
Using the large-$x$ behavior of $p(x)$, given in Eq. \eqref{px_asym}, we find that, for $x\gg 1$
\begin{equation}
p(x|X)\simeq A_{d,\alpha}  x^{-\nu}~e^{- x/\xi}  \,,
\end{equation}
where
\begin{equation}
\xi=\frac{1}{1+q^*(z)}\,.
\end{equation}
Thus for $X<X_c$, the PDF $p(x|X)$ decays as a function of $x$ on a typical length $\xi>0$. Recalling that $q^*(z)\to -1$ for $z\to z_c$, we find that $\xi$ diverges when the system approaches the phase transition. In particular, it is possible to show that $\xi$ diverges, for $z\to z_c$ from below, as
\begin{equation}
\xi\sim
\begin{cases}
(z_c-z)^{-1/(\nu-2)}
 &\text{  for  }2<\nu<3\\
\\
(z_c-z)^{-1/2}
 &\text{  for  }\nu>3\,.\\
\end{cases}
\end{equation}

\vskip 0.3cm

\noindent{{\bf {Critical phase ($X=X_c$):}}} Exactly at the transition point, the typical length $\xi$ diverges and therefore the single-jump distribution develops a power-law tail for large $x$
\begin{equation}
p(x|X)\simeq A_{d,\alpha}  x^{-\nu}\,.
\label{critical}
\end{equation}
Note that the result in Eq. \eqref{critical} is only valid for $x\ll O(N)$. This is because when computing the integral in Eq. \eqref{marginal1} we have assumed that the factor $e^{-qx}$ does not contribute to the saddle point equation, which is only true if $x\ll O(N)$. As we will show below, configurations with $x\sim O(N)$ are exponentially rare when $X=X_c$.

We note that the behavior of our system in the subcritical phase and at the critical point is somewhat reminiscent of a standard phase transition such as in the Ising model in $d\geq2$. The subcritical phase ``corresponds'' to the paramagnetic phase of the Ising model. The marginal distribution $p(x|X)$ plays an ``analogous'' role as the spin-spin correlation function in the Ising model. In the case of the Ising model, the correlation function decays exponentially with distance with a characteristic correlation length that diverges as one approaches the critical point. Similarly, here there is a characteristic run length $\xi$ characterizing the exponential decay of $p(x|X)$ with $x$ on the subcritical side, with $\xi$ diverging as one approaches the critical point.

\vskip 0.3cm

\noindent{{\bf {Supercritical phase ($X=X_c$):}}} Above the transition, $q^*(z)$ freezes to the value $-1$ and therefore the typical length $\xi$ remains infinite. Indeed, for $X>X_c$, we expect that the condensate develops as a bump in the tail of the PDF $p(x|X)$. The location of this bump is related to the fraction of $X$ that is contained in the condensate. Moreover, the area under the bump is the probability that a particular single-run displacement becomes the condensate. In the presence of a single condensate, this area should therefore be $1/N$. Finally, as we will show, the shape of the bump is related to the order of the phase transition.
Since we expect the condensate to contain a finite fraction of $X$, we need to investigate configurations where $x\sim O(N)$. Thus, it is useful to define the scaled variable $y=x/N$, which is of order one when $x\sim O(N)$, and to rewrite $p(x|X)$ as
\begin{equation}
p(x=yN|X=zN)=p(yN)~\frac{Z(N(z-y),N-1)}{Z(Nz,N)}\,.
\label{marginal3}
\end{equation}
Let us first investigate the exponential part of $p(x|X)$.

\begin{figure}
\centering
\includegraphics[width=0.7\textwidth]{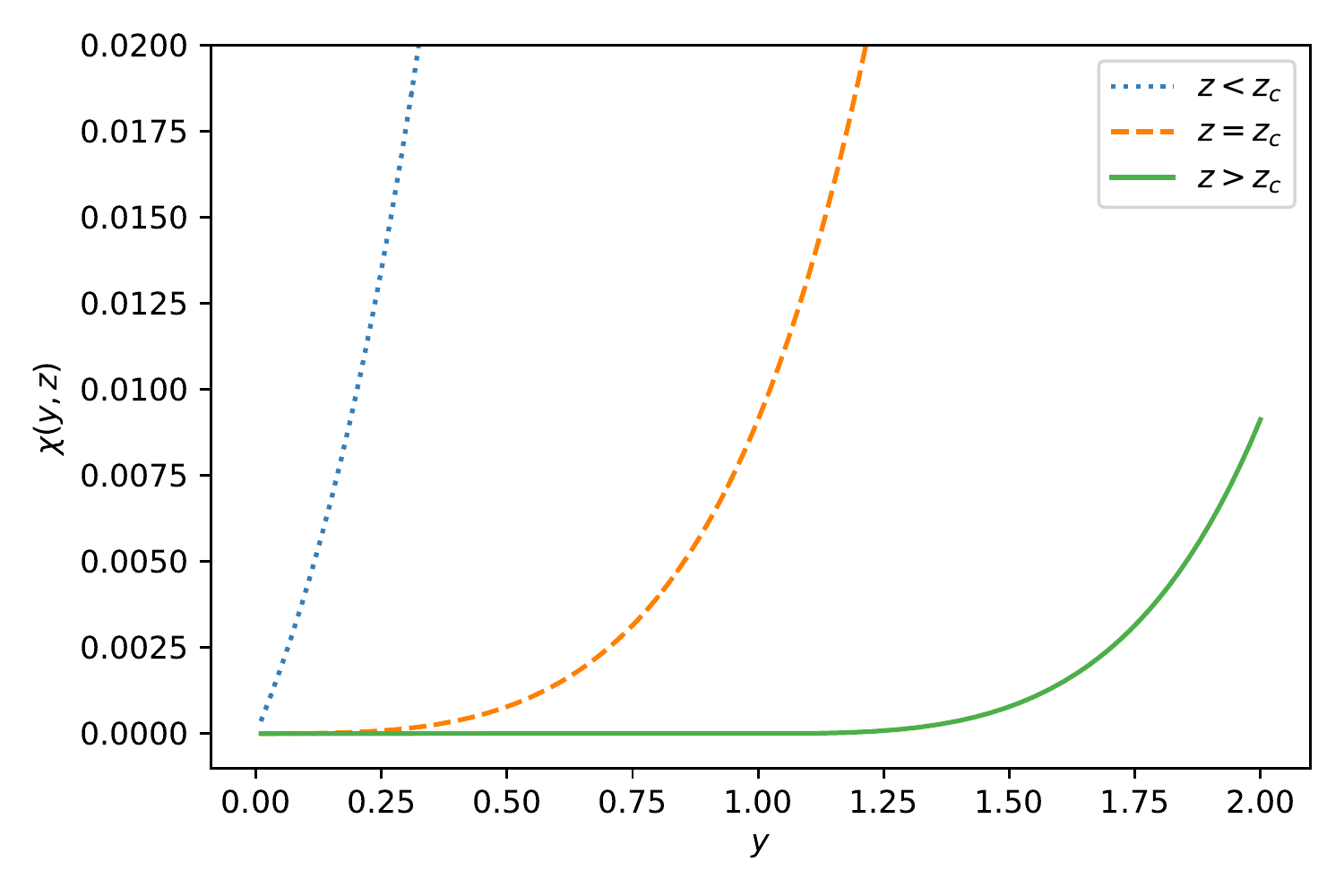} 
\caption{\label{fig:chi} Rate function $\chi(y,z)$ versus $y$, for different values of $z$. The curves are obtained from the exact result in Eq. \eqref{Azy}, for $d=6$ and $\alpha=0$. For $z<z_c$, we observe that $\chi(y,z)>0$ for any $y>0$. Conversely, for $z>z_c$, the rate function $\chi(y,z)$ vanishes for $y<z_{\rm ex}=z-z_c$ (in this case $z_{\rm ex}=1$).
}
\end{figure}

Plugging the large-deviation form of $Z(X,N)$, given in Eq. \eqref{eq:d<5} and the large-$x$ behavior of $p(x)$, given in Eq. \eqref{px_asym}, into Eq. \eqref{marginal3}, we find
\begin{equation}
p(x|X)\sim \exp\left[-N \chi\left(\frac{x}{ N},\frac{ X}{ N}\right)\right]\,,
\end{equation}
where
\begin{equation}
\chi(y,z)=\psi_{d,\alpha}(z-y)+y-\psi_{d,\alpha}(z)\,,
\label{Azy}
\end{equation}
and $\psi_{d,\alpha}(z)$ is given in Eq. \eqref{psi_d_cond}. Thus, in the regime where $x\sim O(N)$, the PDF $p(x|X)$ assumes a large deviation form with rate function $\chi(y,z)$. The rate function $\chi(y,z)$ is shown in Fig. \ref{fig:chi} as a function of $y$, for different values of $z$.

Using the expression of $\psi_{d,\alpha}(z)$ in Eq. \eqref{psi_d_cond}, it is easy to show that, for $y>z_{\rm ex}$, one has $\chi(y,z)>0$, where we recall that $z_{\rm ex}=z-z_c$ (see Fig. \ref{fig:chi}). Thus, configurations where $x>X_{\rm ex}$ become exponentially rare for large $x$. On the other hand, for $y<z_{\rm ex}$, we find that $\chi(y,z)=0$. This means that configurations with $y<z_{\rm ex}$ are not forbidden and that a bump can arise in the tail of $p(x|X)$. Note however that where $\chi(y,z)=0$ the large-deviation description fails and that we need to carefully consider the full distribution, and not just the exponential part.

Therefore, we focus on the region $0<y<z_{\rm ex}$, where we have just shown that the exponential part of the distribution of $p(x|X)$ vanishes. We use the expression in Eq. \eqref{eq:PXN_limit6} to approximate both the numerator and the denominator of Eq. \eqref{marginal3}. Using also the asymptotic expression of $p(x)$ for large $x$, given in Eq. \eqref{px_asym}, we obtain
\begin{equation}
p(x|X)\simeq\frac{A_{d,\alpha}}{N^{\nu}y^{\nu}}\frac{1}{(1-y/z_{\rm ex})^{\nu}}\,,
\end{equation}
where $A_{d,\alpha}$ is given in Eq. \eqref{A_d_alpha} and we recall that $z_{\rm ex}=z-z_c$. Going back to the original variables $x=y/N$ and $X=z/N$, we find that in the regime where both $x$ and $X$ scale linearly with $N$, with $0<x<X_{\rm ex}$, 
\begin{equation}
p(x|X)\simeq\frac{A_{d,\alpha}}{x^{\nu}}\frac{1}{(1-x/X_{\rm ex})^{\nu}}\,.
\end{equation}
Note that this approximate expression breaks down when $x\to X_{\rm ex}$. As we will see, at $x\sim X_{\rm ex}$ the condensate bump appears in the tail of the distribution of $p(x|X)$.

Let us now focus on the region where $x\sim X_{\rm ex}$, where we expect the bump to appear. In this region, the numerator of Eq. \eqref{marginal3} can be approximated using the expression in Eq. \eqref{Z_cond}. On the other hand, the denominator can be approximated with the expression in Eq. \eqref{eq:PXN_limit6}. Using also the large-$x$ expansion of $p(x)$, given in Eq. \eqref{px_asym}, we obtain, to leading order
\begin{equation}
p(x|X)\simeq  \frac{1 }{N}p_{\rm cond}(x-X_{\rm ex},N)\,,
\end{equation}
where $p_{\rm cond}(y,N)$ is given in Eq. \eqref{p_cond_anomalous} for $2<\nu<3$ and in Eq. \eqref{p_cond_normal} for $\nu>3$. Thus, for $z>z_c$, a condensate bump appears at $x\sim X_{\rm ex}$. For $2<\nu<3$, the condensate bump has an anomalous shape described by the function $V_{\nu}(y)$ in Eq. \eqref{V}, with fluctuations of order $O(N^{1/(\nu-1)})$. For $\nu>3$, the condensate has a Gaussian shape in the vicinity of its peak, with fluctuations of order $ O(\sqrt{N})$. In both cases, we observe that the bump width vanishes relative to its location, since $X_{\rm ex}\sim O(N)$. Moreover, the area under the bump corresponds to the probability that the condensate appears in a particular running-phase. Since $p_{\rm cond}(y,N)$ is normalized to one, we find that this area is $1/N$, signaling the presence of a single condensate. 
The results above are summarized in Fig. \ref{fig:scales} and are in agreement with the numerical simulations, presented in Fig. \ref{fig:marginal}. The asymptotic behaviors of $p_{\rm cond}(y,N)$ are given in Eqs. \eqref{p_cond_asympt1} and \eqref{p_cond_asympt2}.

Finally, it is also instructive to investigate the condensate fraction $m_c$, defined as the fraction of the total displacement that is carried by the condensate. In this section we have shown that, for $\nu>2$, the condensate is located at $x=X_{\rm ex}$, with sublinear fluctuations around this value. Thus, for $N\to \infty$, the condensate fraction converges to
\begin{equation}
m_c=\frac{X - X_{c}}{X}\,,
\end{equation}
or, in terms of the scaled variable $z$,
\begin{equation}
m_c=\frac{z-z_{c}}{z}\,.
\label{mc}
\end{equation}
This quantity $m_c$ is the natural order parameter of the system, with $z$ being the corresponding control parameter. Indeed, for $z<z_c$ no condensate can form and thus $m_c=0$. On the other hand, above the transition, the condensate fraction becomes positive.

\section{Fixed-$T$ ensemble}
\label{sec:T}

In this section, we consider a single RTP in the fixed-$T$ ensemble. As explained in Section \ref{sec:model}, according to this alternative model, the total duration $T$ of the RTP trajectory is fixed and the number $N$ of running phases is a random variable. While the fixed-$N$ ensemble can be easily mapped into a discrete-time random walk, the fixed-$T$ ensemble is a truly continuous-time process and is often taken as the standard model for RTPs. The goal of this section is to show that the results of this paper can be extended also to the fixed-$T$ ensemble. The technique that we apply in the following is based on a mapping of the continuous-time trajectory of the RTP to a discrete-time random walk in Laplace space. This method has been used to compute several observables, e.g. the survival probability, of a fixed-$T$ RTP \cite{MLDM20a,MLDM20,BMS2021,LM20}. For the sake of simplicity, we will henceforth focus on the model where the speed $v_0$ of the particle is kept fixed. We recall that this corresponds to taking the limit $\alpha\to 0$. It is easy to extend the computations of this section to generic $\alpha>0$.

By definition, we consider the starting point to be a tumbling event, thus $N\geq 1$ indicates also the total number of tumblings. Let us denote by $\tau_i$ the duration of the $i$-th running phase, i.e., the running phase after the $i$-th tumbling. We also recall that $x_i$ denotes the $x$-component displacement of the particle during the $i$-th running phase. Since we are assuming that the tumblings happen with a constant rate $\gamma$, for the first $N-1$ running phases the PDF of $\tau_i$ is given by
\begin{equation}
P(\tau_i)=\gamma e^{-\gamma \tau_i}\,.
\end{equation}
However, since we are fixing the total time $T$, the last running phase $\tau_N$ is yet to be completed and thus its probability weight is given by
\begin{equation}
P(\tau_N)=\int_{\tau_N}^{\infty}dt\, \gamma ~ e^{-\gamma t}=e^{-\gamma \tau_N}\,.
\end{equation}
Thus, the joint probability of the running times $\{\tau_i\}=\tau_1\,,\ldots,\tau_N$ and of the number $N$ of tumblings, fixing the total time $T$, is given by
\begin{equation}
P(\{\tau_i\},N,T)=\left[\prod_{i=1}^{N-1}\gamma e^{-\gamma \tau_i}\right]e^{-\gamma\tau_N}\delta\left(\sum_{i=1}^{N}\tau_i-T\right)\,,
\label{eq:joint_tau}
\end{equation}
where the delta function constrains the total time to be $T$. Note that in the expression in Eq. \eqref{eq:joint_tau}, while $\{\tau_i\}$ and $N$ are random variables, the total time $T$ is a fixed parameter of the problem. We will use this convention for the rest of this section. We now want to write the joint PDF of the $x$-direction displacements $\{x_i\}=x_1\,,\ldots\,,x_N$, of the running times $\{\tau_i\}$ and of the number $N$ of tumblings, given the total fixed time $T$. This probability can be written as
\begin{equation}
P(\{x_i\},\{\tau_i\},N,T)=P(\{x_i\}|\{\tau_i\})P(\{\tau_i\},N,T)\,,
\label{eq:joint_x_tau}
\end{equation}
where $P(\{x_i\}|\{\tau_i\})$ denotes the probability density of the displacements $\{x_i\}$, conditioned on the running times $\{\tau_i\}$. This joint probability factorizes as
\begin{equation}
P(\{x_i\}|\{\tau_i\})=\prod_{i=1}^{N}P(x_i|\tau_i)\,.
\label{eq:joint_x|tau}
\end{equation}
This PDF $P(x_i|\tau_i)$ can then be computed as follows. During the $i$-th running phase the particle moves with constant velocity $v_0$. Thus, denoting by $\vec{l_i}$ the displacement in the $d$-dimensional space during the $i$-th running phase, we know that the norm $l_i=|\vec{l_i}|$ is simply given by $l_i=v_0 \tau_i$. We also know that the direction of $\vec{l_i}$ is uniformly distributed. 
One can show (see Appendix \ref{app:p(x)}) that the distribution of the $x$-component $x$ of a $d$-dimensional vector $\vec{\ell}$ with random direction and norm $\ell$ is given by
\begin{equation}
p(x|l)=\frac1l f_d\left(\frac{x}{l}\right)\,,
\label{eq:PDF_x|ell}
\end{equation}
where $f_d(z)$ is given in Eq. \eqref{fd}. Plugging this result into Eq. \eqref{eq:joint_x|tau}, we obtain
\begin{equation}
P(\{x_i\}|\{\tau_i\})=\prod_{i=1}^{N}\frac{1}{v_0\tau_i} f_d\left(\frac{x_i}{v_0\tau_i}\right)\,.
\label{eq:joint_x|tau2}
\end{equation}
Plugging the expressions for $P(\{\tau_i\},N,T)$  and $P(\{x_i\}|\{\tau_i\})$, given in Eqs. \eqref{eq:joint_tau} and \eqref{eq:joint_x|tau2} respectively, into Eq. \eqref{eq:joint_x_tau}, we find that
\begin{equation}
P(\{x_i\},\{\tau_i\},N,T)=\frac{1}{\gamma}\left[\prod_{i=1}^{N}\gamma e^{-\gamma \tau_i}\frac{1}{v_0\tau_i} f_d\left(\frac{x_i}{v_0\tau_i}\right)\right]\delta\left(\sum_{i=1}^{N}\tau_i-T\right)\,.
\label{eq:joint_x_tau2}
\end{equation}
Integrating over the variables $\{\tau_i\}$ we finally obtain the joint PDF of the displacements $\{x_i\}$ and of the number $N$ of tumblings, given the total time $T$, 
\begin{equation}
P(\{x_i\},N,T)=\frac{1}{\gamma}\int_{0}^{\infty}d\tau_1\,\ldots\int_{0}^{\infty}d\tau_N\,\left[\prod_{i=1}^{N}\gamma e^{-\gamma \tau_i}\frac{1}{v_0\tau_i} f_d\left(\frac{x_i}{v_0\tau_i}\right)\right]\delta\left(\sum_{i=1}^{N}\tau_i-T\right)\,.\\
\label{eq:joint_x}
\end{equation}

The PDF $Z(X,T)$ of the final position $X$ at fixed time $T$ can then be written as
\begin{equation}
Z(X,T)=\sum_{N=1}^{\infty}\int_{-\infty}^{\infty}dx_1\,\ldots\int_{-\infty}^{\infty}dx_N\,P(\{x_i\},N,T)\delta\left(\sum_{i=1}^{N}x_i-X\right)\,,
\label{eq:PDF_X}
\end{equation}
where the delta function constraints the final position to be $X$. Note that in Eq. \eqref{eq:PDF_X} we integrate out the displacement variables $\{x_i\}$ and we sum over the total number $N$ of tumblings in order to obtain the marginal PDF of $X$. Finally, plugging the expression for $P(\{x_i\},N,T)$, given in Eq. \eqref{eq:joint_x}, into Eq. \eqref{eq:PDF_X}, we obtain
\begin{equation}
Z(X,T)=\frac{1}{\gamma}\sum_{N=1}^{\infty}\prod_{i=1}^{N}\int_{-\infty}^{\infty}dx_i \int_{0}^{\infty} d\tau_i\,\gamma e^{-\gamma \tau_i}\frac{1}{v_0\tau_i} f_d\left(\frac{x_i}{v_0\tau_i}\right)\delta\left(\sum_{i=1}^{N}\tau_i-T\right)\delta\left(\sum_{i=1}^{N}x_i-X\right)\,.
\label{eq:PDF_X2}
\end{equation}
It is useful to rewrite the delta functions in Eq. \eqref{eq:PDF_X2} using 
\begin{equation}
\delta(X)=\frac{1}{2\pi i}\int_{\Gamma}dq~e^{-qX}\,,
\end{equation}
and
\begin{equation}
\delta(T)=\frac{1}{2\pi i}\int_{\Gamma'}ds~e^{-sT}\,,
\end{equation}
where $\Gamma$ and $\Gamma'$ are imaginary-axis Bromwich contours in the complex $q$ and $s$ plane, respectively. This yields
\begin{equation}
Z(X,T)=\frac{1}{\gamma}\sum_{N=1}^{\infty}\prod_{i=1}^{N}\int_{-\infty}^{\infty}dx_i \int_{0}^{\infty} d\tau_i\,\gamma e^{-\gamma \tau_i}\frac{1}{v_0\tau_i} f_d\left(\frac{x_i}{v_0\tau_i}\right)\frac{1}{2\pi i}\int_{\Gamma}dq~e^{-q\sum_{i=1}^{N}x_i+qX}\frac{1}{2\pi i}\int_{\Gamma'}ds~e^{-s\sum_{i=1}^N \tau_i+sT}\,.
\end{equation}
The variables $\{x_i\}$ and $\{\tau_i\}$ are now fully decoupled and the expression above can be rewritten as
\begin{equation}
Z(X,T)=\frac{1}{\gamma}\frac{1}{2\pi i}\int_{\Gamma}dq~e^{qX}\frac{1}{2\pi i}\int_{\Gamma'}ds~e^{sT}\sum_{N=1}^{\infty}\left[\hat{p}(q,s)\right]^{N}=\frac{1}{\gamma}\frac{1}{2\pi i}\int_{\Gamma}dq~e^{qX}\frac{1}{2\pi i}\int_{\Gamma'}ds~e^{sT}\frac{\hat{p}(q,s)}{1-\hat{p}(q,s)}\,,
\label{eq:PDF_X3}
\end{equation}
where
\begin{equation}
\hat{p}(q,s)=\int_{-\infty}^{\infty}dx~e^{-qx} \int_{0}^{\infty} d\tau\,e^{-s\tau}\gamma e^{-\gamma \tau}\frac{1}{v_0\tau} f_d\left(\frac{x}{v_0\tau}\right)\,.
\label{eq:hat_p}
\end{equation}
Plugging the expression of $f_d(z)$, given in Eq. \eqref{fd}, into Eq. \eqref{eq:hat_p}, we obtain, after few steps of algebra, 
\begin{equation}
\hat{p}(q,s)=\frac{\gamma}{\gamma+s} ~{}_2F_1\left(\frac12,1,\frac{d}{2},\left(\frac{v_0 ~ q}{\gamma+s}\right)^2\right)\,,
\label{eq:hat_p2}
\end{equation}
where ${}_2F_1\left(a,b,c,z\right)$ is the ordinary hypergeometric function. This Fourier-Laplace transform of the effective jump distribution was also computed in Ref. \cite{MAD12}, but explicit expressions were given only for $d=1,2,$ and $3$. It is easy to check (using explicit expressions for the hypergeometric function) that our exact expression in Eq. \eqref{eq:hat_p2}, valid for all $d$, coincides with those of Ref. \cite{MAD12} for $d=1,2,$ and $3$. Plugging this result for $\hat{p}(q,s)$ into Eq. \eqref{eq:PDF_X3}, we obtain
\begin{equation}
Z(X,T)=\frac{1}{2\pi i}\int_{\Gamma}dq~e^{qX}\int_{\Gamma'}ds~e^{sT}\frac{ {}_2F_1\left(\frac12,1,\frac{d}{2},\left(\frac{v_0 ~ q}{\gamma+s}\right)^2\right)}{\gamma+s-\gamma~ {}_2F_1\left(\frac12,1,\frac{d}{2},\left(\frac{v_0 ~ q}{\gamma+s}\right)^2\right)}\,.
\label{eq:PDF_X_transf2}
\end{equation}
It is useful to perform the change of variable $q\to \tilde{q}=-v_0 q/(\gamma+s)$, which yields
\begin{equation}
Z(X,T)=\frac{1}{v_0}\frac{1}{2\pi i}\int_{\Gamma'}d\tilde{q}\,e^{-\gamma  X \tilde{q}/v_0} ~{}_2F_1\left(\frac12,1,\frac{d}{2},\tilde{q}^2\right)\frac{1}{2\pi i}\int_{\Gamma}ds\,e^{s (T-X\tilde{q}/v_0)} \frac{(\gamma+s)}{\gamma+s-\gamma ~{}_2F_1\left(\frac12,1,\frac{d}{2},\tilde{q}^2\right)}\,.
\label{eq:PDF_X_final2b}
\end{equation}
We remark that up to now no approximation has been made and that the expression for $Z(X,T)$ in Eq. \eqref{eq:PDF_X_final2b} is exact for any $X$ and $T$. For simplicity, we now set $\gamma=v_0=1$, and we obtain
\begin{equation}
Z(X,T)=\frac{1}{2\pi i}\int_{\Gamma'}d\tilde{q}\,e^{-  X \tilde{q}} ~{}_2F_1\left(\frac12,1,\frac{d}{2},\tilde{q}^2\right)\frac{1}{2\pi i}\int_{\Gamma}ds\,e^{s (T-X\tilde{q})} \frac{(1+s)}{1+s- ~{}_2F_1\left(\frac12,1,\frac{d}{2},\tilde{q}^2\right)}\,.
\label{eq:PDF_X_final2}
\end{equation}

\subsection{Typical regime}
We now focus on the late time limit $T\to \infty$. To investigate this limit, we expand for small $s$ on the right-hand side of Eq. \eqref{eq:PDF_X_final2} and we obtain
\begin{equation}
Z(X,T)\simeq\frac{1}{2\pi i}\int_{\Gamma'}dq\,e^{-  X q}~{}_2F_1\left(\frac12,1,\frac{d}{2},q^2\right) \frac{1}{2\pi i}\int_{\Gamma}ds\,e^{s (T-Xq)} \frac{1}{s- \left[{}_2F_1\left(\frac12,1,\frac{d}{2},q^2\right)-1\right]}\,.
\label{eq:PDF_X_limit1}
\end{equation}
Now the integral over $s$ can be easily computed and one obtains
\begin{equation}
Z(X,T)\simeq\frac{1}{2\pi i}\int_{\Gamma'}dq\,{}_2F_1\left(\frac12,1,\frac{d}{2},q^2\right) \exp\left[-X q~ {}_2F_1\left(\frac12,1,\frac{d}{2},q^2\right)-T\left( 1-{}_2F_1\left(\frac12,1,\frac{d}{2},q^2\right)\right) \right] \,.
\label{eq:PDF_X_limit2}
\end{equation}
In the typical regime the variable $X$ scales for large $T$ as $\sqrt{T}$. It is useful to define the scaled variable $y=X/\sqrt{T}$ and to perform the change of variable $q\to q/\sqrt{ T}$, yielding
\begin{equation}
Z(X=y\sqrt{T},T)\simeq \frac{1}{2\pi i}\frac{1}{\sqrt{ T}}\int_{\Gamma'}dq\,{}_2F_1\left(\frac12,1,\frac{d}{2},\frac{q^2}{ T}\right)\exp\left[-yq~{}_2F_1\left(\frac12,1,\frac{d}{2},\frac{q^2}{ T}\right)- 	T\left( 1-{}_2F_1\left(\frac12,1,\frac{d}{2},\frac{q^2}{ T}\right)\right) \right] \,.
\label{eq:PDF_X_typ1}
\end{equation}
Expanding to leading order for large $T$, we find
\begin{equation}
Z(X=y\sqrt{T},T)\simeq \frac{1}{2\pi i}\frac{1}{\sqrt{ T}}\int_{\Gamma'}dq\,e^{-qy+q^2/d}\,.\nonumber
\label{eq:PDF_X_typ2}
\end{equation}
Performing the integral over $q$, we finally find that in the typical regime where $X\sim \sqrt{T}$ and $T\gg 1$,
\begin{equation}
Z(X,T)\simeq \frac{1}{\sqrt{2\pi D T}}e^{-X^2/(4DT)}
\label{eq:PDF_X_typ3}
\end{equation}
where
\begin{equation}
D=\frac{1}{d}\,.
\end{equation}
Thus, in the large-$T$ limit, the typical shape of the PDF $Z(X,T)$ of $X$ is Gaussian. The typical regime is therefore indistinguishable from a passive Brownian motion with diffusion constant $D$ and no sign of the activity of the RTP is present at this scale. Note that this effective diffusion coefficient $D$ is equal to the one that we have computed for the fixed-$N$ ensemble (see Eq. \eqref{D}). To observe any signs of the active nature of the process one needs to study the shape of the position distribution $Z(X,T)$ in the large-deviation regime where $X$ scales linearly with $T$.

\subsection{Large-deviation regime}

\begin{figure*}[t]
\centering
\includegraphics[width=1\textwidth]{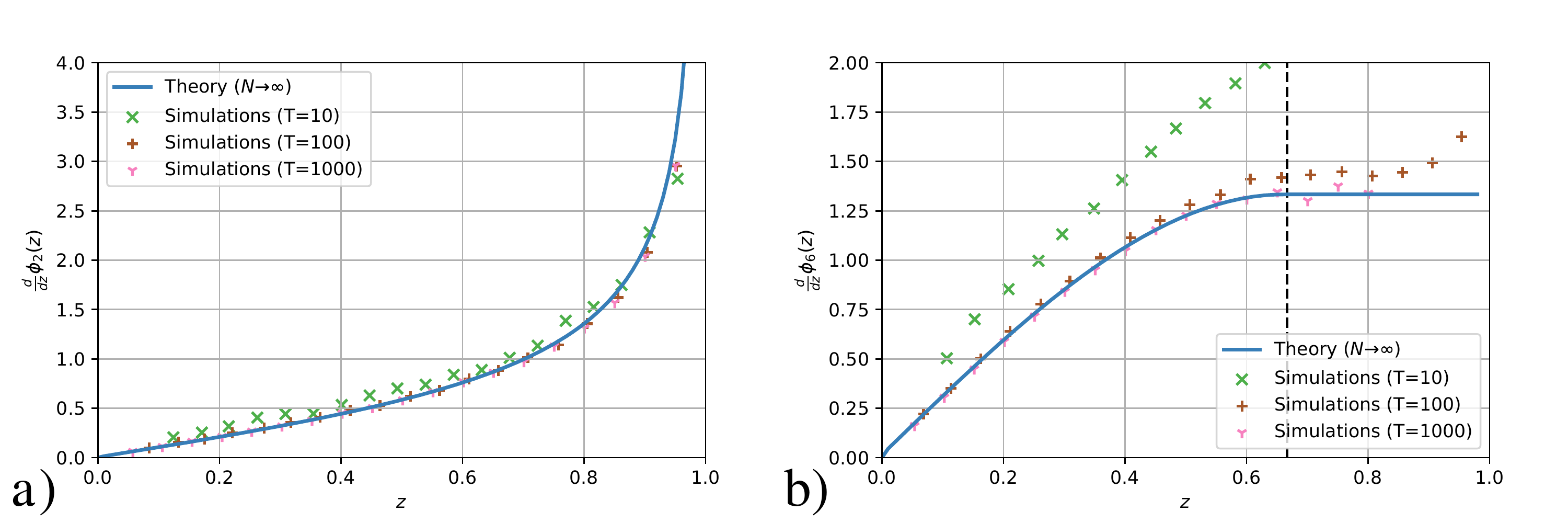} 
\caption{\label{fig:phi} {\bf a)} First derivative of the rate function $\phi_{d}(z)$ versus $z$, for $d=2$. The continuous blue line corresponds to the exact result in Eq. \eqref{eq:phi2}, valid in the limit $N\to\infty$. In this case, no transition occurs.
{\bf b)} First derivative of the rate function $\phi_d(z)$ versus $z$, for $d=6$. The continuous blue line corresponds to the exact result in Eq. \eqref{eq:phi6}. The vertical dashed line signals the critical point $z_c$ at which the phase transition occurs. For $z>z_c$, the rate function $\phi_d(z)$ becomes exactly linear in $z$. In both panels, the symbols are the results of numerical simulations obtained at finite $N$, as described in Section \ref{sec:numerics}.
}
\end{figure*}

We now focus on the large-deviation regime, where $X\sim T$. Some of the results of this section have already been derived in \cite{PTV2020}, where the authors compute the rate function of the position of a discrete-time persistent random walk and then take the continuous-time limit to study the RTP. Here, we first present a different and more general technique to compute the rate function for an RTP in $d$ dimensions. Note that our technique can be easily generalized to more complicated RTP models, e.g., with random velocities. Then, we interpret these results in light of the new findings, presented in the previous sections, and we characterize the nature of the phase transition.

It is useful to introduce the scaled variable $z=X/T$. Note that, since $|X|$ cannot exceed the value $T$, corresponding to a straight $x$-direction run with no tumbling, we have $|z|\leq 1$. From Eq. \eqref{eq:PDF_X_limit2}, we obtain
\begin{equation}
Z(X= z T,T)\simeq \frac{1}{2\pi i}\int_{\Gamma}dq\,{}_2F_1\left(\frac12,1,\frac{d}{2},q^2\right) \exp\left[-	TS_d(q,z) \right] \,,
\label{eq:PDF_X_limit3}
\end{equation}
where
\begin{equation}
S_d(q,z)=1-(1+qz)\, {}_{2}F_1\left(\frac12,1,\frac{d}{2},q^2\right)\,.
\label{eq:Sqz}
\end{equation}
First, we try to compute the integral in Eq. \eqref{eq:PDF_X_limit3} by saddle point approximation. Note that the function ${}_{2}F_1\left(\frac12,1,\frac{d}{2},q^2\right)$ has two branch cuts in the complex-$q$ plane, for real $q$ and $|q|>1$. Thus, we need to solve the following saddle point equation
\begin{equation}
z=g_d(q)\,
\label{eq:cond2}
\end{equation}
for $|q|<1$, where
\begin{equation}
g_d(q)=-\frac{(2q/d) ~{}_{2}F_1\left(3/2,2,1+d/2,q^2\right)}{(2q^2/d )~{}_{2}F_1\left(3/2,2,1+d/2,q^2\right)+{}_{2}F_1\left(1/2,1,d/2,q^2\right)}\,.
\label{eq:g}
\end{equation}
For any $d$, $g_d(q)$ is a decreasing odd function of $q$. We also notice that its maximum value is reached at $q=-1$. To compute this value $g_d(-1)$, we use the following 
asymptotic expansion for the ordinary hypergeometric function 
\cite{Gradshteyn_book}
\begin{equation}
{}_2F_1(\alpha,\beta,\gamma,q)\simeq\
\begin{cases}
\frac{\Gamma(\gamma)\Gamma(\gamma-\alpha-\beta)}{\Gamma(\gamma-\alpha)\Gamma(\gamma-\beta)} & \text{for } \gamma>\alpha+\beta,\\
\\
\frac{\Gamma(\alpha+\beta)}{\Gamma(\alpha)\Gamma(\beta)}\log\left(\frac{1}{1-q}\right) & \text{for } \gamma=\alpha+\beta,\\
\\
\frac{\Gamma(\gamma)\Gamma(\alpha+\beta-\gamma)}{\Gamma(\alpha)\Gamma(\beta)} (1-q)^{\gamma-\alpha-\beta} & \text{for } \gamma<\alpha+\beta\,,
\end{cases}
\label{eq:hypergeom1}
\end{equation}
and we obtain
\begin{equation}
g_d(-1)=\
\begin{cases}
1 & \text{for }d\leq 5 ,\\
\\
2/(d-3)<1 & \text{for } d>5,\\
\end{cases}
\label{eq:g_cases}
\end{equation}

Thus, recalling that $|z|\leq 1$ and focusing on the case $z>0$, for $d\leq 5$ the condition in Eq. \eqref{eq:cond2} is always satisfied for some value $q^*(z)>-1$. Thus, for $d<5$ we find that
\begin{equation}
Z(X,T)\sim \exp\left[- T \phi_d\left( z=\frac{X}{T}\right)\right]\,,
\label{PXT}
\end{equation}
where
\begin{equation}
\phi_d(z)=S_d(q^*(z),z)\,,
\label{phi_below}
\end{equation}
$S_d(q,z)$ is given in Eq. \eqref{eq:Sqz} and $q^*(z)$ is the unique solution of Eq. \eqref{eq:cond2}. On the other hand, for $d>5$, the saddle point equation \eqref{eq:cond2} admits a solution only up to some critical value $z=z_c=2/(d-3)$, where the condensation transition occurs. For $z>z_c$, Eq. \eqref{eq:cond2} has no real solution and the maximum of $S_d(q,z)$ is reached at $q=-1$, independently of $z$ and $\phi_d(z)=S_d(-1,z)$. Thus, for $d>5$ we find that the large-deviation form in Eq. \eqref{PXT} is still valid, with
\begin{equation}
\phi_d(z)=
\begin{cases}
S_d(q^*(z),z) & \text{for } z<z_c,\\
\\
-\frac{1}{d-3}+\frac{d-2}{d-3}z & \text{for } z>z_c\,,
\end{cases}
\label{phi_transition}
\end{equation}
where $S_d(q,z)$ is given in Eq. \eqref{eq:Sqz} and $q^*(z)$ is the unique solution of Eq. \eqref{eq:cond2}. Notably, for $d>5$, the rate function $\phi_d(z)$ is non-analytic at $z=z_c$. This indicates the presence of the condensation phase-transition. Comparing this result with the one obtained for the fixed-$N$ ensemble, we find that the criterion for condensation is the same for the two models. Indeed, recalling that we are considering $\alpha=0$, here we observe condensation for $\nu=(d-1)/2>2$, exactly as for the fixed-$N$ case.

In the special cases $d=1,2,4,6$, the expression of $\phi_d(z)$ becomes simple and is given by
\begin{eqnarray}
\phi_1(z)&=&\frac{1-\sqrt{1-z^2}}{2}\,\label{eq:phi1}\\
\phi_2(z)&=&1-\sqrt{1-z^2}\,\label{eq:phi2}\\
\phi_4(z)&=&z^2\,,\label{eq:phi3}\\
\phi_6(z)&=&
\begin{cases}
\frac32 z^2-\frac{9}{16}z^4 & \text{for } |z|<z_c,\\
\\
\frac{4}{3}z-\frac13 & \text{for } |z|>z_c\,.\label{eq:phi6}
\end{cases}
\end{eqnarray}
These results for $\phi_d(z)$ match with the ones derived in Ref. \cite{PTV2020}. The result in Eq. \eqref{eq:phi2}, valid for $d=2$, had already been obtained in \cite{SBS20} solving the Fokker-Plank equation associated to the system. The rate function of the position distribution of a fixed-$T$ RTP has also been derived in dimension $d=1$ in the presence of a constant drift \cite{DM12}.

Note that here we have provided explicit results for the rate function for a specific velocity distribution, namely, when the direction is chosen isotropically and the speed $v=v_0$ is a constant. In fact, this rate function, when it exists, can be derived for generic velocity distribution $P(\vec v)$, as shown in Appendix \ref{app:large_dev}.

\subsection{Order of the transition}

We now investigate, for $d>5$, the order of the phase transition. Just below the transition, i.e. in the limit $z\to z_c$, we expect the solution $q^*(z)$ of the saddle point equation \eqref{eq:cond2} to be close to $-1$. Thus, we expand $S_d(q,z)$ in Eq. \eqref{eq:Sqz} with $q=-1+s$ for small $s$. In the case $5<d<7$, using the asymptotic behavior of the hypergeometric function close to unit value \cite{buhring}, we find
\begin{equation}
S_d(q=-1+s,z)\simeq 1+\frac{d-2}{d-3}(1-z)+\frac{d-2}{d-5}\left(\frac{2}{d-3}-z\right)s-\frac{1}{\sqrt{\pi}}2^{(d-3)/2}\Gamma\left(\frac{d}{2}\right)\Gamma\left(-\frac{d-3}{2}\right)(1-z)s^{(d-3)/2}\,.
\end{equation}
Minimizing this expression with respect to $s$ and then expanding for $z\to z_c=2/(d-3)$, we obtain
\begin{equation}
\phi_d(z)\simeq -\frac{1}{d-3}+\frac{d-2}{d-3}z+c_d~(z_c-z)^{(d-3)/(d-5)}\,,
\end{equation}
where $c_d$ is a $d$-dependent constant.
Comparing this result with the expression for $\phi_d(z)$ in the case $z>z_c$ in Eq. \eqref{phi_transition}, we conclude that, in this case, the phase transition is of order
\begin{equation}
n=\left\lceil\frac{d-3}{d-5}\right\rceil\,.
\end{equation}

On the other hand, in the case $d>7$, the function $S_d(q,z)$ can be expanded as
\begin{equation}
S_d(q=-1+s,z)\simeq 1+\frac{d-2}{d-3}(1-z)+\frac{d-2}{d-5}\left(\frac{2}{d-3}-z\right)s+\frac{(d-2)\left[3z(d-3)-d-5\right]}{(d-7)(d-5)(d-3)}s^2\,.
\end{equation}
Minimizing this expression with respect to $s$ and then expanding for $z\to z_c$, we obtain
\begin{equation}
\phi_d(z)\simeq -\frac{1}{d-3}+\frac{d-2}{d-3}z+\frac{(d-7)(d-3)(d-2)}{4(d-5)(d-1)}(z-z_c)^2\,.
\end{equation}
Thus, in this case the order of the transition is $n=2$. Comparing these results with those of Section \ref{sec:position}, we notice that the order of the phase transition at given $d$ is the same for the fixed-$N$ and fixed-$T$ ensembles. In Section \ref{sec:marginal}, we have shown that the order of the phase transition in the fixed-$T$ ensemble is related to the nature of the condensate itself. In particular, for $5<d<7$, we expect the condensate to have an anomalous shape, with anomalous fluctuations of order $T^{2/(d-3)}$. On the other hand, for $d>7$ the transition becomes of order two and, in analogy with what we observe for the fixed-$N$ ensemble, we expect the condensate to have a Gaussian shape, with fluctuations of order $\sqrt{T}$.

\subsection{Asymptotics of $\phi_d(z)$}

Here, we investigate the asymptotic behavior of $\phi_d(z)$. Let us first consider, for generic $d$, the limit $z\to 0$. In a small region around $z=0$ the rate function $\phi_d(z)$ is always given by
\begin{equation}
\phi_d(z)=S_d(q^*(z),z)\,,
\end{equation}
where $S_d(q,z)$ is given in Eq. \eqref{eq:Sqz} and $q^*(z)$ is the unique solution of Eq. \eqref{eq:cond2}. It is easy to check that for small $z$, $q^*(z)$ is also small. Thus, expanding the right-hand side of Eq. \eqref{eq:cond2} for small $q$, we obtain
\begin{equation}
z\simeq-\frac{2}{d}q\,.
\end{equation}
Therefore, we find that, at leading order, $q^*(z)=-(d/2)z$. Plugging this value into the expression of $S(q,z)$, given in Eq. \eqref{eq:Sqz}, and expanding for small $z$, we find that
\begin{equation}
\phi_d(z)\simeq \frac{d}{4}z^2\,.
\label{phi_small}
\end{equation}
Plugging this expansion into the expression for $Z(X,T)$ in Eq. \eqref{PXT}, we find that, for small $z=X/T$,
\begin{equation}
Z(X,T)\sim \exp\left[- \frac{~d}{4 T}X^2 \right]\,.
\end{equation}
Thus, the small-$z$ behavior of the rate function $\phi_d(z)$ matches smoothly to the typical Gaussian behavior (see Eq. \eqref{eq:PDF_X_typ3}).

Next, we consider the limit $z\to 1$. For $d>5$, we already know from Eq. \eqref{phi_transition} that, in the limit $z\to 1$, one has
\begin{equation}
\phi_d(z)=1+\frac{d-2}{d-3}(z-1)\,.
\end{equation}
On the other hand, for $d< 5$, the rate function $\phi_d(z)$ is given in Eq. \eqref{phi_below}. Thus, we have to solve Eq. \eqref{eq:cond2} for $z\to 1$. From Eq. \eqref{eq:g_cases} we know that $g_d(q=-1)=1$, for $d< 5$. Thus, we expect the solution $q^*(z)$ of Eq. \eqref{eq:cond2} to be of the type $q^*(z)=-1+s$ with $s$ small. It is useful to consider the cases $1<d<3$ and $3<d<5$ separately.

In the case $1<d<3$, we plug $q^*(z)=-1+s$ in the condition in Eq. \eqref{eq:cond2}. Expanding for small $s$ (using Eq. \eqref{eq:hypergeom1}), we obtain
\begin{equation}
z\simeq 1-\frac{d-1}{3-d}s\,,
\end{equation}
which yields
\begin{equation}
q^*(z)\simeq-1+\frac{3-d}{d-1}(1-z)\,.
\end{equation}
Plugging this expression for $q^*(z)$ in Eq. \eqref{phi_below} and expanding for $z\to 1$ from below, we obtain
\begin{equation}
\phi_d(z)\simeq1-\frac{1}{\sqrt{\pi}}\Gamma\left(\frac{d}{2}\right)\Gamma\left(\frac{3-d}{2}\right)\frac{2}{d-1}\left(2\frac{3-d}{d-1}\right)^{(d-3)/2}~(1-z)^{(d-1)/2}\,.\label{expansion2}
\end{equation}

We now consider the case $3<d<5$. Plugging $q^*(z)=-1+s$ into Eq. \eqref{eq:cond2} and expanding for small $s$, we find
\begin{equation}
z\simeq 1-\sqrt{\pi}\,\frac{d(d-2)}{4(d-3)}\,\frac{1}{\Gamma\left(1+\frac{d}{2}\right)\Gamma\left(\frac{5-d}{2}\right)}2^{(5-d)/2}s^{(5-d)/2}\,,
\end{equation}
which implies
\begin{equation}
1+q^*(z)\sim (1-z)^{2/(5-d)}\,.
\end{equation}
Plugging this value in Eq. \eqref{phi_below} and expanding for $z\to 1$, we obtain
\begin{equation}
\phi_d(z)\simeq 1-\frac{d-2}{d-3}~(1-z)\,.
\label{expansion4}
\end{equation}

To summarize, we have shown that, in the limit $z\to 1$,
\begin{equation}
\phi_d(z)=
\begin{cases}
1-\tilde{c}_d~(1-z)^{(d-1)/2}+o((1-z)^{(d-1)/2}) & ~~~~~\text{for } 1<d<3\,,\\
\\
1-\frac{d-2}{d-3}~(1-z)+o((1-z)) &~~~~~ \text{for } 3<d<5\,,\\
\\
1-\frac{d-2}{d-3}~(1-z) & ~~~~~\text{for } d>5\,,\\
\\
\end{cases}
\end{equation}
where $\tilde{c}_d$ is a $d$-dependent constant. Thus, for $d<3$, $\phi_d(z)$ approaches the limit value $1$, for $z\to 1$, with an exponent $(d-1)/2$ which depends continuously on $d$. For $3<d<5$, the rate function $\phi_d(z)$ becomes locally linear for $z\to 1$ and for $d>5$ one has a full region $z_c<z<1$ where $\phi_d(z)$ is exactly linear.

\section{Numerical simulations}

In this section we describe the numerical techniques that we have used to verify our theoretical results for the rate function $\psi_{d,\alpha}(z)$, which is defined as
\begin{equation}
\psi_{d,\alpha}(z)=-\lim_{N\to \infty}\frac{\log\left[Z(X=zN,N)\right]}{N}\,,
\label{psi_num}
\end{equation}
where $Z(X,N)$ is the distribution of the total $x$-component displacement $X$ after $N$ runs. Let us first recall that, in order to probe the typical Gaussian regime $X\sim \sqrt{N}$, one can use direct sampling. Indeed, a single $x$-direction displacement $x$ can be written as
\begin{equation}
x=v\, \tau\,u \,\,,
\end{equation}
where $v>0$ is the speed of the RTP, with distribution $W(v)$, $\tau>0$ is the running time, exponentially distributed with fixed rate $\gamma=1$ and $-1<u<1$ is the $x$-component of the $d$-dimensional unit vector that represents the direction of the RTP. Since this direction is uniformly distributed in space, one can show that $u$ is distributed according to $f_d(u)$, given in Eq. \eqref{fd} \cite{MLDM20,MLDM20a}. Thus, for each of the $N$ running phases, one can generate, by standard sampling techniques, three random number $v_i$, $\tau_i$ and $u_i$ and then one can obtain the $x$-component displacement $x_i$ by multiplying these variables. Finally, one can obtain $X$ using the definition
\begin{equation}
X=\sum_{i=1}^N x_i\,.
\end{equation}
This simple and direct procedure is useful to probe the typical regime $X\sim \sqrt{N}$. However, it is unfeasible to adopt such a strategy to compute numerically the large-deviation tails of $Z(X,N)$. For instance, extracting $10^6$ samples, one is only able to access events with probabilities of the order $10^{-6}$ or higher. How can one simulate events that happen with very small probability, say of order $10^{-100}$?

In order to compute numerically the large-deviation tails of the PDF $Z(X,N)$ we adopt a technique based on a constrained Markov Chain Monte Carlo (MCMC) algorithm, similar to the one proposed in \cite{GM2019,NMV10,NMV11}. The configuration $\mathcal{C}$ of our system is specified by the set of numbers  $\mathcal{C}=\{(v_1,\tau_1,u_1)\,,\ldots\,,(v_N,\tau_N,u_N)\}$ and we are interested in the rate function of 
\begin{equation}
X(\mathcal{C})=\sum_{i=1}^{N}v_i\tau_i u_i\,
\end{equation}
in the large deviation regime where $X(\mathcal{C})\sim O(N)$. With this goal in mind, we implement a MCMC dynamics in the space of RTP configurations. Let us remark that the MCMC dynamics is defined in configuration space and has nothing to do with the real RTP dynamics. Since we are interested in configurations that correspond to a very large $X(\mathcal{C})$, we impose the constraint $X(\mathcal{C})>X^*$, where $X^*$ is some fixed $O(N)$ parameter. We choose an initial condition $\mathcal{C}_0$ that satisfies the constraint and then we evolve the system using the Metropolis rule. In other words, assuming that at a given step the current configuration is $\mathcal{C}=\{(v_1,\tau_1,u_1)\,,\ldots\,,(v_N,\tau_N,u_N)\}$, we choose one of the running phases $i$ at random and we propose the move $(v_i,\tau_i,u_i) \to ({v_i}^{\rm new},{\tau_i}^{\rm new},u_i^{\rm new})$, where
\begin{eqnarray}
{v_i}^{\rm new}=v_i+\delta v\,,\\
{\tau_i}^{\rm new}=\tau_i +\delta \tau\,,\\\nonumber
{u_i}^{\rm new}=u_i +\delta u\,,\nonumber
\end{eqnarray}
where $\delta v$, $\delta \tau$, and $\delta u$ are drawn from a uniform distribution. The new configuration is then simply 
\begin{equation}
\mathcal{C}^{\rm new}=\{(v_1,\tau_1,u_1)\,,\ldots\,,({v_{i-1}},{\tau_{i-1}},u_{i-1})\,,({v_i}^{\rm new},{\tau_i}^{\rm new},u_i^{\rm new})\,,\,,({v_{i+1}},{\tau_{i+1}},u_{i+1})\,,\ldots\,,(v_N,\tau_N,u_N)\}
\end{equation}
If $\mathcal{C}^{\rm new}$ does not satisfy the constraint $X(\mathcal{C}^{\rm new})>X^*$, then the move is immediately rejected. If instead $X(\mathcal{C}^{\rm new})>X^*$, the move is accepted with probability
\begin{equation}
p_{\rm acc}=\min\left[1,\frac{ e^{- \tau_i^{\rm new}}~W(v_i^{\rm new})f_d(u_i^{\rm new})}{ e^{- \tau_i}~W(v_i)f_d(u_i)}\right]\,,
\end{equation}
and rejected otherwise. This particular choice of $p_{\rm acc}$, which correspond to the Metropolis-Hastings algorithm, guarantees that the RTP configurations are sampled with the right statistical weight. If the move is accepted we update the current position $\mathcal{C}\to \mathcal{C}^{\rm new}$. Initially, we let the system evolve for $10^7$ sweeps (by sweep we denote $N$ move proposals), in order to forget the initial condition. We measure $X(\mathcal{C})$ every $10^2$ sweeps, to avoid sample correlations.

\label{sec:numerics}
\begin{figure*}[t]
\centering
\includegraphics[width=1\textwidth]{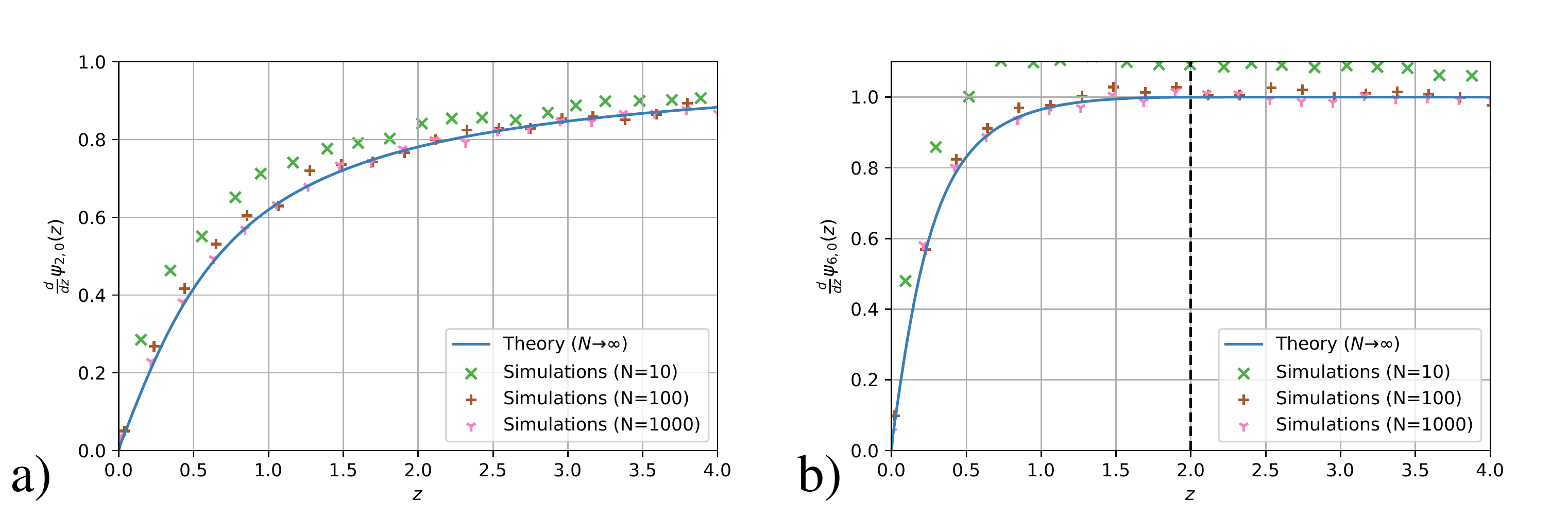} 
\caption{\label{fig:psi} {\bf a)} First derivative of the rate function $\psi_{d,\alpha}(z)$ versus $z$, for $d=2$ and $\alpha=0$. The continuous blue line corresponds to the exact result in Eq. \eqref{psi_2_0}, valid in the limit $N\to\infty$. For this choice of the parameters $d$ and $\alpha$, no transition occurs.
{\bf b)} First derivative of the rate function $\psi_{d,\alpha}(z)$ versus $z$, for $d=6$ and $\alpha=0$. The continuous blue line corresponds to the exact result in Eq. \eqref{psi_d_cond}. The vertical dashed line signals the critical point $z_c$ at which the phase transition occurs. In both panels, the symbols are the results of numerical simulations obtained at finite $N$, as described in Section \ref{sec:numerics}.
}
\end{figure*} 

\begin{figure*}[t]
\centering
\includegraphics[width=0.6\textwidth]{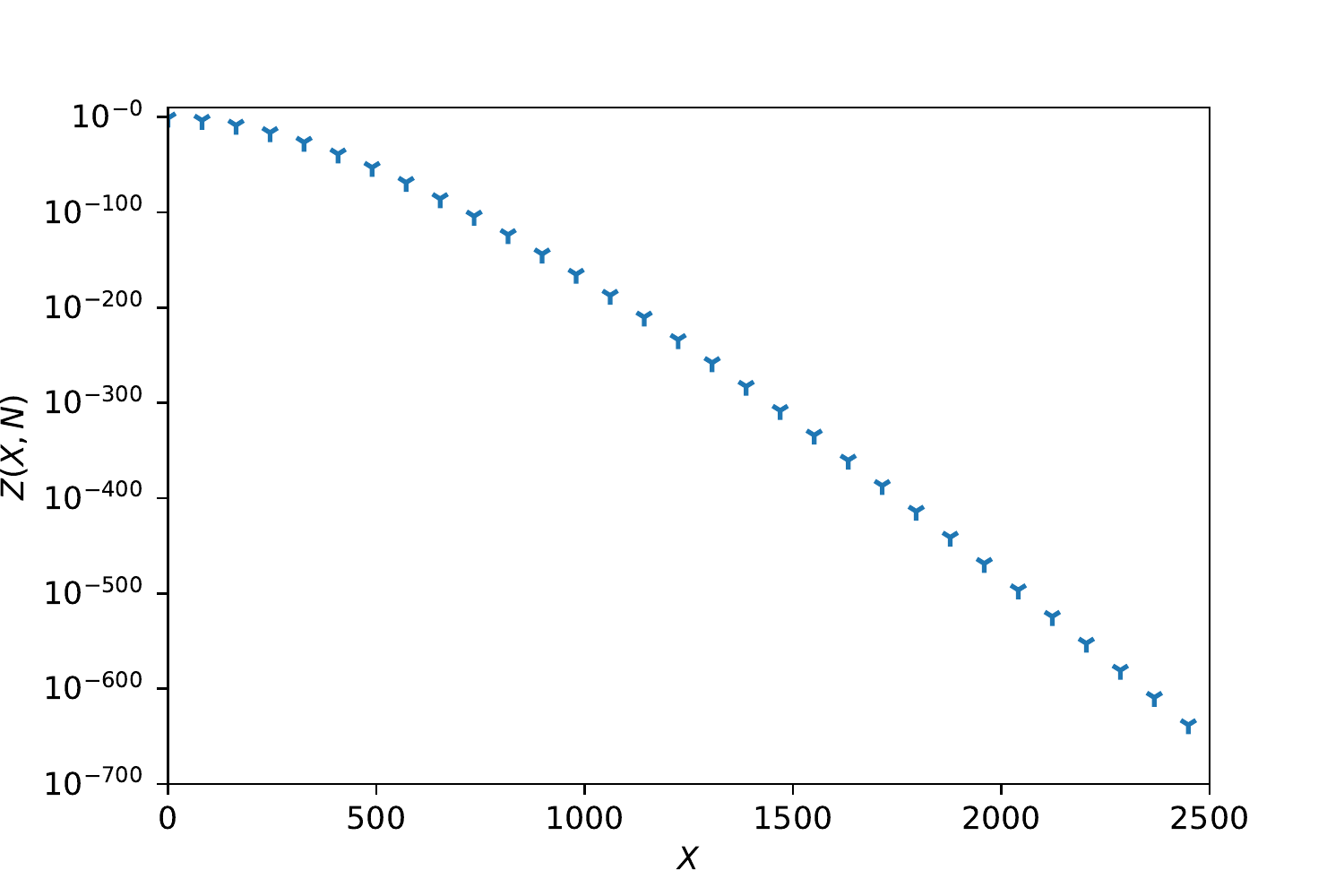} 
\caption{\label{fig:Znum} Numerical curve of the PDF $Z(X,N)$ as a function of $X$, for $d=2$, $\alpha=0$ and $N=1000$, obtained with the constrained Markov Chain Monte Carlo algorithm described in the text.
}
\end{figure*}

With this procedure, we can build an histogram which approximates the PDF $P(X,N|X>X^*)$ of $X$, conditioned on $X>X^*$. This quantity can be written as, for $X>X^*$,
\begin{equation}
P(X,N|X>X^*)=\frac{P(X,N)}{P(X>X^*)}\,.
\end{equation}
Taking the natural logarithm of both sides and using the notation $Z(X,N)=P(X,N)$, we find
\begin{equation}
\log\left[P(X,N|X>X^*)\right]= \log\left[Z(X,N)\right]-\log\left[P(X>X^*)\right]\,.
\end{equation}
Note that the last term is independent of $X$. Using the definition of $\psi_{d,\alpha}(z)$, given in Eq. \eqref{psi_num}, we find, for large $N$,
\begin{equation}
\log\left[P(X,N|X>X^*)\right]=-N\psi_{d,\alpha}\left(\frac{ X}{ N}\right)-\log\left[P(X>X^*)\right]\,,
\end{equation}
Finally, taking a derivative with respect to $X$ on both sides, we get
\begin{equation}
\psi_{d,\alpha}'\left(\frac{ X}{ N}\right)=
-\frac{d}{dX}\log\left[P(X,N|X>X^*)\right]\,,
\end{equation}
where $\psi_{d,\alpha}'(z)=\frac{d}{dz}\psi_{d,\alpha}(z)$. Thus, we are able to compute numerically the first derivative of the rate function. Then, one can obtain $\psi_{d,\alpha}(z)$ via numerical integration. Note however that, with the method described above, one can only probe a small region $(X^*,X^*+\Delta)$, where $\Delta>0$ is a small number compared to $X^*$. Therefore, one has to use several values of $X^*$ in order to sample the large-deviation regime. In our case, we used $20$ different values of $X^*$. Our numerical estimate of $\frac{d}{dz}\psi_{d,\alpha}(z)$ is shown in Fig. \ref{fig:psi} for $\alpha=0$, $d=2,6$ and for different values of $N$. In the case $N=10^4$, the numerical curves are in excellent agreement with the theory, both in the fluid and in the condensed phases. Integrating $\frac{d}{dz}\psi_{d,\alpha}(z)$ numerically, we also compute $\psi_{d,\alpha}(z)$, which is shown in Fig. \ref{fig:psi_int} and is in excellent agreement with the theory. Similarly, one can also compute the PDF $Z(X,N)$, which is shown, for $d=2$ and $\alpha=0$, in Fig. \ref{fig:Znum} with precision smaller than $10^{-100}$.

In the case of the fixed-$T$ ensemble, a similar algorithm can be applied. The only complication is that the number of running phases is not fixed anymore. Therefore, it might happen that, proposing a move, the total simulation time becomes shorter than $T$. In such a case, we simply add a new running phase at the end of the trajectory. In this way, one obtains the first derivative of $\phi_d(z)$, which is shown in Fig. \ref{fig:phi} for $d=2,6$ and for different values of $N$. Note that for the fixed-$T$ ensemble one has $z<1$, since the maximal distance that one can travel in a time $T$ is $v_0 T$. For this reason, sampling the region close to $z=1$ becomes increasingly complicated. Nevertheless, in Fig. \ref{fig:phi}, we observe a good agreement between the numerics and the theoretical curve.

With the technique described above, one can also sample the marginal probability distribution $p(x|X)$ of a single-run displacement. This can be achieved by using the MCMC algorithm described above, keeping $X^*$ fixed. During the Monte Carlo dynamics, one has access to the full configuration $\mathcal{C}$ of the RTP. Thus, one can sample the single-run displacement $x$, for instance choosing $x=x_1$. To be precise, since $X$ has to satisfy the constraint $X>X^*$ (and not $X=X^*$), in this way one would compute the marginal probability $p(x|X>X^*)$, conditioned on the event $X>X^*$. However, since in practice the system is always in a small region $(X^*,X^*+\Delta)$, one has that $p(x|X>X^*)$ is a good approximation of $p(x|X^*)$ (see Fig. \ref{fig:marginal}).

Finally, let us mention that a similar method, based on an exponentially biased MCMC algorithm, has been proven useful to simulate large deviations of the RTP model \cite{DMS21,HMS20}. However, to the best of our knowledge, such techniques cannot be used to simulate the RTP model in the condensed phase.

\section{Conclusions}
\label{sec:conclusion}

In this paper, we have investigated the late-time position of a single RTP in $d$ dimensions, with velocity distribution $W(v)$ and tumbling rate $\gamma$. First, we have focused on the fixed-$N$ ensemble, i.e., we have considered the number $N$ of running phases to be fixed. We have shown that due to the isotropy of the process, it is sufficient to study the distribution $Z(X,N)$ of the displacement of the particle in the $x$-component after $N$ running phases. We have observed that, even if in the typical regime where $X\sim \sqrt{N}$ the PDF $Z(X,N)$ has a Gaussian shape, its large-deviation tails still carry the signatures of the active nature of the process. Moreover, we have shown that for several choices of $d$ and $W(v)$, the system undergoes a dynamical condensation transition in the large-deviation regime. This transition is signaled by a singularity of the rate function of $Z(X,N)$. Below the transition, all running phases contribute to the total displacement $X$ by roughly the same amount. On the other hand, above some critical value $X=X_c$, a condensate emerges in the form of a single run which dominates the RTP trajectory. Using a grand-canonical argument, we have identified a precise criterion for condensation.

In the special case $W(v)= \alpha (1-v)^{\alpha-1}$, we have exactly computed the rate function $\psi_{d,\alpha}(z)$, where $z=X/ N$. We have shown that condensation happens only if $\nu=(d+2\alpha-1)/2>2$. In particular, for $\nu>3$, we have observed that $\psi_{d,\alpha}(z)$ has a second-order singularity at some critical value $z_c$, which we have computed exactly, while for $2<\nu<3$ the order of the transition depends continuously on $\nu$. Moreover, we have investigated the precise nature of the condensate, studying the marginal probability of a single-run displacement. For $\nu>3$, we have observed that the condensate size has Gaussian fluctuations of order $\sqrt{N}$. On the other hand, for $2<\nu<3$, we have shown that the condensate has an anomalous shape, with large fluctuations of order $N^{1/(\nu-1)}$. We have also extended our results to the fixed-$T$ ensemble, where the total duration $T$ of the trajectory is fixed. In the case of fixed velocity ($\alpha=0$), we have computed the rate function of the total displacement $X$ of the particle for arbitrary $d$. We have observed that an analogous condensation transition occurs also for this model above some critical value $X_c$ of $X$. Moreover, we have employed a constrained Markov chain Monte Carlo technique to verify our large-deviation result, probing events with probability smaller than $10^{-100}$. Our numerical simulations are in excellent agreement with our theoretical results.

In this paper, we have shown that condensation transitions are a general feature of the RTP model. In future works, it would be interesting to investigate other RTP models that satisfy the condensation criterion. We have shown that a second-order transition is linked to a normal condensate, while a higher-order transition corresponds to an anomalous condensate. Therefore, it would be relevant to investigate what happens in models that display a first-order condensation transition, see e.g. \cite{GM2019}.

Another interesting open problem is related to the criterion for condensation. The argument we 
have presented is based on a grand canonical description of the system, which fails if the 
single-run distribution $p(x)$ decays slower than any exponential. In this case,
we have conjectured that a condensation transition will occur if $p(x)$ decays faster than 
$1/|x|^3$ for large $|x|$. It would be interesting to prove this conjecture.

\section*{Acknowledgement}

We thank G. Gradenigo and A. Rosso for useful discussions.

\appendix

\section{Relation between $P(\vec{R},N)$ and $Z(X,N)$}
\label{app:map}

In this appendix, we derive the relation in Eq. \eqref{relation} between the PDF $P(\vec{R},N)$ of the position $\vec{R}$ of the RTP after $N$ steps and the distribution $Z(X,N)$ of the $x$-component $X$ of $\vec{R}$. Moreover, we show that $Z(X,N)$ and $P(\vec{R},N)$ share the same rate function $\psi(z)$ in the large-deviation regime where $X$ and $R=|\vec{R}|$ scale linearly with $N$. As a consequence of the isotropy of the process, the PDF $P(\vec{R},N)$ depends only on the magnitude $R$ of $\vec{R}$ and not on its orientation. In other words, the orientation of $\vec{R}$ is distributed uniformly at random. Given a vector $\vec{R}$ with fixed norm $R$ and random orientation, it is possible to show that the PDF of the $x$-component $X$ of $\vec{R}$ is (see Appendix A of \cite{MLDM20})
\begin{equation}
P(X|R)=\frac{1}{R} f_d\left(\frac{X}{R}\right)\,,
\end{equation}
where the function $f_d(z)$ can be computed for any $d$ and is given in Eq. \eqref{fd}. Thus, the PDF of $X$ can be written as
\begin{equation}
Z(X,N) = \int_{\mathbb{R}^d}d\vec{R}~ \frac{1}{R} f_d\left(\frac{X}{R}\right) P( R, N) \,,
\label{relation_app}
\end{equation}
where we integrate over all possible values of $\vec{R}$, weighted by the PDF $P(\vec R, N)= P( R, N) $. This is precisely the relation given in Eq. \eqref{relation}.

We now want to show that in the large-deviation regime where $|X|\sim O(N)$ and $R\sim O(N)$, $Z(X,N)$ and $P(\vec{R},N)$ share the same rate function $\psi(z)$. We can perform the integral in Eq. \eqref{relation_app} in the radial coordinate and we obtain
\begin{equation}
Z(X,N) = \frac{2\pi^{d/2}}{\Gamma(d/2)}\int_{0}^{\infty} dR~R^{d-2}  f_d\left(\frac{X}{R}\right) P(R, N) \,.
\end{equation}
Using the expression of $f_d(z)$, given in Eq. \eqref{fd}, we obtain
\begin{equation}
Z(X,N) = \frac{2\pi^{(d-1)/2}}{\Gamma((d-1)/2)}\int_{|X|}^{\infty} dR~R^{d-2}  \left(1-\frac{X^2}{R^2}\right)^{(d-3)/2}P(R, N) \,,
\end{equation}
and making the change of variable $R\to u=R/X$, we obtain
\begin{equation}
Z(X,N) = \frac{2\pi^{(d-1)/2}}{\Gamma((d-1)/2)}|X|^{d-1}\int_{1}^{\infty} du~u^{d-2}  \left(1-u^{-2}\right)^{(d-3)/2}P(R=u|X|, N) \,.
\label{rel2}
\end{equation}

Let us now focus on the regime where $X\sim O(N)$. Plugging the scaled variable $z=X/N$ in Eq. \eqref{rel2}, we find
\begin{equation}
Z(X=zN,N) = \frac{2\pi^{(d-1)/2}}{\Gamma((d-1)/2)}|zN|^{d-1}\int_{1}^{\infty} du~u^{d-2}  \left(1-u^{-2}\right)^{(d-3)/2}P(R=u|z|N, N) \,.
\label{rel3}
\end{equation}
In the large deviation regime where $R\sim O(N)$, we expect $P(R,N)\sim \exp\left[-N\psi(R/N)\right]$, where $\psi(z)$ is the rate function associated to $P(R,N)$. Plugging this expression in Eq. \eqref{rel3}, we find
\begin{equation}
Z(X=zN,N) \sim \int_{1}^{\infty} du~u^{d-2}  \left(1-u^{-2}\right)^{(d-3)/2}e^{-N\psi(u|z|)} \,.
\end{equation}
In the limit of large $N$, this integral is dominated by values close to the lower limit $u=1$. Thus, we obtain
\begin{equation}
Z(X=zN,N) \sim e^{-N\psi(|z|)} \,,
\end{equation}
which can be written as
\begin{equation}
Z(X,N) \sim e^{-N\psi(|X|/N)} \,.
\end{equation}
Therefore, $Z(X,N)$ and $P(\vec{R},N)$ have the same rate function $\psi(z)$.

\section{Distribution of the $x$-direction displacements}
\label{app:p(x)}

In this appendix, we want to compute the distribution of the $x$-direction displacements $x_1\,,\ldots\,,x_N$ of the RTP associated to the $N$ running phases. For $1\leq i\leq N$, $x_i$ is the $x$-component of the $d$-dimensional vector $\vec{\ell}_i$. These vectors $\vec{\ell}_i$ are i.i.d. random variables, their direction is drawn uniformly at random and their magnitude is given by $\ell_i=v_i\,\tau_i$, where $v_i>0$ is drawn from the speed distribution $W(v)$ and $\tau_i$ is an exponential random variable with rate $\gamma$. Thus, the distribution of the magnitude $\ell$ of one of these displacement vectors is
\begin{equation}
P(\ell)=\int_{0}^{\infty}d\tau\,\int_{0}^{\infty}dv~W(v)~\gamma e^{-\gamma \tau}\delta\left(\ell-v \tau\right)\,.
\end{equation}
Performing the integral over $\tau$, we get
\begin{equation}
P(\ell)=\int_{0}^{\infty}dv~\frac{1}{v}W(v)~\gamma e^{-\gamma(\ell/v)}\,.
\label{Pell_app}
\end{equation}
One can then show that the distribution of the $x$-component of a $d$-dimensional vector $\vec{l}$ with fixed norm and uniformly distributed direction is given by
\begin{equation}
p(x|l)=\frac1l f_d\left(\frac{x}{l}\right)\,,
\label{eq:PDF_x|l_app}
\end{equation}
where 
\begin{equation}
f_d(z)=\frac{\Gamma\left(d/2\right)}{\sqrt{\pi}\Gamma\left((d-1)/2\right)}(1-z^2)^{(d-3)/2} \theta(1-|z|)\,,
\label{eq:fd_app}
\end{equation}
$\Gamma(y)$ is the Gamma function and $\theta(y)$ is the Heaviside theta function. For a derivation of this result in Eq. \eqref{eq:fd_app} see Appendix A of \cite{MLDM20}. Thus, the distribution $p(x)$ of the displacement $x$ of the particle during a single running phase is given by, using Eqs. \eqref{Pell_app} and \eqref{eq:PDF_x|l_app},
\begin{equation}
p(x)=\int_{0}^{\infty}dv~\frac{1}{v}W(v)\int_{0}^{\infty}d\ell~\frac{1}{\ell}f_d\left(\frac{x}{\ell}\right)~\gamma e^{-\gamma(\ell/v)}\,.
\label{Px_app}
\end{equation}
Note that, since $f_d(z)$ is symmetric around $z=0$, the PDF $p(x)$ is also symmetric around $x=0$.

\section{Large-$|x|$ behavior of $p(x)$}
\label{app:asymp}
The result in Eq. \eqref{Px_app} is valid for any distribution $W(v)$. We now consider the special case where
\begin{equation}
W(v)=\frac{\alpha}{v_0}\left(1-\frac{v}{v_0}\right)^{\alpha-1}~\theta(v)~\theta(v_0-v)\,.
\label{W_app}
\end{equation}
In particular, we are interested in computing the large-$x$ behavior of $p(x)$. Using the expression for $W(v)$ given in Eq. \eqref{W_app}, we find
\begin{equation}
p(x)=\frac{\Gamma\left(d/2\right)}{\sqrt{\pi}\Gamma\left((d-1)/2\right)}\frac{\alpha}{v_0}\gamma \int_{x}^{\infty}d\ell~\frac{1}{\ell}\left(1-\frac{x^2}{\ell^2}\right)^{(d-3)/2}\int_{0}^{v_0}dv~\frac{1}{v}\left(1-\frac{v}{v_0}\right)^{\alpha-1}~ e^{-\gamma \ell/v}\,.
\label{Px_app2}
\end{equation}
Performing the changes of variable $\ell\to u= x/\ell$ and $v\to w=v/v_0$, we find
\begin{equation}
p(x)=\frac{\Gamma\left(d/2\right)}{\sqrt{\pi}\Gamma\left((d-1)/2\right)}\frac{\alpha}{v_0}\gamma \int_{0}^{1}du~\frac{1}{u}\left(1-u^2\right)^{(d-3)/2}\int_{0}^{1}dw~\frac{1}{w}\left(1-w\right)^{\alpha-1}~ e^{-\gamma x/(v_0 w u)}\,.
\label{Px_app3}
\end{equation}
It is useful to perform the changes of variable  $u\to t=1-u^2$ and $w\to s=1-w$, and we obtain
\begin{equation}
p(x)=\frac12 \frac{\Gamma\left(d/2\right)}{\sqrt{\pi}\Gamma\left((d-1)/2\right)}\frac{\alpha}{v_0}\gamma \int_{0}^{1}dt~\frac{1}{1-t}t^{(d-3)/2}\int_{0}^{1}ds~\frac{1}{1-s}s^{\alpha-1}~ \exp\left(-\frac{\gamma}{v_0}\frac{x}{ (1-s)\sqrt{1-t}}\right)\,.
\label{Px_app4}
\end{equation}
For $x\gg 1$, the integral is dominated by small values of $s$ and $t$, thus, expanding for small $s$ and $t$, we find
\begin{equation}
p(x)\simeq\frac12 \frac{\Gamma\left(d/2\right)}{\sqrt{\pi}\Gamma\left((d-1)/2\right)}\frac{\alpha}{v_0}\gamma e^{-\gamma x/v_0} \int_{0}^{1}dt~e^{-\gamma x t/(2v_0)}t^{(d-3)/2}\int_{0}^{1}ds~e^{-\gamma x s/v_0}s^{\alpha-1}~ \,.
\label{Px_app5}
\end{equation}
Computing the integrals, we find
\begin{equation}
p(x)\simeq \frac{\Gamma\left(d/2\right)}{\sqrt{\pi}\Gamma\left((d-1)/2\right)}\frac{\alpha}{v_0}\gamma e^{-\gamma x/v_0}2^{(d-3)/2} \left(\frac{v_0}{\gamma x}\right)^{(d+2\alpha-1)/2}\Gamma\left(\frac{d-1}{2}\right)\Gamma\left(\alpha\right)~ \,.
\label{Px_app6}
\end{equation}
Finally, using the symmetry of $p(x)$, we find that for $|x|\gg 1$,
\begin{equation}
p(x)\simeq A_{d,\alpha} \frac{\gamma}{v_0} e^{-\gamma |x|/v_0} \left(\frac{v_0}{\gamma|x|}\right)^{(d+2\alpha-1)/2}~ \,,
\label{Px_app7}
\end{equation}
where 
\begin{equation}
A_{d,\alpha}= \frac{\Gamma\left(d/2\right)\alpha \Gamma\left(\alpha\right) }{\sqrt{\pi}}2^{(d-3)/2}\,.
\label{A_d_alpha_app}
\end{equation}

\section{Range of validity of the Central Limit Theorem}
\label{app:CLT}

Consider $N$ i.i.d. random variables $\{x_1,x_2,\cdots, x_N\}$
each drawn from a normalised distribution $p(x)$. The distribution
of their sum $X$ can be expressed as
\begin{equation}
Z(X,N)=\int_{-\infty}^{\infty}dx_1\,\ldots\int_{-\infty}^{\infty}dx_N
~\left[\prod_{i=1}^{N}p(x_i)\right]\delta\left(X-\sum_{i=1}^{N}x_i\right)\,.
\label{ZXN.1}
\end{equation}
Taking a Fourier transform factorises the $N$-fold integrals
\begin{equation}
\int_{-\infty}^{\infty} Z(X,N)\, e^{i\, k\, X}\, dX= 
\left[\hat{p}(k)\right]^N \, ,
\label{FoTr.1}
\end{equation}
where 
\begin{equation}
\hat{p}(k)=\int_{-\infty}^{\infty}p(x)\, e^{i\, k\, x}\, dx
\label{pq_transform}
\end{equation}
is the Fourier transform of $p(x)$. Finally, inverting the Fourier transform
in Eq. (\ref{FoTr.1}) one gets the integral representation
\begin{equation}
Z(X,N)=\int_{-\infty}^{\infty} \frac{dk}{2\pi}\, e^{-i\, k\, X}\,
\left[\hat{p}(k)\right]^N\,.
\label{Zapp}
\end{equation}
Note that this expression is valid for all $X$ and all $N$ and arbitrary
$p(x)$. Motivated by the RTP problem, we focus on $p(x)$'s that
are symmetric with a finite second moment $\sigma^2$. In this case,
the central limit theorem (CLT) is valid for large $N$ which predicts
that in the region up to $|X|\sqrt{N}$, the distribution
$Z(X,N)$ converges to a Gaussian shape for large $N$
\begin{equation}
Z(X,N)\simeq \frac{1}{\sqrt{2\pi\, \sigma^2\,N}}e^{-X^2/(2\,\sigma^2\,N)}\, .
\label{Gaussian_CLT}
\end{equation}
One may ask whether this Gaussian shape remains valid over a larger 
range, outside the region $X\sim \sqrt{N}$ of the validity of the CLT. 

To answer this question, we start from the integral representation
of $Z(X,N)$ in Eq. (\ref{Zapp}) which can 
be rewritten as
\begin{equation}
Z(X,N)=\int_{-\infty}^{\infty} \frac{dk}{2\pi}\,
e^{-i\, k\, X +N\,\log(\hat{p}(k))}\,.
\label{Zapp2}
\end{equation}
In order to probe the Gaussian regime where $X\sim \sqrt{N}$, 
we first set $y=X/\sqrt{N}$. 
Performing the change of variable $k\to k/\sqrt{N}$ in 
Eq. \eqref{Zapp} gives
\begin{equation}
Z(X,N)=\frac{1}{\sqrt{N}}\int_{-\infty}^{\infty}
\frac{dk}{2\pi}\, e^{-i\, k\, y+N\,\log(\hat{p}(k/\sqrt{N}))}\,.
\label{ZXN.2}
\end{equation}
Thus large $N$ limit probes the small $k$ behavior of $\hat{p}(k)$
defined in Eq. (\ref{pq_transform}).
We next assume that $\hat{p}(k)$ has the small $k$ expansion
\begin{equation}
\hat{p}(k)\simeq 1 - \frac{\sigma^2\, k^2}{2}+ c\, |k|^{\beta}+\cdots
\label{small_q.1}
\end{equation}
where $2<\beta\le 4$. Since we assumed $p(x)$ to be normalised to unity, the
first term is unity. Moreover, since $p(x)$ is symmetric, there is
no linear term in $\hat{p}(k)$ in the small $k$ expansion.
The second term is automatic since the variance $\sigma^2$ is finite
and the third correction term must appear with exponent $\beta>2$.
We also assume that $\beta\le 4$.
The prefactor $c$ of $|k|^\beta$ is just an unimportant nonzero constant.

Substituting the small $k$ expansion \eqref{small_q.1} in \eqref{ZXN.2} 
in the large $N$ limit we get, keeping only leading order terms up to 
$O(|k|^\beta)$ (note that $2<\beta\le 4$),
\begin{equation} 
Z(X,N)\simeq 
\frac{1}{\sqrt{N}}\int_{-\infty}^{\infty} \frac{dk}{2\pi}\, ~e^{-i\, k\, 
y- \sigma^2 k^2/2+c\, N^{1-\beta/2}\, |k|^{\beta}}\,, 
\label{ZXN.3} 
\end{equation} 
If $\beta<4$, then the prefactor of the third term inside 
the exponent is $c$. However, if $\beta=4$, then this prefactor will be 
slightly modified from $c$ since the expansion of the logarithm will give 
rise to a term of $O(k^4)$ also. But in any case, we just need that $c$ 
is some nonzero constant in Eq. (\ref{ZXN.3}). It is possible to check 
that in the case of the RTP model considered in Section 
\ref{sec:position}, one has $\beta=4$. Since $\beta>2$, the term $c 
N^{1-\beta/2}\, |k|^{\beta}$ is small for large $N$, and we can expand the 
exponential as 
\begin{equation}
Z(X,N)\simeq\frac{1}{2\pi}
\frac{1}{\sqrt{N}}\int_{-\infty}^{\infty} dk~e^{-i\, k\, y-\sigma^2 k^2/2}
\left(1+c\, N^{1-\beta/2}\,|k|^{\beta}\right)\,.
\label{ZXN.4}
\end{equation}
Performing the first integral gives the leading Gaussian term and
rearranging the second term slightly gives
\begin{equation}
Z(X,N)\simeq \frac{1}{\sqrt{2\pi N 
\sigma^2}}e^{-y^2/(2\sigma^2)}\left[1+\sqrt{2\pi\sigma^2}c 
N^{1-\beta/2}\int_{-\infty}^{\infty}dk~e^{-(\sigma^2/2) 
(k-iy/\sigma^2)^2 }|k|^{\beta}\right]\,. 
\label{ZXN.5}
\end{equation} 
Performing the 
change of variable $k\to k+iy/\sigma^2$, we obtain 
\begin{equation} 
Z(X,N)\simeq \frac{1}{\sqrt{2\pi N 
\sigma^2}}e^{-y^2/(2\sigma^2)}\left[1+\sqrt{2\pi\sigma^2}c 
N^{1-\beta/2}\int_{-\infty}^{\infty}dk~e^{-(\sigma^2/2) k^2 
}|k+iy/\sigma^2|^{\beta}\right]\,. 
\label{ZXN.6}
\end{equation} 

When $y$ is of order 
one, i.e. when $X\sim O(\sqrt{N})$, the correction term vanishes as 
$N^{1-\beta/2}$ and we obtain the leading Gaussian term, as predicted by the 
CLT. On the other hand, for $y\gg 1$, the second integral over $k$ can be 
approximated to leading order for large $y$ as 
\begin{equation} 
Z(X,N)\simeq \frac{1}{\sqrt{2\pi N 
\sigma^2}}e^{-y^2/(2\sigma^2)}\left[1+\tilde{c} 
N^{1-\beta/2}y^{\beta}\right]\,, 
\label{ZXN.7}
\end{equation} where $\tilde{c}$ is just a 
constant. The correction term can be neglected when 
$N^{1-\beta/2}y^{\beta}\ll 1$ and therefore the CLT is valid for any $y$ 
such that $y\ll N^{(\beta-2)/(2\beta)}$. Recalling that $y=X/\sqrt{N}$, 
we obtain that the CLT is valid up to a wider range than $\sqrt{N}$, namely, up to 
\begin{equation}
|X|\ll N^{(\beta-1)/\beta} \, . 
\label{range.1}
\end{equation}
For 
instance, for the RTP model considered in Section \ref{sec:position}, we 
have $\beta=4$ and thus the CLT is valid for $|X|\ll N^{3/4}$.

\section{Large deviation for the position distribution in the $x$ direction} 

\label{app:large_dev}

In this Appendix we give a formula for the large deviation of the $x$ coordinate 
in the fixed $T$ ensemble of the RTP, valid for a model with an arbitrary distribution of velocity, using
an equivalent but slightly different method as in the text. 
Let us recall that the displacements $x_i$ along the $x$ axis, and the durations $\tau_i$ associated to the
$i$-th running phase are i.i.d. variables, except for the last run which is incomplete and hence has a different distribution. Their distribution is $P(x,\tau)= p(x|\tau) \gamma e^{-\gamma \tau}$, 
with $p(x|\tau) = \int d^d \vec v P(\vec v) \delta(x-v_1 \tau) = \langle \delta(x-v_1 \tau) \rangle$,
where $v_1=\vec v \cdot e_x$ denotes the first component of the velocity, and here and below
$\langle \cdots \rangle$ denotes an average with respect to the distribution of the velocity $\vec v$ (here assumed to be quite general). Thus one can write 
\be
\hat p(q,s)= \int_{-\infty}^{+\infty} dx \int_0^{+\infty} d\tau~ e^{- q x - s \tau } \langle \delta(x-v_1 \tau) \rangle 
\gamma e^{-\gamma \tau} = \left\langle \frac{\gamma}{\gamma + s + q v_1}  \right\rangle\,.
\ee
For an isotropic distribution with $|\vec v|=v_0$ in dimension $d$, the explicit form is given in 
\eqref{eq:hat_p2} in the text.

Let us start from Eq. \eqref{eq:PDF_X3} and set $\gamma=1$ for simplicity. Eq. \eqref{eq:PDF_X3} can be written as
\be
\int_0^{+\infty} dT \int_{-\infty}^{+\infty} dX e^{-s T - q X } Z(X,T) = \frac{\hat p(q,s)}{1- \hat p(q,s)}\,. \label{FT} 
\ee
Let us first assume that $Z(X,T)$ admits a large deviation form $Z(X,T) \sim e^{- T \phi_d(z=X/T)}$. Inserting this form on the left-hand side of Eq. \eqref{FT}, we get 
\be
\int_0^{+\infty} dT~ T \int_{-\infty}^{+\infty} dz ~e^{-s T - q T z - T \phi_d(z)} \sim
\int_0^{+\infty} dT ~T e^{-s T - T \min_{z \in \mathbb{R}} (q z +  \phi_d(z)) } \,,~
\ee
where we used a saddle point estimate in the integral over $z=X/T$. This integral becomes divergent
when $s$ decreases and reaches
\be  \label{rel1n} 
s= s(q) := - \min_z [ q z + \phi_d(z) ] \,.
\ee  
Now looking at  the right-hand side of Eq. \eqref{FT} we see that we expect a singularity when $\hat p(q,s)=1$. We will 
surmise that these singularities are the same 
\be
\hat p(q,s)=1 \quad \Leftrightarrow \quad s = s(q) \,.
\ee 
More explicitly the function $s(q)$ is the root $s=s(q)$ of the equation
\be \label{equas} 
\left\langle \frac{1}{1 + s + q v_1}  \right\rangle = 1 \quad \Leftrightarrow ~~\left\langle \frac{s+ q v_1}{1 + s + q v_1} \right\rangle = 0\,.
\ee
Once $s(q)$ is known, the inversion of \eqref{rel1n} determines the large deviation function
\be \label{inversion} 
 \phi_d(z) = \max _q (- q z - s(q))\,.
\ee 

These formulae allow to easily determine the small $z$ behavior of the large deviation function as
a function of the moments of the random variable $v_1$, assuming that they exist. Expanding 
\eqref{equas} in powers of $q$ to second order one obtains $\phi_d(z)$ to quadratic order
\be
s(q) = - \langle v_1 \rangle q + \langle (v_1 - \langle v_1 \rangle)^2 \rangle q^2 + O(q^3) \quad \Rightarrow \quad \phi_d(z) = \max _q (- q z - s(q)) = \frac{(z+ \langle v_1 \rangle)^2}{4 \langle v_1^2 \rangle} + O(z^3) 
\ee
where we have not assumed any symmetry of the distribution of $\vec v$. When the distribution of 
$v_1$ is symmetric in $v_1 \to - v_1$, it is convenient to symmetrize \eqref{equas} and 
rewrite it as $\langle \frac{s(1+s) - q^2 v_1^2}{(1+s)^2 - q^2 v_1^2} \rangle =0$. One 
obtains using Mathematica $s(q)=c_2 q^2 + (c_4-2 c_2^2) q^4
+ (c_6 - 6 c_2 c_4 + 7 c_2^3) q^6 + O(q^8)$ and
\be
\phi_d(z) = 
\frac{z^2}{4 c_2}-\frac{\left(c_4-2 c_2^2\right) z^4}{16 c_2^4}+\frac{\left(9 c_2^4-10
   c_4 c_2^2-c_6 c_2+4 c_4^2\right) z^6}{64 c_2^7}+O\left(z^8\right) \quad , \quad c_n := \langle v_1^n \rangle \,.
   \ee
Note that for an isotropic distribution of $\vec v$, with $\langle \vec v^2 \rangle=1$,
$c_2 = 1/d$ and $\phi_d(z) =  \frac{d}{4} z^2 + O(z^4)$, as in Eq. \eqref{phi_small}.

The saddle point equation \eqref{equas} can be conveniently rewritten by performing the change of variable
$w=q/(1+s)$ and introducing the function
\be
F(w) := \left\langle \frac{1}{1 + v_1 w} \right
\rangle \,.
\label{Fw}
\ee
Simple manipulations then lead to $\phi_d(z) = \max _q (- q z - s(q))  = \max_w ( 1 - (1+ z w) F(w) )$.
The function $\phi_d(z)$ can then be obtained in a parametric form (by eliminating $w$)
\be
z = - \frac{F'(w)}{F(w) + w F'(w)} \quad , \quad \phi_d(z) = 1 - (1+ z w) F(w) = 1 - \frac{F(w)^2}{F(w) + w F'(w)} 
\ee
where one can alternatively use the simpler formula $\phi_d'(z)= - w F(w)$.

For an isotropic distribution of velocities with $|\vec v|=1$ in dimension $d$, one has 
$F(w)= {}_2F_1\left(\frac{1}{2},1;\frac{d}{2};w^2\right)$ and one recovers the
formula \eqref{eq:Sqz}, \eqref{eq:cond2} and 
\eqref{eq:g} given in the text (where the variable $w$ is $q$ there). They are valid as
long as $|w|<1$, beyond which the saddle point value freezes at $w=\pm 1$, as discussed in the text. 

We have assumed so far that the function $F(w)$, defined in Eq. \eqref{Fw}, exists. Of 
course there are some distributions $P(\vec v)$ for which the average in Eq. \eqref{Fw} may 
not exist. In fact, for any distribution $P(\vec v)$ which is nonzero at $v_1=-1/w$, the
average in Eq. (\ref{Fw}) is divergent. 
An example of this is simply the Gaussian distribution $P(\vec v)=e^{-{\vec v}^2/2}/(2\pi )^{d/2}$. 
Recall that in the main text we have chosen the direction isotropically and taken the speed 
distribution $W(v)=\alpha(1-v)^{\alpha-1}$, which has a finite support $v\in (0,1)$. In this 
example the average in Eq. \eqref{Fw} is well defined. In cases where $F(w)$ in Eq. 
\eqref{Fw} does not exist, it indicates that the distribution does not admit a 
large-deviation form on a scale $X\sim O(T)$, as assumed. In this case, a condensation 
may still occur, but at a smaller scale $X\sim T^{\gamma}$ with $1/2<\gamma<1$.

\end{document}